\documentclass[notitlepage,tightenlines,nofootinbib,superscriptaddress,aps,pra,twocolumn]{revtex4-2}
\pdfoutput=1

\usepackage[T2A]{fontenc}
\usepackage[utf8]{inputenc}
\usepackage[russian,english]{babel}
\newcommand{\cyrchar}[1]{\foreignlanguage{russian}{#1}}

\usepackage{amsthm}

\usepackage{newpxtext,newpxmath}

\usepackage{amsthm,amsmath,amssymb,amsbsy}
\usepackage{mathtools,mathrsfs,bbding}
\usepackage{bbm}
\usepackage{braket}
\usepackage{enumerate}
\usepackage[shortlabels]{enumitem}
\usepackage{graphicx}
\usepackage{stackengine}
\usepackage{scalerel}
\usepackage{tensor}       
\usepackage[left=.73in,right=.73in,top=.7in,bottom=1in]{geometry}
\usepackage{graphicx}
\usepackage[dvipsnames,svgnames]{xcolor}
\usepackage{array}
\usepackage{makecell}
\newcolumntype{x}[1]{>{\centering\arraybackslash}p{#1}}
\usepackage{tikz}
\usepackage{pgfplots}
\usepackage{booktabs}
\usepackage{xfrac}
\usepackage{siunitx}
\usepackage{centernot}
\usepackage{comment}
\usepackage{chngcntr}

\usepackage[pdftex]{hyperref}
\hypersetup{
    bookmarksnumbered=true, 
    breaklinks=true,
    unicode=false, 
    pdfstartview={FitH}, 
    pdfnewwindow=true, 
    colorlinks=true, 
    allcolors=MidnightBlue!70!black!70!TealBlue,
    pdftitle={Beyond Hoeffding and Chernoff: Trading conclusiveness for advantages in quantum hypothesis testing}
}

\theoremstyle{plain}

\newtheorem{theorem}{Theorem}
\newtheorem{conjecture}[theorem]{Conjecture}

\newtheorem{proposition}[theorem]{Proposition}
\newtheorem{corollary}[theorem]{Corollary}
\newtheorem{lemma}[theorem]{Lemma}

\theoremstyle{definition}

\newtheorem*{remark}{Remark}

\let\oldproofname\proofname
\renewcommand{\proofname}{\rm\bf{\oldproofname}}

\newcommand{\bb}{\begin{equation}\begin{aligned}\hspace{0pt}}
\newcommand{\bbb}{\begin{equation*}\begin{aligned}}
\newcommand{\ee}{\end{aligned}\end{equation}}
\newcommand{\eee}{\end{aligned}\end{equation*}}
\newcommand\floor[1]{\lfloor#1\rfloor}

\renewcommand{\epsilon}{\varepsilon}

\newcommand{\id}{\mathbbm{1}}

\newcommand{\ve}{\varepsilon}

\DeclareMathOperator{\Tr}{Tr}

\DeclareMathOperator{\supp}{supp}

\let\nc\newcommand
\renewcommand{\S}{\mathcal{S}}
\nc{\R}{\mathcal{R}}
\nc{\W}{\mathcal{W}}
\nc{\T}{\mathcal{T}}
\nc{\B}{\mathcal{B}}

\newcommand{\E}{\mathcal{E}}
\nc{\EE}{\mathbb{E}}
\nc{\K}{\mathcal{K}}
\nc{\V}{\mathcal{V}}
\newcommand{\X}{\mathcal{X}}
\nc{\F}{\mathcal{F}}
\nc{\D}{\mathcal{D}}
\nc{\Y}{\mathcal{Y}}
\renewcommand{\P}{\mathcal{P}}
\nc{\M}{\mathcal{M}}

\nc{\I}{\mathcal{I}}
\nc{\Q}{\mathbb{Q}}
\nc{\RR}{\mathbb{R}}
\nc{\CC}{\mathbb{C}}

\nc{\HH}{\mathbb{H}}
\nc{\MM}{\mathbb{M}}
\nc{\DD}{\mathbb{D}}
\newcommand{\NN}{\mathbb{N}}

\nc{\J}{\mathcal{J}}

\newcommand{\lsmatrix}{\left(\begin{smallmatrix}}
\newcommand{\rsmatrix}{\end{smallmatrix}\right)}

\newcommand{\deff}[1]{\textbf{\emph{#1}}}

\stackMath
\newcommand\xxrightarrow[2][]{\mathrel{%
  \setbox2=\hbox{\stackon{\scriptstyle#1}{\scriptstyle#2}}%
  \stackunder[5pt]{%
    \xrightarrow{\makebox[\dimexpr\wd2\relax]{$\scriptstyle#2$}}%
  }{%
   \scriptstyle#1\,%
  }%
}}

\newcommand{\tendsn}{\xxrightarrow[\! n\rightarrow \infty\!]{}}
\newcommand{\tendsp}[1]{\xxrightarrow[\! n\rightarrow \infty\!]{#1}}

\stackMath

\makeatletter
\newcommand*\rel@kern[1]{\kern#1\dimexpr\macc@kerna}
\newcommand*\widebar[1]{%
  \begingroup
  \def\mathaccent##1##2{%
    \rel@kern{0.8}%
    \overline{\rel@kern{-0.8}\macc@nucleus\rel@kern{0.2}}%
    \rel@kern{-0.2}%
  }%
  \macc@depth\@ne
  \let\math@bgroup\@empty \let\math@egroup\macc@set@skewchar
  \mathsurround\z@ \frozen@everymath{\mathgroup\macc@group\relax}%
  \macc@set@skewchar\relax
  \let\mathaccentV\macc@nested@a
  \macc@nested@a\relax111{#1}%
  \endgroup
}

\newcommand{\fakepart}[1]{
 \par\refstepcounter{part}
  \sectionmark{#1}
}

\counterwithin*{equation}{part}
\counterwithin*{theorem}{part}
\counterwithin*{figure}{part}
\counterwithin*{section}{part}

\allowdisplaybreaks

\let\nc\newcommand
\nc{\proj}[1]{\ket{#1}\!\bra{#1}}
\renewcommand{\bar}{\;\rule{0pt}{9.5pt}\right|\;}
\nc{\lset}{\left\{\left.}
\nc{\rset}{\right\}}
\nc{\lsetr}{\left\{\,}
\nc{\rsetr}{\right.\right\}}
\nc{\barr}{\;\rule{0pt}{9.5pt}\left|\;}

\newcommand{\pbeta}{\overline{\beta}}
\newcommand{\palpha}{\overline{\alpha}}

\newcommand{\pconc}{\pi_n}

\nc{\wt}{\widetilde}

\usepackage{tikz}
\usetikzlibrary{decorations.pathmorphing}
\usetikzlibrary{arrows.meta}

\newcommand{\widetildearrow}[1]{%
  \smash{%
  \overset{%
    \begin{tikzpicture}[baseline=0ex]
      \draw[{Stealth[length=1.5pt, width=3pt, inset=0.3pt]}-{Stealth[length=1.5pt, width=3pt, inset=0.3pt]},
            line width=0.5pt,  
            decorate, 
            decoration={snake, amplitude=0.5pt, segment length=6pt}] 
        (0,0) -- (0.76em,0);
    \end{tikzpicture}%
  }{#1}%
  }%
  \vphantom{\widetilde{#1}}
}
\nc{\rev}[1]{{\widetildearrow{#1}}}

\nc{\pavg}{\pi^{({\rm avg})}}
\nc{\erravg}{\operatorname{err}^{(\rm avg)}}
\nc{\errmin}{\operatorname{err}^{(\rm max)}}
\nc{\perravg}{\overline{\operatorname{err}}^{(\rm avg)}}
\nc{\perrmin}{\overline{\operatorname{err}}^{(\rm max)}}

\nc{\s}{\acute s}

\newcommand{\type}[1]{t_{#1}}

\nc{\norm}[2]{\left\lVert#1\right\rVert_{\,#2}}
\renewcommand{\proj}[1]{\ket{#1}\!\bra{#1}}
\nc{\pro}[1]{#1 #1^\dagger}
\nc{\lnorm}[2]{\left\lVert#1\right\rVert_{\ell_{#2}}}

\renewcommand{\Pr}{\mathbb{P}}

\let\textgeq\relax
\let\texteq\relax

\newcommand{\texteq}[1]{\stackrel{\mathclap{\mbox{\scriptsize #1}}}{=}}

\newcommand{\textgeq}[1]{\stackrel{\mathclap{\mbox{\scriptsize #1}}}{\geq}}

\makeatletter
\renewenvironment{proof}[1][\proofname]{\par
\pushQED{\qed}%
\normalfont \topsep6\p@\@plus6\p@\relax
\trivlist
\item\relax
{\bfseries  
#1\@addpunct{.}}\hspace\labelsep\ignorespaces 
}{%
\popQED\endtrivlist\@endpefalse
}
\makeatother

\makeatletter 
\renewcommand\onecolumngrid{
\do@columngrid{one}{\@ne}%
\def\set@footnotewidth{\onecolumngrid}
\def\footnoterule{\kern-6pt\hrule width 1.5in\kern6pt}%
}
\makeatother 

\usepackage[most,breakable]{tcolorbox}
\renewenvironment{boxed}[1][white]%
  {\expandafter\ifstrequal\expandafter{#1}{filled}{\begin{tcolorbox}[colback=MidnightBlue!70!black!70!TealBlue!2!white,colframe=MidnightBlue!70!black!70!TealBlue!30!white,breakable,enhanced,left=5.75pt,right=5.75pt,grow sidewards by=10pt]}{\begin{tcolorbox}[colback=white,colframe=gray!15,breakable,enhanced,left=5.75pt,right=5.75pt,grow sidewards by=10pt]}}%
  {\end{tcolorbox}}

\newcommand{\mclose}{\mathclose{}}
\newcommand{\fleft}{\mathopen{}\left}
\newcommand{\fright}{\aftergroup\mclose\right}

\newcommand{\Hsc}{H^*}

\makeatletter
\patchcmd{\@ssect@ltx}
    {\addcontentsline{toc}{#1}{\protect\numberline{}#8}}
    {}
    {}
    {}
\def\l@subsection#1#2{}
\def\l@subsubsection#1#2{}
\makeatother

\usepackage{epigraph}

\begin{document}


\title{%
\texorpdfstring{Beyond Hoeffding and Chernoff:\\Trading conclusiveness for advantages in quantum hypothesis testing}{Beyond Hoeffding and Chernoff: Trading conclusiveness for advantages in quantum hypothesis testing}%
}

\author{Kaiyuan Ji}
\email{kj264@cornell.edu}
\affiliation{School of Electrical and Computer Engineering, Cornell University, Ithaca, New York 14850, USA}

\author{Bartosz Regula}
\email{bartosz.regula@gmail.com}
\affiliation{Mathematical Quantum Information RIKEN Hakubi Research Team, RIKEN Pioneering Research Institute (PRI) and RIKEN Center for Quantum Computing (RQC), Wako, Saitama 351-0198, Japan}


\begin{abstract}
The ultimate limits of quantum state discrimination are often thought to be captured by asymptotic bounds that restrict the achievable error probabilities, notably the quantum Chernoff and Hoeffding bounds.
Here we study hypothesis testing protocols that are permitted a probability of producing an inconclusive discrimination outcome, and investigate their performance when this probability is suitably constrained.
We show that even by allowing an arbitrarily small probability of inconclusiveness, the limits imposed by the quantum Hoeffding and Chernoff bounds can be significantly exceeded. This completely circumvents the conventional trade-offs between error exponents in hypothesis testing while incurring only a vanishingly small overhead over conventional approaches. 
Such improvements over standard state discrimination are robust and can be obtained even when an exponentially vanishing probability of inconclusive outcomes is demanded. 
Relaxing the constraints on the inconclusive probability can enable even larger advantages, but this comes at a price.  We show a `strong converse' property of this setting: targeting error exponents beyond those achievable with vanishing inconclusiveness necessarily forces the probability of inconclusive outcomes to converge to one. By exactly quantifying the rate of this convergence, we give a complete characterisation of the trade-offs between error exponents and rates of conclusive outcome probabilities.
Overall, our results provide a comprehensive asymptotic picture of how the allowance for inconclusive measurement outcomes reshapes optimal quantum hypothesis testing.
\end{abstract}

\maketitle


\let\oldaddcontentsline\addcontentsline
\renewcommand{\addcontentsline}[3]{}


\section{Introduction}\label{sec:intro}


Binary quantum hypothesis testing, or more simply quantum state discrimination, is one of the most fundamental problems in quantum information that underlies both the theory of quantum information processing as well as many of its applications. Beyond direct technological applications such as quantum sensing and metrology~\cite{helstrom_1969,lloyd_2008,giovannetti_2006}, the precise understanding of hypothesis testing is key to quantifying the performance and reliability of quantum communication~\cite{hayashi_2003,hayashi_2016,wang_2012,leung_2015,anshu_2019-1,cheng_2023-1,khatri_2020}, and it precisely characterises the rates of quantum thermodynamic work extraction~\cite{brandao_2013,horodecki_2013,yungerhalpern_2016} or optimal schemes for the distillation of resources such as quantum entanglement~\cite{vedral_1998,brandao_2010-1,liu_2019,regula_2020,lami_2024-1}.

The performance of quantum hypothesis testing is typically studied in terms of asymptotic error exponents: given an increasing number of copies of the quantum state to be discriminated, how fast can we make the discrimination error probability go to zero?~\cite{hiai_1991,ogawa_2000,ogawa_2004}  Many bounds exist that impose restrictions on the achievable discrimination power. Two prominent and frequently employed examples of such bounds are the quantum Hoeffding bound~\cite{audenaert_2007,nussbaum_2009,hayashi_2007,nagaoka_2006,audenaert_2008}, which says that there is an inherent trade-off between the errors in asymmetric hypothesis testing, and the quantum Chernoff bound~\cite{audenaert_2007,nussbaum_2009}, which limits the asymptotic errors achievable in symmetric hypothesis testing. These bounds are often considered to constitute the ultimate restrictions on the performance of hypothesis testing~\cite{hayashi_2016,khatri_2020}.

In this work, we put the fundamental character of these restrictions into question by asking: what does it truly cost to beat  these bounds? As we will see, there exist many ways to exceed Hoeffding and Chernoff bounds while incurring only vanishingly small overheads, and even more extensive improvements are possible by accepting larger costs, which we precisely characterise. Our approach 
relies on a simple modification of the conventional setting of hypothesis testing: we allow 
for discrimination protocols that may, with a suitably constrained probability, fail to conclusively distinguish  the two hypotheses, resulting in an abstention rather than making a guess. 
We introduce a general and comprehensive framework to characterise the landscape of advantages that can be obtained through this approach, quantifying exactly how much of this `inconclusiveness' is needed to achieve desired ranges of error exponents, establishing optimal trade-offs, and delineating the limits of possible improvements.

Our analysis can be broadly divided into two regimes. The first is that of low inconclusiveness, where a very high probability of obtaining a conclusive outcome is required and thus minimal costs are incurred. 
Crucially, we show that even an arbitrarily small probability of inconclusive outcomes already suffices to surpass the limitations 
of the Chernoff and Hoeffding bounds. 
In particular, the trade-offs imposed by the the Hoeffding bound are completely overcome in this setting, with both of the discrimination errors reaching their optimal decay rates simultaneously. 
What this means is that one does not need to trade conclusiveness of the protocol for higher performance on equal terms: an utmost improvement in the achievable error exponents is attained while making only an infinitesimally small sacrifice in conclusiveness, meaning that inconclusive outcomes are vanishingly unlikely and thus the overhead incurred (e.g., by requiring repeated experiments due to the inconclusive results) is negligible compared to the improvement it brings.
Perhaps even more strikingly, we demonstrate protocols that only allow for an exponentially small probability of inconclusiveness, while still exhibiting higher performance than conventional hypothesis testing.
These results reveal that the traditional bounds are not insurmountable limits, but rather artifacts of requiring perfect decisiveness --- just a touch of inconclusiveness is enough to exceed them in a number of ways. 

To study the limits of such advantages and establish the optimality of our results, we then investigate the opposite regime of high inconclusiveness. This can be understood as quantifying
the cost of demanding even higher performance in state discrimination, inspired by the analysis of strong converse results in quantum information theory~\cite{ogawa_2000,mosonyi_2015}. 
In this context, 
we first show that additional advantages can be gained by only sacrificing the conclusiveness of one of the two hypotheses, while retaining an arbitrarily low inconclusive probability under the other hypothesis.
More generally, we introduce an exact and complete characterisation of the trade-offs between the error exponents that can be achieved when the protocols are conclusive and the 
rate at which the conclusive probability must decay.
We show that these trade-offs are governed by a class of divergences based on sandwiched R\'enyi relative entropies~\cite{muller-lennert_2013,wilde_2014}, revealing new applications of these quantities.

\subsection{Prior works}

The discrimination of quantum states with inconclusive outcomes has attracted attention before~\cite{ivanovic_1987,dieks_1988,peres_1988,chefles_1998, fiurasek_2003, rudolph_2003, croke_2006, herzog_2009, bagan_2012, zhuang_2020, barnett_2009, bae_2015}, although it was mostly studied only in non-asymptotic settings, motivated in particular by the problem of unambiguous state discrimination.
Problems related to the setting of asymptotic, probabilistic hypothesis testing that we employ here have been studied in classical statistics and information theory under various names, most commonly hypothesis testing with rejection or with abstention~\cite{nikulin_1989,gutman_1989,grigoryan_2011,sason_2012}. 
Our setting also shares similarities with sequential hypothesis testing~\cite{wald_1945,martinezvargas_2021,li_2022}, where the number of samples required 
by a test in the discrimination protocol is not fixed and can be adjusted adaptively. 
Remarkably, it was previously shown that sequential methods also allow one to exceed conventional limitations on the error exponents of quantum hypothesis testing~\cite{martinezvargas_2021,li_2022}.
We will indeed exploit this connection with sequential hypothesis testing for some of our achievability results. 
However, our approach is often amenable to simpler analysis and will allow us to gain additional insights and significantly more general extensions through the different regimes studied in this work. 
Parts of our investigation also share features with `almost-fixed-length hypothesis testing'~\cite{lalitha_2016} studied in classical information theory, which interpolates between conventional and sequential hypothesis testing; the precise approach here is however more general.
In particular, to the best of our knowledge, the connections with strong converse exponents have not been explored even classically.

We emphasise also that our results differ from some recent works which showed that conventional restrictions on hypothesis testing do not apply when postselection is allowed for free, i.e.\ when the probability of inconclusive outcomes is not taken into consideration when evaluating the performance of a postselected protocol~\cite{regula_2024,regula_2024-1}. Those approaches are inherently different from conventional hypothesis testing, which complicates direct comparisons of achievable rates.
Here we show an improvement over conventional bounds even when the overhead resulting from the probabilistic character of the protocols is appropriately constrained and accounted for, and in particular demonstrate that major advantages can be obtained with \emph{no} postselection needed whatsoever. Furthermore, our strong-converse--style results apply in the complementary regime of high inconclusiveness, which more closely connects with the postselected setting. These results refine the understanding of postselected hypothesis testing, characterising exactly the rate of vanishing conclusiveness --- in a sense, the `amount of postselection' --- needed to achieve very high error~exponents, allowing for an interpolation between postselected and conventional approaches.


\subsection{Structure of the paper}

In the main text, we provide an introduction to the setting of hypothesis testing with inconclusive outcomes, a detailed discussion of the difficulties associated with understanding its asymptotic error exponents, and an overview of our methods and main findings. We divide our discussion into two parts. First, in Section~\ref{sec:asymmetric} we discuss asymmetric hypothesis testing (Hoeffding setting), where one is interested in studying the trade-offs between the different kinds of errors one can make in distinguishing quantum states; in particular, Section~\ref{subsec:setting} serves as an introduction to the setting of our work and clarifies its connection with conventional hypothesis testing. 
Then, in Section~\ref{sec:symmetric} we discuss symmetric hypothesis testing (Chernoff setting), where one instead bounds the asymptotic behaviour of the average error probability.

The technical statements in the main text are sometimes simplified for the sake of providing intuitive explanations. Full details and rigorous proofs can be found in the \hyperlink{app}{Appendix}.


\section{Asymmetric hypothesis testing}\label{sec:asymmetric}


\subsection{Setting}\label{subsec:setting}

In the setting of quantum hypothesis testing, we are given a source which emits independent and identically distributed (i.i.d.) copies of an unknown quantum state that could be either $\rho$ or $\sigma$. It is our task to determine which of the two states is the true hypothesis as efficiently as possible.

In conventional hypothesis testing, the discrimination is done using two-outcome measurements (tests): given $n$ copies of the unknown state, we perform a global measurement defined by the POVM elements $(M_n, N_n)$ where $N_n = \id - M_n$. Upon obtaining the first outcome of the measurement we guess $\rho$, and upon obtaining the second outcome we guess $\sigma$. There are then two types of errors that can be made here: either incorrectly guessing $\sigma$, which is known as the \emph{type I error} and whose probability is
\begin{equation}\begin{aligned}
    \alpha_n(M_n, N_n) 
    \coloneqq \Tr N_n \rho^{\otimes n},
\end{aligned}\end{equation}
or incorrectly guessing $\rho$, which is known as the \emph{type II error}, denoted 
\begin{equation}\begin{aligned}
    \beta_n(M_n, N_n) 
    \coloneqq \Tr M_n \sigma^{\otimes n}.
\end{aligned}\end{equation}
We will omit the measurement in the notation and simply use $\alpha_n$ and $\beta_n$ to refer to the two error probabilities. 
When more and more copies of the unknown state are available, we expect the errors to vanish asymptotically: $\alpha_n \sim \exp(-n A)$ and $\beta_n \sim \exp(-n B)$ for some exponents $A$ and $B$.\footnote{We use the notation $\alpha_n \sim \exp(-n A)$ to denote the fact that $-\frac1n \log \alpha_n = A + o(1)$.} Determining these exponents and understanding the optimal trade-offs between them is the key goal of the study of hypothesis testing.

The key modification that we will consider here is to allow a third measurement outcome which represents an inconclusive discrimination --- in other words, an abstention instead of making a guess. We then consider three-outcome tests of the form $(M_n, N_n, \id - M_n - N_n)$.
Here, a guess is only made when the measurement yields a conclusive outcome, and thus a natural figure of merit is the probability of error when the third outcome is \emph{not} obtained.  This is quantified by the conditional error probabilities
\begin{equation}\begin{aligned}
    \palpha_n(M_n, N_n) &\coloneqq \frac{\Tr N_n \rho^{\otimes n}}{\Tr (M_n + N_n) \rho^{\otimes n}},\\
    \pbeta_n(M_n, N_n) &\coloneqq \frac{\Tr M_n \sigma^{\otimes n}}{\Tr (M_n + N_n) \sigma^{\otimes n}}.
\end{aligned}\end{equation}

Such a figure of merit highlights a difference between this setting and conventional hypothesis testing, as the latter is always conclusive. 
In order to compare the performance of such inconclusive protocols with conventional ones on equal ground, one can easily turn any discrimination scheme with inconclusiveness into a fully conclusive one: 
 perform the measurement and, if an inconclusive outcome is obtained, simply collect more samples and repeat the measurement. 
Of course, the need for this repetition means that additional overheads are incurred in the number of copies of states used in the discrimination procedure. To understand the implementation cost of such a probabilistic discrimination scheme, we then need to precisely characterise the behaviour of the probability of the measurement yielding a conclusive outcome, which is
\begin{equation}\begin{aligned}
    \pconc(\rho) &\coloneqq \Tr (M_n + N_n) \rho^{\otimes n},\\
    \pconc(\sigma) &\coloneqq \Tr (M_n + N_n) \sigma^{\otimes n},
\end{aligned}\end{equation}
where we again omit the measurement $(M_n, N_n)$ from the notation and treat the dependence on the measurements as implicit.
Consider for instance a case where the conditional discrimination error scales as $\palpha_n \sim \exp(-n A)$ and the corresponding probability of conclusiveness satisfies $\pconc(\rho) = p$ for some constant $p \in (0,1)$. 
Repeating each inconclusive attempt entails having to run each step $\frac{1}{p}$ times on average before obtaining a conclusive outcome, resulting in an expected total of $n' = n/p$ copies of the unknown state being consumed.  Since the error probability then scales as $\alpha_n=\palpha_n\pi_n(\rho)\sim\exp( - n' p A )$, the de facto rate at which it decays with respect to the overall number of copies $n'$ is given by $p A$.  We refer to this rate, namely the product of the conclusive probability and the conditional error exponent, as the \emph{effective error exponent} of a hypothesis testing protocol.  The effective error exponent enables a practically fair comparison between protocols with inconclusive outcomes and conventional, deterministic protocols, as it takes into account all costs directly or indirectly induced by the discrimination procedure before a conclusion is eventually made.

An important observation here is that, when the probability of conclusiveness is sufficiently high, there is effectively no difference between the two settings in terms of their practical implementation: if we can take $p \to 1$, then 
 the difference between the effective error exponent $p \palpha_n$ and the conditional error exponent $\palpha_n$ disappears, and therefore the latter already faithfully characterizes the performance of the protocol --- it can be implemented with an overhead cost that vanishes asymptotically.
As we will see, this is indeed often possible in practice.


\subsection{Error exponent trade-offs}

A fundamental result in conventional hypothesis testing is Stein's lemma~\cite{stein_unpublished,chernoff_1956} and its quantum generalisation~\cite{hiai_1991,ogawa_2000}. It tells us that if the type I error probability is assumed to be some small constant, $\alpha_n \leq \ve$, then the optimal type II error probability satisfies $\beta_n \sim \exp(-nD(\rho\|\sigma))$, where $D(\rho\|\sigma)$ is the (Umegaki) quantum relative entropy~\cite{umegaki_1962}.  Conversely, constraining the type II error probability by a constant $\ve$ means that the best achievable type I error probability is $\alpha_n \sim \exp(-nD(\sigma\|\rho))$.

The constant $\ve$ here can be made as small as desired, ensuring that both of the errors are in a sense small. 
This does not, however, guarantee an exponential decay of both errors, which is the more desirable behaviour in an asymptotic setting where more and more copies of the quantum states to be discriminated are available.
Crucially, should such exponential decay be required, it is no longer possible in conventional hypothesis testing to simultaneously achieve exactly the exponents given by the relative entropies $D(\rho\|\sigma)$ and $D(\sigma\|\rho)$: there is necessarily a trade-off between the achievable exponents. 

Specifically, say that we ask that $\alpha_n \sim \exp(-n A)$ and $\beta_n \sim \exp(-n B)$. What values of $(A, B)$ are achievable in this case? This trade-off is characterised by the Hoeffding bound~\cite{hoeffding_1967,blahut_1974,hayashi_2007,nagaoka_2006,audenaert_2008}, which tells us that hypothesis testing is possible if and only if $B \leq H_A(\sigma\|\rho)$ for a fixed $A$, or equivalently if $A \leq H_B(\rho \| \sigma)$ with $B$ fixed, where $H$ denotes the \emph{quantum Hoeffding divergence}
\begin{equation}\begin{aligned}\label{eq:hoeffding_div}
    H_A(\rho \| \sigma) \coloneqq \sup_{s \in (0,1)} \frac{s-1}{s} \Big( A - D_s(\rho  \| \sigma) \Big).
\end{aligned}\end{equation}
Here $D_s(\rho\|\sigma) = \frac{1}{s-1} \log \Tr \rho^{s} \sigma^{1-s}$ is the Petz--R\'enyi divergence~\cite{petz_1986}.
For any $A > 0$, we have $H_A(\sigma\|\rho) < D(\rho\|\sigma)$, enforcing a strict trade-off. The constraint arising from this is shown in Figure~\ref{fig:hoeffding}.


\subsection{Beyond Hoeffding: classical case}\label{sec:asymmetric_classical}

To see intuitively how the limitation of Hoeffding's bound can be overcome with just a little bit of inconclusiveness, it is instructive to first look at the classical case, where the quantum states $\rho$ and $\sigma$ are replaced by probability distributions $P$ and $Q$. When multiple i.i.d.\ samples of such distributions are available, their behaviour can be fully determined by their empirical distributions, also known as \emph{types}~\cite{csiszar_2011}. The type $t_{x^n}$ of a sequence $x^n$ of $n$ symbols from a finite alphabet is defined as $t_{x^n}(x) = \frac{1}{n} N(x | x^n)$, where $N(x|x^n)$ is the number of times that the symbol $x$ appears in the sequence $x^n$. 
For any probability distribution $P$ and parameter $\delta \geq 0$, the set of $\delta$-typical sequences $T_{P,n}^\delta$ consists of all sequences $x^n$ whose type is $\delta$-close to the distribution $P$.

Consider now the simple three-outcome test $(M_n, N_n, \id - M_n - N_n)$ given by
\begin{equation}\begin{aligned}\label{eq:classical_test}
    M_n = \Pi_{P,n}^{\delta}, \qquad  N_n = \Pi_{Q,n}^{\delta},
\end{aligned}\end{equation}
which are the projectors (indicator functions) corresponding to the $\delta$-typical sets $T_{P,n}^\delta$ and $T_{Q,n}^\delta$. For sufficiently small $\delta$, the $\delta$-typical sets of different distributions are disjoint, meaning that this forms a valid measurement. 
It is then an elementary exercise in the theory of types (see Appendix, Section~\ref{app:typicality}) to verify that with this choice of tests, 
the hypothesis testing errors  scale as
\begin{equation}\begin{aligned}
    \palpha_n \sim \exp(-n D(Q\| P)), \qquad \pbeta_n \sim \exp(- n D(P \| Q)).
\end{aligned}\end{equation}
That is, \emph{both} of the relative entropy exponents can be achieved simultaneously, going significantly beyond the constraints of conventional hypothesis testing. And most importantly, this can be done with \emph{arbitrarily high} probability of conclusiveness: the properties of types tell us that $\pconc(P) \to 1$ and $\pconc(Q)\to 1$. This remarkable phenomenon has an intuitive explanation: we really only need to worry about distinguishing the $\delta$-typical sequences --- either $\delta$-typical for $P$, or for $Q$. Since any non-typical sequence is extremely unlikely to occur, even if we simply declare an inconclusive outcome upon seeing such a sequence, this has a negligible impact on the probabilities of conclusiveness.
This choice of measurements is indeed optimal here. The finding then motivates the investigation of whether an analogous claim can be made in the quantum case: can the error exponents given by $D(\sigma\|\rho)$ and $D(\rho\|\sigma)$ be simultaneously achieved with vanishingly little inconclusiveness?



\subsection{Quantum difficulties}\label{sec:difficulties}

A na\"ive attempt to extend the construction of the classical test from~\eqref{eq:classical_test} immediately encounters a major problem: because the two states $\rho$ and $\sigma$ need not commute, the equivalents of typical projectors in quantum information~\cite{schumacher_1995,hayashi_2001,bjelakovic_2003} do not generally commute either,  and hence even for small $\delta$ the $\delta$-typical projectors may not be orthogonal. It is then not obvious how to ensure that a test defined by combining the two is a valid quantum measurement, in the sense that $M_n + N_n \leq \id$. 

\begin{figure*}[t]
\includegraphics[width=.85\textwidth]{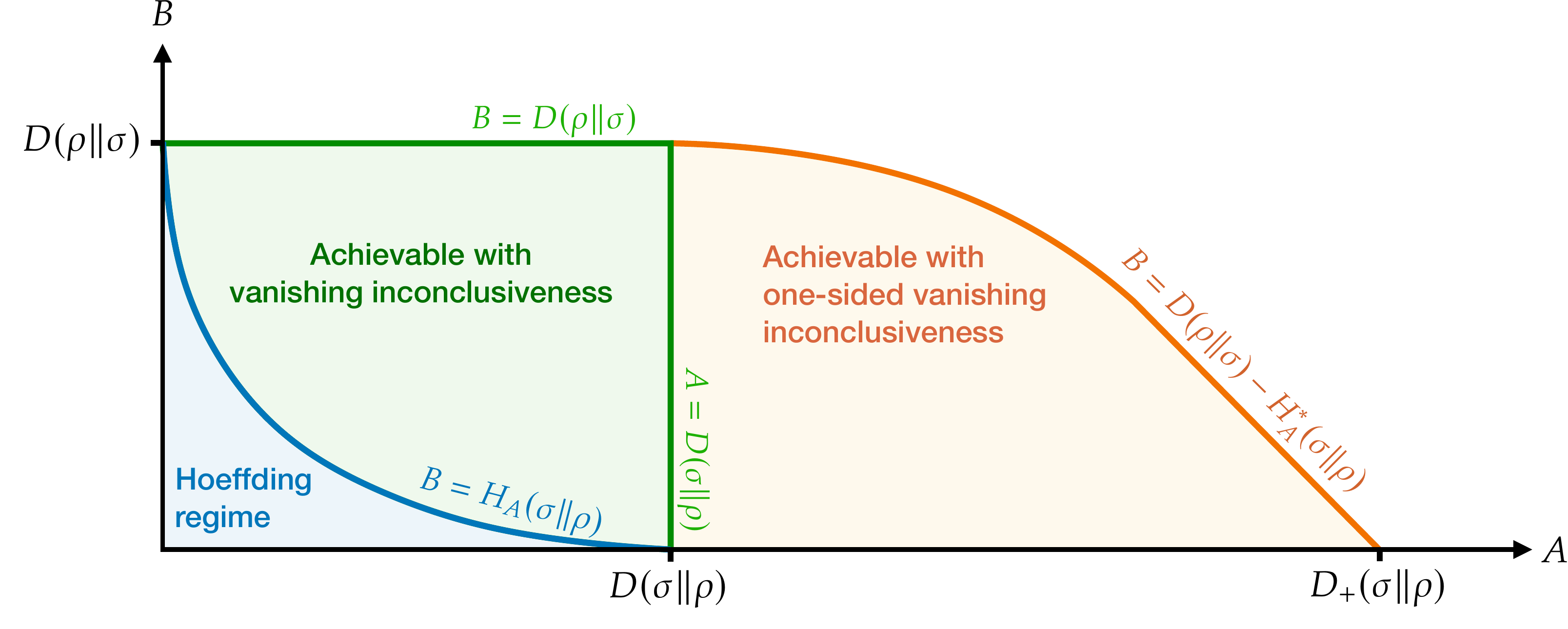}
\caption{%
\textbf{Beyond Hoeffding with low inconclusiveness.} The schematic plot shows the achievable ranges of exponents of the conditional errors of quantum hypothesis testing, $\palpha_n \sim \exp(-nA)$ and $\pbeta_n \sim \exp(-nB)$, in the different regimes studied in this work. The blue region demarcated by the curve $B = H_A(\sigma\|\rho)$, where $H_A$ is the Hoeffding divergence, shows all of the exponents that can be achieved in conventional hypothesis testing using deterministic protocols. All error exponents in the green region, delineated by $B = D(\rho\|\sigma)$ and $A = D(\sigma\|\rho)$, can be achieved with 
an arbitrarily small --- or asymptotically vanishing --- probability of inconclusive outcomes (Proposition~\ref{prop:highprob_exponents}).  The orange region below the curve defined by $B = D(\rho\|\sigma) - \Hsc_A(\sigma\|\rho)$, with $\Hsc_A$ denoting the Han--Kobayashi anti-divergence, can be achieved by tests that are highly conclusive only for the null hypothesis $\rho$ (Proposition~\ref{prop:onesided}).
}%
\label{fig:hoeffding}
\end{figure*}

One simple way to try to mitigate this is as follows. 
From the results of the quantum Stein's lemma, we know that there exists a sequence of two-outcome tests $(M_n, \id - M_n)$ such that the type II error satisfies $\beta_n \sim \exp(- n D(\rho\|\sigma))$ while the type I error is constant, and also that there exists a sequence of tests $(\id-N_n,N_n)$ such that the type I error satisfies $\alpha_n \sim \exp(-n D(\sigma\|\rho))$ with constant type II error probability. We can then define a simple construction of a probabilistic test as
\begin{equation}\begin{aligned}
    \big(\delta M_n, \, (1-\delta) N_n, \, \id - \delta M_n - (1-\delta) N_n \big),
\end{aligned}\end{equation}
for an arbitrary constant $\delta \in (0,1)$. Rescaling the POVM elements by the factors $\delta$ and $1-\delta$ is sufficient to ensure that they can be completed to a valid measurement, as their sum does not exceed identity.  
These constant factors do not affect the conditional error exponents, and we obtain that the errors scaling as $\palpha_n \sim \exp(-n D(\sigma\|\rho))$ and $\pbeta_n \sim \exp(-n D(\rho\|\sigma))$ are again both achievable. However, the inclusion of the normalisation factors $\delta$ means that we necessarily sacrifice the probability of conclusive discrimination: once we account for this, only the effective exponents $A = \delta D(\sigma\|\rho)$ and $B = (1-\delta) D(\rho\|\sigma)$ can be considered to be realistically achievable 
in a way that allows for a fair comparison with conventional hypothesis testing (cf.\ discussion in Section~\ref{subsec:setting}).
Interestingly, even when accounting for the diminished probability of conclusiveness, this is already enough to improve over the quantum Hoeffding bound: in the error exponent plot of Figure~\ref{fig:hoeffding}, any pair of exponents lying below the straight line joining the points $(0, D(\rho\|\sigma)$ and $(D(\sigma\|\rho),0)$ can be achieved in this way. However, this does not match the performance of the optimal classical tests.

Let us then look more closely at how tests that underlie results such as the quantum Stein's lemma are constructed. Many asymptotic results from classical information theory can be `lifted' to the quantum setting by replacing types with the empirical distributions resulting from Schur--Weyl duality~\cite{hayashi_2001,keyl_2001}, or equivalently through the approach based on asymptotic spectral pinching~\cite{hiai_1991,hayashi_2002}. Both of these techniques effectively force the two states to commute with each other in a suitable way, which then allows for a direct application of classical techniques. Once again, though, when applying this approach to the setting of our work, a key issue emerges: both Schur--Weyl measurements and pinching-based methods require the choice of a basis, typically the eigenbasis of one of the two hypotheses. Indeed, denoting by $\E_{\sigma^{\otimes k}}$ the pinching map with respect to $\sigma^{\otimes k}$, it is known that~\cite{hayashi_2002}
\begin{equation}\begin{aligned}
    \lim_{k\to\infty} \frac1k D\!\left(\left. \E_{\sigma^{\otimes k}}(\rho^{\otimes k}) \right\| \sigma^{\otimes k} \right) = D(\rho \| \sigma).
\end{aligned}\end{equation}
This fundamental result underlies a large part of modern quantum information theory and is one way to motivate the Umegaki quantum relative entropy $D(\rho\|\sigma)$ as the operational equivalent of the classical Kullback--Leibler divergence~\cite{hiai_1991}. 
If, however, we instead pinch the second argument of the relative entropy --- namely, $\sigma^{\otimes k}$ in the eigenbasis of $\rho^{\otimes k}$ --- then~\cite{lipka-bartosik_2024}
\begin{equation}\begin{aligned}\label{eq:Dstar}
    \lim_{k\to\infty} \frac1k D\!\left( \rho^{\otimes k} \left\| \E_{\rho^{\otimes k}}(\sigma^{\otimes k}) \right.\right) = D^\star(\rho \| \sigma),
\end{aligned}\end{equation}
where $D^\star$ denotes a divergence obtained as a limit of a family of reverse sandwiched R\'enyi divergences~\cite{audenaert_2015}, and it can be strictly smaller than $D(\rho\|\sigma)$~\cite{audenaert_2015,lipka-bartosik_2024,hayashi_2024-1}. Precisely,
\begin{equation}\begin{aligned}
    D^\star(\rho\|\sigma) = \lim_{s\to1^-} \frac{s}{1-s} \wt{D}_{1-s}(\sigma \| \rho),
\end{aligned}\end{equation}
where $\wt{D}_{1-s}(\sigma\|\rho) = \frac{1}{-s} \log \Tr \left(\rho^{\frac{s}{2(1-s)}} \sigma \rho^{\frac{s}{2(1-s)}}\right)^{1-s}$ denotes the sandwiched R\'enyi divergence~\cite{muller-lennert_2013,wilde_2014}. 
A closed-form expression for $D^\star$ can be obtained through a judicious use of dark magic~\cite[Section~3]{audenaert_2015}.

An application of pinching to the classical construction of~\eqref{eq:classical_test} then either gives us the achievable exponents
\begin{equation}\begin{aligned}
    A \leq D(\sigma \| \rho), \qquad B \leq D^\star(\rho\|\sigma),
\end{aligned}\end{equation}
which are achieved by first pinching with $\E_{\rho^{\otimes k}}$, then applying the classical result of Eq.~\eqref{eq:classical_test} with the choice of $P = \rho^{\otimes k}$ and $Q = \E_{\rho^{\otimes k}}(\sigma^{\otimes k})$, and finally taking the limit $k\to\infty$, or the exponents
\begin{equation}\begin{aligned}
    A \leq D^\star(\sigma \| \rho), \qquad B \leq D(\rho\|\sigma)
\end{aligned}\end{equation}
by choosing the opposite pinching. This discrimination strategy already shows that a major improvement over the Hoeffding bound can be obtained for all quantum states: 
since $D^\star(\sigma \| \rho) \geq \wt{D}_{1/2}(\rho \| \sigma) = - \log F(\rho, \sigma)$ where $F$ denotes the fidelity~\cite{audenaert_2015,hayashi_2024-1}, the fact that $F(\rho, \sigma) < 1$ whenever $\rho\neq\sigma$ implies that it is possible to achieve one of the optimal error exponents given by the quantum relative entropy while still ensuring an exponentially fast decrease of the other error, all the while the probability of conclusive discrimination converges to one. Alas, it again does not lead to a simultaneous achievability of the two exponents given by the Umegaki relative entropies.

To better understand the difficulty in reaching the optimal performance for quantum states, let us consider another way that the quantum relative entropy can be asymptotically characterised. Closely related to the quantum Stein's lemma is the fact that the relative entropy can be achieved by measurements: for any pair of states, there exists a sequence of POVMs with corresponding measurement channels $(\M_k)_k$ such that~\cite{hiai_1991,hayashi_2001}
\begin{equation}\begin{aligned}
    \lim_{k\to\infty} \frac1k D(\M_k(\rho^{\otimes k}) \| \M_k(\sigma^{\otimes k})) = D(\rho \| \sigma).
\end{aligned}\end{equation}
However, the sequence of measurements that attains $D(\rho \| \sigma)$ need not be the same as the one that attains $D(\sigma \| \rho)$,  while if we were to apply the classical approach outlined in the previous section, we would need to use the same sequence of measurements for both type I and type II error exponents. This in fact fully characterises the exponents achievable with high conclusiveness: we can show that the conditional errors $\palpha_n \sim \exp(-n A)$ and $\pbeta_n \sim \exp(-n B)$ can be achieved with probability of conclusive outcomes $\pi_n(\rho),\pi_n(\sigma) \to 1$ if and only if
\begin{equation}\begin{aligned}\label{eq:joint_measurement}
    A \leq \liminf_{k\to\infty} \frac1k D(\M_k(\sigma^{\otimes k}) \| \M_k(\rho^{\otimes k})),\\
    B \leq \liminf_{k\to\infty} \frac1k D(\M_k(\rho^{\otimes k}) \| \M_k(\sigma^{\otimes k}))
\end{aligned}\end{equation}
for a single sequence $(\M_k)_k$ of measurement channels. (See Appendix, Section~\ref{app:typical_measurements}).

The understanding of whether the exponents $A = D(\sigma\|\rho)$ and $B = D(\rho\|\sigma)$ can be achieved in this setting thus reduces to understanding the question of whether there exists a sequence $(\M_k)_k$ of measurements that achieves both of the relative entropies simultaneously. 
This resembles questions that were previously asked in the setting of sequential quantum hypothesis testing~\cite{martinezvargas_2021,li_2022}; coupled with the known connections between sequential and probabilistic hypothesis testing in classical information theory~\cite{lalitha_2016}, this motivates us to look into that setting.


\subsection{Optimal quantum hypothesis testing with arbitrarily low inconclusiveness from sequential hypothesis testing}\label{subsec:sequential}

In sequential quantum hypothesis testing~\cite{martinezvargas_2021,li_2022}, the samples of the unknown quantum state are provided and processed one at a time, and once the test reaches a certain threshold of confidence about the state --- which may require more than $n$ samples 
for the $n$th test of the  protocol,
unlike in conventional hypothesis testing --- a guess of either $\rho$ or $\sigma$ is made. 
In this context, a clever testing protocol was proposed in Ref.~\cite{li_2022} as a modification of the classical sequential probability ratio test~\cite{wald_1945,wald_1948}, itself a generalisation of the standard log-likelihood ratio test. 
The approach of~\cite{li_2022} performs a random walk that adaptively decides whether the best measurement sequence to use is the one that achieves $D(\rho\|\sigma)$, or the one that achieves $D(\sigma\|\rho)$. Crucially, the protocol indeed achieves both of the quantum relative entropies as the error exponents simultaneously.
While the setting we study here is distinct from sequential hypothesis testing, a simple adaptation of the protocol of~\cite{li_2022} can be used to turn it into a probabilistic one with an inconclusive outcome, and an analysis of its performance
allows us to show the desired achievability also in our setting. Together with a matching converse bound showing optimality, this leads to a complete characterisation of the exponents of inconclusive quantum hypothesis testing achievable with arbitrarily low inconclusiveness.

\begin{proposition}\label{prop:highprob_exponents}
Let $\palpha_n \sim \exp(-n A)$ and $\pbeta_n \sim \exp(-n B)$ be the conditional errors of quantum hypothesis testing. For any quantum states $\rho$ and $\sigma$, there exists a sequence of tests that achieves the error exponents $A$ and $B$ with arbitrarily high probability of conclusive outcomes, 
i.e.\ $\pi_n(\rho), \pi_n(\sigma) \geq 1-\ve$ for all $n$ and for any desired $\ve \in (0,1)$,
if and only if
\begin{equation}\begin{aligned}
    A &\leq D(\sigma \| \rho),\\
    B &\leq D(\rho \| \sigma).
\end{aligned}\end{equation}
Equivalently, the probability of conclusive outcomes can be taken to satisfy $\pi_n(\rho), \pi_n(\sigma) \to 1$ as $n\to\infty$.
Furthermore, these bounds on $A$ and $B$ satisfy a strong converse property: neither can be exceeded even if one allows a lower probability of conclusiveness $\pi_n(\rho), \pi_n(\sigma) \geq c$ for some constant $c > 0$.
\end{proposition}

This shows that, as long as the experimenter is willing to tolerate an arbitrarily small --- or, indeed, asymptotically vanishing --- probability of obtaining an inconclusive outcome, the relative entropy exponents can be simultaneously achieved for both the type I and type II errors in quantum hypothesis testing.  In other words, the seemingly unavoidable trade-off between the errors implied by the quantum Hoeffding bound can be completely circumvented, and this incurs almost no operational costs, as the event of obtaining an inconclusive outcome can be made arbitrarily unlikely and thus negligible for practical purposes.

We stress that this does not require an inherently different setting or interpretation than conventional hypothesis testing --- in particular, no postselection or any other unrealistic assumptions are needed to achieve the exponents given by Proposition~\ref{prop:highprob_exponents}.
This is because the optimal protocol can be adapted to work in a fully conclusive manner,
always returning a guess of either $\rho$ or $\sigma$ and never terminating with an inconclusive discrimination.
The way to implement this mirrors our discussion in Section~\ref{subsec:setting}: 
upon obtaining an inconclusive outcome, the experimenter may simply request additional copies of the unknown state and repeat the same experiment. Although such additional repetitions certainly do not come for free, as they involve more copies of the unknown state being consumed and thus have an adverse effect on the performance, the associated overhead costs can be made arbitrarily small and even completely vanish in the asymptotic limit. 
The expected number of repetitions needed to guarantee conclusive discrimination of $n$ copies 
is exactly $\pi_n(\omega)^{-1}$, where $\omega \in \{\rho, \sigma\}$ is the unknown state, and so, in order to measure the efficiency of this protocol with the expense of additional repetitions accounted for, one may turn to the effective error exponent --- namely, the rate at which the error decays with respect to the expected number of consumed copies across all repetitions.
The effective error exponent defined in this way equals the conditional error exponent (i.e., exponent of $\palpha_n$ or $\pbeta_n$) multiplied by the corresponding conclusive probability ($\pi_n(\rho)$ or $\pi_n(\sigma)$).  In Proposition~\ref{prop:highprob_exponents}, since the conclusive probabilities $\pi_n(\rho)$ and $\pi_n(\sigma)$ asymptotically approach $1$, virtually no expense is incurred on the effective error exponents due to repetition; the effective error exponents are thus equal to the conditional error exponents $A=D(\rho\|\sigma)$ and $B=D(\sigma\|\rho)$, sidestepping the conventional trade-offs on terms that are fully comparable with conventional quantum hypothesis testing.

The details of the protocol that achieves these exponents, including a complete self-contained proof based on~\cite{li_2022}, can be found in the Appendix (Section~\ref{app:sequential}).

An interesting consequence of Proposition~\ref{prop:highprob_exponents} can be obtained by combining it with Eq.~\eqref{eq:joint_measurement}.
\begin{corollary}\label{cor:single_measurement_sequence}
For any two quantum states $\rho$ and $\sigma$, there exists a single sequence $(\M_k)_k$ of measurements such that 
\begin{equation}\begin{aligned}
    \liminf_{k\to\infty} \frac1k D(\M_k(\sigma^{\otimes k}) \| \M_k(\rho^{\otimes k})) = D(\sigma \| \rho),\\
    \liminf_{k\to\infty} \frac1k D(\M_k(\rho^{\otimes k}) \| \M_k(\sigma^{\otimes k})) = D(\rho\|\sigma).
\end{aligned}\end{equation}
\end{corollary}
What this result means is that the optimal choices of the quantities on the right-hand side of Eq.~\eqref{eq:joint_measurement} are simply the Umegaki relative entropies $D(\sigma\|\rho)$ and $D(\rho\|\sigma)$. 
The question of whether such a choice of measurements is possible was raised earlier in~\cite{martinezvargas_2021,li_2022}, motivated by sequential quantum hypothesis testing. In fact, we believe that the resolution of the problem could already be deduced through a slight extension of the main results of~\cite{li_2022}, but this does not seem to have been observed in that work. While here we do not study sequential hypothesis testing,  Corollary~\ref{cor:single_measurement_sequence} has immediate consequences also for that setting:
it shows that the asymptotic performance of adaptive and non-adaptive protocols in sequential hypothesis testing is actually the same, provided that one considers the large-block limit studied in~\cite{li_2022}.
This contrasts with what the numerical example in \cite{li_2022} suggests in the small-block case, where adaptive protocols were observed to have an advantage over non-adaptive ones.


\subsection{Going further beyond: one-sided conclusiveness}\label{sec:onesided}

The fact that such a major advantage over conventional hypothesis testing is possible with an arbitrarily small probability of inconclusiveness makes one wonder: would it be possible to obtain even higher advantages by allowing more inconclusiveness, trading conclusiveness for distinguishability?
An important feature of the bound in Proposition~\ref{prop:highprob_exponents} is its \emph{strong converse} character: as long as the probability of conclusive outcomes is bounded away from zero, beating this bound is impossible. That is, achieving any higher rates must involve at least one of the conclusive probabilities asymptotically converging to \emph{zero}. However, as we will now show, there is a significant range of error exponents that can be achieved while maintaining arbitrarily high conclusiveness for one of the two hypotheses.

The study of hypothesis testing in the asymmetric setting is sometimes motivated by one of the hypotheses having a higher importance than the other --- for example in cases where a false negative could have much worse consequences than a false positive --- which justifies aiming to make one of the errors as small as possible even at the expense of the other. Following this reasoning, one may allow for relaxed requirements for the achievable conclusive probabilities: if the `more important' hypothesis $\rho$ is true, we want our probabilistic hypothesis testing protocol to yield a conclusive outcome with a very high probability; but if the other hypothesis $\sigma$ is true, conclusive discrimination is not crucial, and we will simply not impose any constraints on its probability. We can exactly characterise the optimal performance of quantum hypothesis testing also in this setting, showing that it allows us to exceed the restrictions of the Hoeffding bound even more strongly, although at the price of losing control of the overhead of this process on one of the two hypotheses.

\begin{proposition}\label{prop:onesided}
Let $\palpha_n \sim \exp(-n A)$ and $\pbeta_n \sim \exp(-n B)$ be the conditional errors of quantum hypothesis testing. 
For any quantum states $\rho$ and $\sigma$, there exists a sequence of tests that achieves the error exponent $A$ with an arbitrarily high probability of conclusively distinguishing $\rho$, that is $\pi_n(\rho)\to 1$, while also achieving the conditional error exponent $B$, 
if and only if
\begin{equation}\begin{aligned}
    B \leq D(\rho\|\sigma) - \Hsc_A(\sigma \| \rho), \label{eq:onesided}
\end{aligned}\end{equation}
where
\begin{equation}\begin{aligned}
    \Hsc_A(\sigma \| \rho) \coloneqq \sup_{s > 1} \frac{s-1}{s} \Big( A - \wt{D}_s(\sigma \| \rho) \Big) \label{eq:strong-converse}
\end{aligned}\end{equation}
is the quantum Han--Kobayashi anti-divergence~\cite{han_1989,mosonyi_2015}.
\end{proposition}
Here, $\wt{D}_s(\sigma\|\rho) = \frac{1}{s-1} \log \Tr \left(\rho^{\frac{1-s}{2s}} \sigma \rho^{\frac{1-s}{2s}}\right)^s$ is again the sandwiched R\'enyi divergence~\cite{muller-lennert_2013,wilde_2014}. The Han--Kobayashi anti-divergence $\Hsc_A(\sigma \| \rho)$ is known to characterise the strong converse exponent of conventional quantum hypothesis testing~\cite{han_1989,mosonyi_2015}, that is, the  smallest possible exponent with which the type II error probability converges to \emph{one} when the type I error probability decays at a given rate $A$. The appearance of this quantity in an achievability result like Proposition~\ref{prop:onesided}, where both of the error probabilities indeed converge to zero, is a rather unique phenomenon.

A key difference in this setting is that the error exponent $A$ can now exceed the standard bound of $D(\sigma\|\rho)$. 
The advantage that this enables is plotted in Figure~\ref{fig:hoeffding}. It is natural here to wonder exactly how large $A$ can get, i.e.\ about its highest achievable value when $B \to 0$. This corresponds to $A = D_+(\sigma\|\rho)$, where
\begin{equation}\begin{aligned}
    D_+(\sigma\|\rho) &\coloneqq \inf_{s > 1}\left(\frac{s}{s-1}D\fleft(\rho\middle\|\sigma\fright)+\wt{D}_s\fleft(\sigma\middle\|\rho\fright)\right)\\
    &\hphantom{:}\geq D(\sigma\|\rho)+D(\rho\|\sigma).
\end{aligned}\end{equation}
While it is not clear to us if the quantity $D_+(\sigma\|\rho)$ admits a more closed-form expression in full generality, in the Appendix (Section~\ref{app:special}) we introduce a simple sufficient condition that ensures a much simpler form of $D_+(\sigma\|\rho) = D_{\max}(\sigma\|\rho) + D(\rho\|\sigma) $ for many quantum states, with $D_{\max} = \wt{D}_\infty$ denoting the max-relative entropy~\cite{datta_2009-2,muller-lennert_2013}.

The error exponents announced in Proposition~\ref{prop:onesided} can be achieved by suitably combining the achievability results for the quantum Stein's lemma~\cite{hiai_1991,ogawa_2000} with those of the strong converse exponent $\Hsc_A(\sigma\|\rho)$~\cite{mosonyi_2015} (see Appendix, Section~\ref{app:special}). The proof of the optimality of this result, however, requires results that go beyond standard converse bounds in hypothesis testing. To investigate such questions, in the next section we will look more closely at the connections between hypothesis testing with inconclusive outcomes and strong converse exponents of standard hypothesis testing.


\subsection{`Strong converse' regime: high inconclusiveness}\label{sec:postselected_asym}

In conventional hypothesis testing --- and indeed in many other information-theoretic tasks --- the strong converse property tells us that exceeding the optimal asymptotic rates necessarily incurs a large error that must asymptotically converge to one~\cite{chernoff_1956,ogawa_2000}. 
The study of strong converse exponents~\cite{han_1989,mosonyi_2015} is dedicated to a precise understanding of this rate of convergence. For instance, in some cases one may be willing to tolerate such a large error, as long as the convergence to 1 happens sufficiently slowly. Seen another way, the strong converse regime can be understood as establishing ultimate limits that cannot be surpassed even when a large error is allowed.

In our setting, this line of reasoning motivates instead the study of exponents of \emph{conclusiveness}. That is, we know from Proposition~\ref{prop:highprob_exponents} that exceeding the error exponent  $\alpha_n \sim \exp(-n D(\sigma\|\rho))$ or $\beta_n \sim \exp(-n  D(\rho\|\sigma))$ forces the inconclusiveness to be large, in the sense that the probability of conclusive outcomes $\pi_n$ must converge to 0. But how fast must this convergence be? In other words, what rate of vanishing conclusiveness is needed to achieve even higher performance?

This brings us into the purview of postselected hypothesis testing~\cite{regula_2024}, which is a setting that focuses purely on the conditional errors $\palpha_n$, $\pbeta_n$ themselves. The previous study of this problem completely discounted the probability $\pconc$ from the rates, effectively allowing an unconstrained probability of inconclusiveness. 
Our results will refine the analysis of this setting by asking exactly how small the probability of conclusive outcomes must be to gain advantages over the other regimes of hypothesis testing. Such questions were previously asked also in the one-shot setting~\cite{gupta_2023}. 

Here we thus initiate a precise study of the trade-offs between the achievable conditional error exponents and probability of conclusive outcomes. 
Our main result is their complete characterisation.  

\begin{proposition}
\label{prop:conclusive}
Let $\palpha_n\sim\exp(-nA)$ and $\pbeta_n\sim\exp(-nB)$ be the conditional errors of quantum hypothesis testing.  There exists a sequence of tests that achieves the error exponents $A$ and $B$ with conclusive probabilities $\pi_n(\rho)\sim\exp(-nK)$ and $\pi_n(\sigma)\sim\exp(-nL)$ if and only if
\begin{align}
	A+K&\leq\inf_{t>1}\left(\frac{t}{t-1}L+\wt{D}_t(\sigma\|\rho)\right), \label{eq:conclusive-1}\\
	B+L&\leq\inf_{s>1}\left(\frac{s}{s-1}K+\wt{D}_s(\rho\|\sigma)\right). \label{eq:conclusive-2}
\end{align}
\end{proposition}

\begin{figure*}[t]
\includegraphics[width=.65\textwidth]{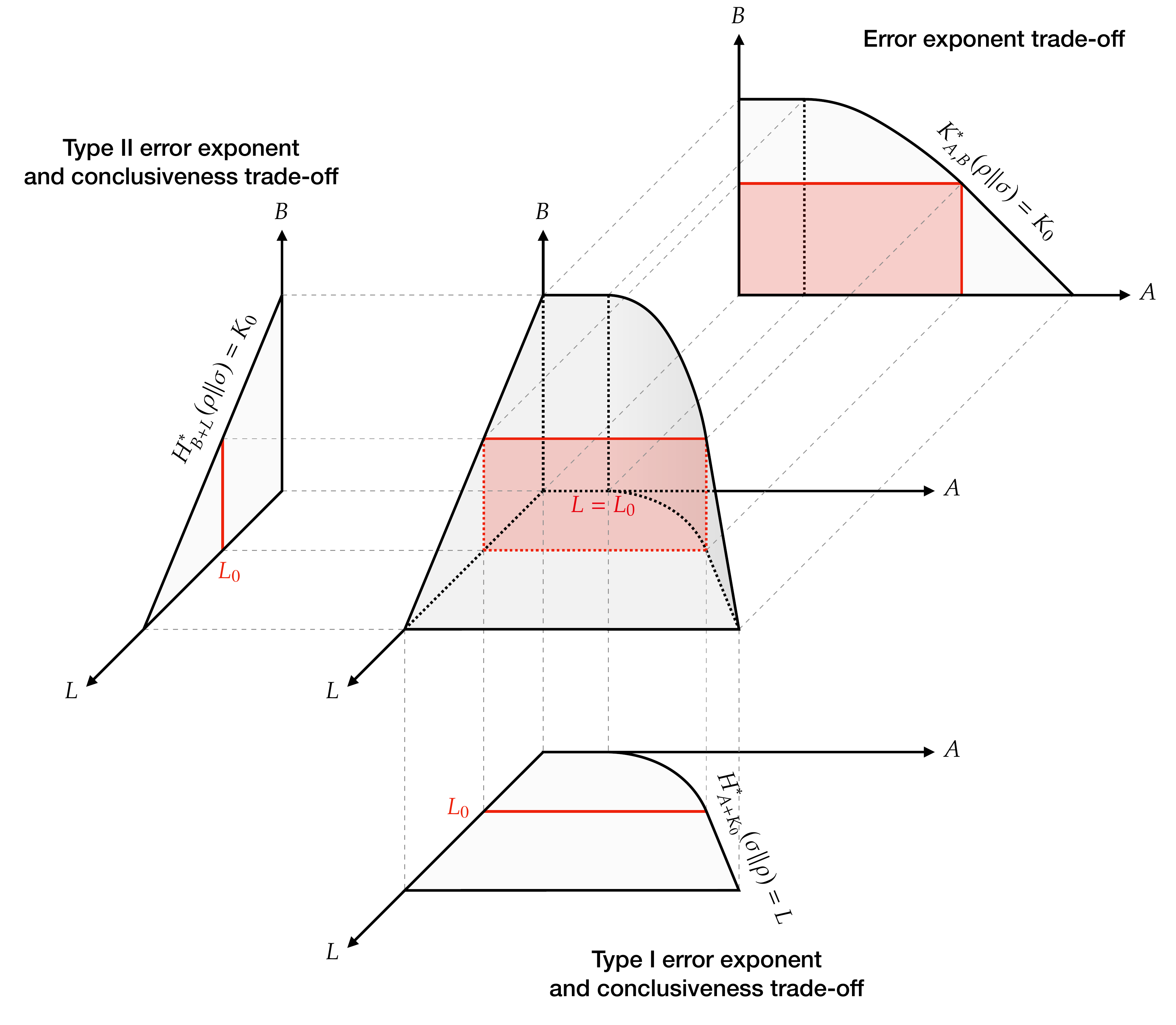}
\caption{%
\textbf{Achievable region of exponents with high inconclusiveness.} The plot shows the relations and trade-offs between exponents of different types characterized by our results in the `strong converse' setting of high inconclusive probability.
Here, $A$ and $B$ are the exponents of the type I and type II conditional errors, $\palpha_n\sim\exp(-nA)$ and $\pbeta_n\sim\exp(-nB)$, and $K$ and $L$ the exponents of the probabilities of conclusive outcomes under the two hypotheses, $\pconc(\rho)\sim\exp(-nK)$ and $\pconc(\sigma)\sim\exp(-nL)$.  The central plot shows the three-dimensional region in which the exponents $A$, $B$, and $L$ are simultaneously achievable 
for some fixed $K = K_0$. 
The red rectangle within the achievable region is its intersection with the plane $L=L_0$, representing the 
achievable error exponents for fixed exponents of conclusiveness, as characterized in  Proposition~\ref{prop:conclusive}.  The three plots surrounding the central one depict  projections of the region, from which one can deduce different trade-offs: between the conditional type II error exponent $B$ and its corresponding conclusive probability $L$ (leftmost plot); between the two conditional errors themselves when $L$ is unconstrained (top right), with $K_{A,B}^*(\rho\|\sigma)$ defined in Eq.~\eqref{eq:kstar}; and between the conditional type I error exponent $A$ and the conclusiveness exponent $L$ (bottom).
}%
\label{fig:KL}
\end{figure*}

Proposition~\ref{prop:conclusive} gives the optimal trade-off between reducing conditional errors and maximising the degrees of conclusiveness in the regime of high inconclusiveness, where the probability of conclusive outcomes decays exponentially fast. 
Note that the trade-off is between multiple objectives (error exponents and conclusiveness exponents), and objectives pertaining to different hypotheses can interfere with each other.  Generally, it can be observed that requiring higher degrees of conclusiveness regarding one hypothesis imposes stronger limitations on reducing the conditional error of the same hypothesis, whereas it creates more room for reducing the conditional error of the other hypothesis.  This interference can be understood as a consequence of the exponentially strong converse properties of conventional hypothesis testing~\cite{mosonyi_2015}, and the connection with the latter is ultimately what allows us to obtain a precise evaluation of the trade-offs. 

We stress that in our discussion here, the \emph{errors} of the discrimination, i.e.\ $\palpha_n$ and $\pbeta_n$, are always asymptotically vanishing; it is only the inconclusiveness that is now allowed to be large.

To demonstrate the intuition behind the connection with conventional strong converses, 
let us suppose that the exponents $A$, $B$, $K$, and $L$ are simultaneously achieved as in Proposition~\ref{prop:conclusive} 
by a sequence of three-outcome tests $(M_n,N_n,\id-M_n-N_n)$. A moment of thought reveals that this imposes that the unconditional success and error probabilities must scale as
\begin{align}
	\Tr M_n\rho^{\otimes n}&=(1-\palpha_n)\pi_n(\rho) \sim\pi_n(\rho) \sim \exp(-nK), \label{eq:conclusive-3}\\
	\Tr M_n\sigma^{\otimes n}&=\pbeta_n\, \pi_n(\sigma) \sim \exp(-n(B+L)), \label{eq:conclusive-4}\\
	\Tr N_n\rho^{\otimes n}&=\palpha_n \, \pi_n(\rho) \sim \exp(-n(A+K)), \label{eq:conclusive-5}\\
	\Tr N_n\sigma^{\otimes n}&=(1-\pbeta_n)\pi_n(\sigma) \sim \pi_n(\sigma) \sim \exp(-nL). \label{eq:conclusive-6}
\end{align}
Note that Eqs.~\eqref{eq:conclusive-3} and \eqref{eq:conclusive-4} can alternatively be interpreted as the type I success probability and the type II error probability, respectively, of a conventional test $(M_n,\id-M_n)$ on the two hypotheses $\rho^{\otimes n}$ and $\sigma^{\otimes n}$.  Under this interpretation, the strong converse analysis of asymmetric hypothesis testing~\cite{mosonyi_2015} tells us that the relevant exponents $K$ and $B+L$ must be subject to the fundamental limitation
\begin{align}
	K&\geq \Hsc_{B+L}(\rho\|\sigma), \label{eq:conclusive-7}
\end{align}
due to the operational interpretation of the quantum Han--Kobayashi anti-divergence $\Hsc_{B+L}(\rho\|\sigma)$ (Eq.~\eqref{eq:strong-converse}) as the strong converse exponent with rate $B+L$.  Applying the same reasoning to Eqs.~\eqref{eq:conclusive-5} and \eqref{eq:conclusive-6} similarly gives
\begin{align}
	L&\geq \Hsc_{A+K}(\sigma\|\rho). \label{eq:conclusive-8}
\end{align}
Invoking the expression for the quantum Han--Kobayashi anti-divergence in Eq.~\eqref{eq:strong-converse}, one can verify with some manipulation that Eqs.~\eqref{eq:conclusive-7} and \eqref{eq:conclusive-8} coincide with Eqs.~\eqref{eq:conclusive-1} and \eqref{eq:conclusive-2}.  
This sketches the converse direction of our proof of Proposition~\ref{prop:conclusive}; to complete it, we also need to show that exponents satisfying Eqs.~\eqref{eq:conclusive-1} and \eqref{eq:conclusive-2} are always simultaneously achievable.  A complete proof including the achievability is provided in the Appendix (Section~\ref{app:conclusive}).

The trade-off characterised in Proposition~\ref{prop:conclusive} has a rich structure, and its implications can be examined through a variety of special cases. First, in the special case of $K=L=0$, Eqs.~\eqref{eq:conclusive-1} and \eqref{eq:conclusive-2} become $A\leq D(\sigma\|\rho)$ and $B\leq D(\rho\|\sigma)$, respectively, 
recovering a weaker version of Proposition~\ref{prop:highprob_exponents}. 
Moreover, consulting the equivalent formulation in Eqs.~\eqref{eq:conclusive-7} and \eqref{eq:conclusive-8}, if $A>D(\sigma\|\rho)$ or $B>D(\rho\|\sigma)$, then at least one of $K$ or $L$ must be strictly positive, thus providing an exponential strong-converse--type statement for the achievability result in Proposition~\ref{prop:highprob_exponents}.  Second, Eqs.~\eqref{eq:conclusive-1} and \eqref{eq:conclusive-2} imply that $A+K\leq L+D_{\max}(\sigma\|\rho)$ and $B+L\leq K+D_{\max}(\rho\|\sigma)$, which tells us that
\begin{align}
	A+B&\leq D_{\max}(\rho\|\sigma) + D_{\max}(\sigma\|\rho) \eqqcolon D_\Omega(\rho\|\sigma) \label{eq:conclusive-9}
\end{align}
for all $K,L\geq0$, 
where 
$D_\Omega$ denotes the Hilbert projective metric~\cite{bushell_1973}.  This entails a strict limit on the trade-off between the two conditional error exponents even when no degree of conclusiveness is demanded whatsoever, which corresponds to the postselected hypothesis testing setting of Ref.~\cite{regula_2024}, refining the findings therein.

One interesting case that can also be deduced from Proposition~\ref{prop:conclusive} is when we only account for the degree of conclusiveness under one of the two hypotheses but impose no restrictions on the conclusiveness under the other hypothesis.  In other words, we are concerned with the following question: to achieve a given pair of exponents in the conditional errors $\palpha_n \sim \exp(-nA)$ and $\pbeta_n \sim \exp(-nB)$, how fast must the conclusiveness decay when the true hypothesis is $\rho$, i.e.\ what is the smallest possible value of the exponent $K$ in $\pconc(\rho) \sim \exp(-nK)$? Some careful manipulation of Eqs.~\eqref{eq:conclusive-1}--\eqref{eq:conclusive-2} shows that the smallest such exponent $K$ given $A$ and $B$ is precisely equal to the following quantity:
\begin{align}
	K^*_{A,B}\fleft(\rho\middle\|\sigma\fright)&\coloneq\sup_{s,t>1}\left(\frac{s}{s-1}-\frac{t-1}{t}\right)^{-1}\left(B-\wt{D}_s\fleft(\rho\middle\|\sigma\fright)\vphantom{+\frac{t-1}{t}\left(A-\wt{D}_t\fleft(\sigma\middle\|\rho\fright)\right)}\right. \notag\\
	&\quad\left.\hphantom{\sup_{s,t>1}}+\frac{t-1}{t}\left(A-\wt{D}_t\fleft(\sigma\middle\|\rho\fright)\right)\right). \label{eq:kstar}
\end{align}
This can indeed be thought of as providing an exponentially strong converse statement for Proposition~\ref{prop:onesided}: if Eq.~\eqref{eq:onesided} is violated, namely if
\begin{align}
	B&>D(\rho\|\sigma)-\sup_{t>1}\frac{t-1}{t}\left(A-\wt{D}_t(\sigma\|\rho)\right),
\end{align}
then $K^*_{A,B}(\rho\|\sigma)$ is necessarily strictly positive, implying that conclusiveness must decay exponentially fast at a rate $K\geq K^*_{A,B}(\rho\|\sigma)>0$.  More generally, the function $K^*_{A,B}(\rho\|\sigma)$ can be deemed as a new anti-divergence that acquires an operational interpretation in quantifying the decay of conclusiveness under fixed conditional error exponents in postselected hypothesis testing.  This provides yet another operational context in which the sandwiched Rényi divergence finds use~\cite{wilde_2014,mosonyi_2015,mosonyi_2017}, in particular with the uncommon construction that combines two such divergences with reversely ordered arguments and potentially distinct parameters $s$ and $t$.


\subsection{Exponentially low inconclusiveness} 
\label{sec:exponentially_good}

As our final variant of probabilistic hypothesis testing, let us return to the setting of low inconclusiveness discussed in Section~\ref{subsec:sequential}, and consider now 
the case where we require the inconclusiveness of the discrimination protocol to be \emph{exponentially} small, or equivalently that the probability of conclusive outcomes converges to 1 exponentially fast. 
This imposes an all-exponential decay of all probabilities associated with not making a correct guess: not only do the conditional error probabilities need to vanish exponentially fast, but so does the probability that the test is inconclusive and no guess is made at all. Just like in conventional hypothesis testing, such a strong requirement will lead to inherent trade-offs, restricting the achievable exponents.

To gain some intuition about this setting, let us first consider the classical case. Here, the setting corresponds to hypothesis testing with rejection~\cite{nikulin_1989,gutman_1989,grigoryan_2011,sason_2012} and is also very closely related to the framework of `almost-fixed-length hypothesis testing' studied in~\cite{lalitha_2016} --- indeed, the methods of those works can also be used to derive this result, although our formulation of the problem is slightly different.

\begin{figure}[t]
\includegraphics[width=.48\textwidth]{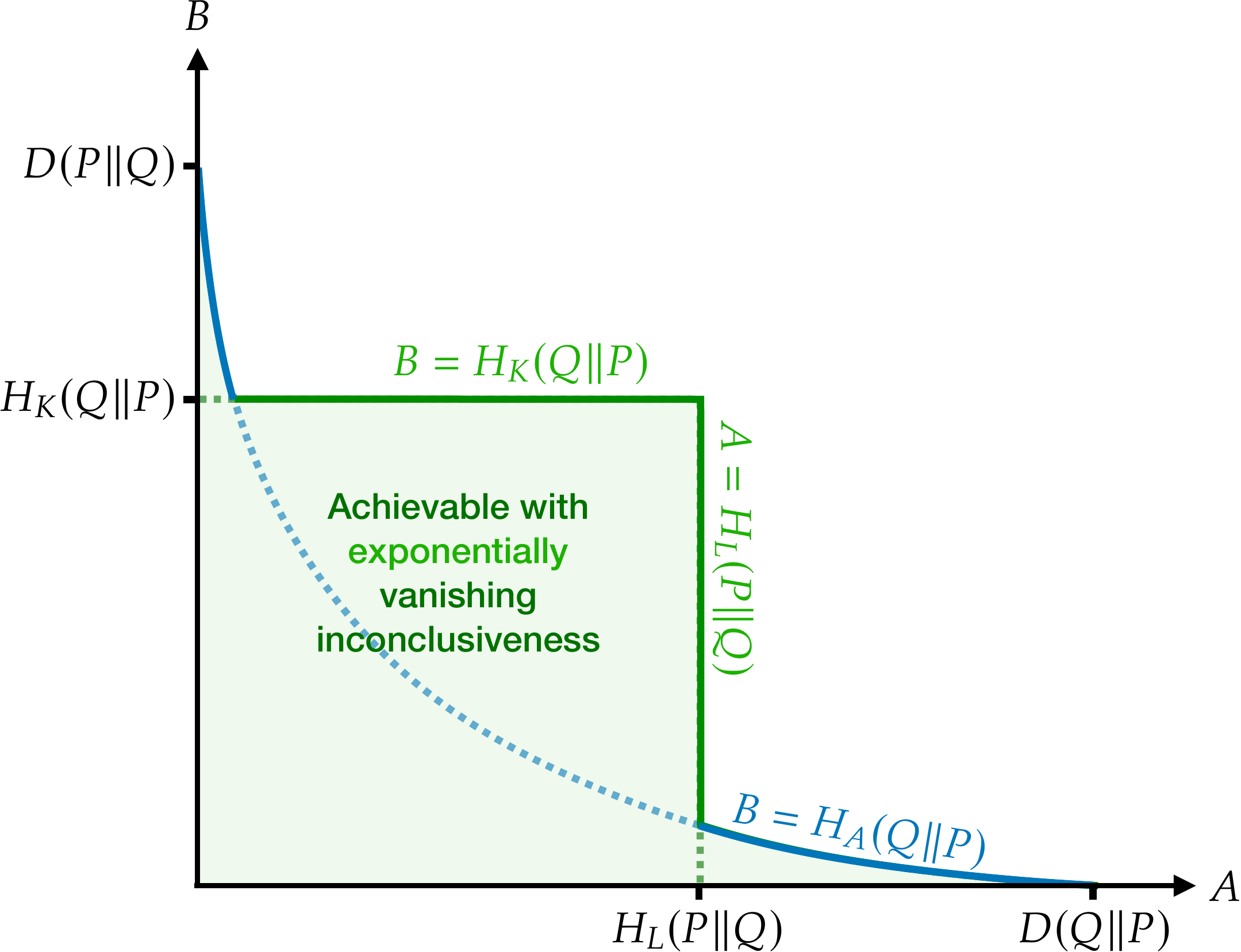}
\caption{%
\textbf{Trade-offs in hypothesis testing with exponentially low inconclusiveness.} The shaded region represents the range of achievable exponents of the conditional errors $\palpha_n \sim \exp(-nA)$ and $\pbeta_n \sim \exp(-nB)$ of hypothesis testing of the classical probability distributions $P$ and $Q$ with probability of inconclusive discrimination converging to $0$ exponentially fast --- with exponent $K$ for the distribution $P$, and exponent $L$ for~$Q$ (Lemma~\ref{lem:exponentially_classical}). 
}%
\label{fig:exponentially}
\end{figure}

\begin{lemma}\label{lem:exponentially_classical}
Consider the hypothesis testing of two classical probability distributions $P$, $Q$ with probabilities of conclusiveness satisfying $1 - \pi_n(P) \sim  \exp(- n K)$ and $1 - \pi_n(Q) \sim  \exp( - n L)$. Then the conditional errors $\palpha_n \sim \exp(-n A)$ and $\pbeta_n \sim \exp(-n B)$ are achievable if and only if
\begin{equation}\begin{aligned}
A &\leq \max \{ H_{B} (P\|Q),\, H_{L} (P\|Q) \},\\
B &\leq \max \{ H_A(Q\|P),\, H_K (Q\|P) \},
\end{aligned}\end{equation}
or equivalently if and only if
\begin{equation}\begin{aligned}
    B \leq \begin{cases} H_A (Q\| P) & \text{ if } A < K \text{ or } A > H_L(P \| Q)\\
                      H_K (Q \| P) & \text{ if } K \leq A \leq H_L(P \| Q).\end{cases}
\end{aligned}\end{equation}
\end{lemma}
Here we recall that $H_A$ denotes the Hoeffding divergence~\eqref{eq:hoeffding_div}. The constraints captured by Lemma~\ref{lem:exponentially_classical} are shown in Figure~\ref{fig:exponentially}.

In the range of exponents satisfying $K \leq A \leq H_L(P \| Q)$, this result improves over the standard Hoeffding bound. The achievability here can be again shown using the method of types: for the choice of the test measurement $M_n$ being the indicator function of the set of sequences $x^n$ satisfying $D(t_{x^n} \| Q) > H_K(Q \| P)$, and $N_n$ being the indicator function of the set of sequences $x^n$ satisfying $D(t_{x^n} \| Q) \leq L$, Sanov's theorem~\cite{sanov_1957}  allows us to deduce that the stated exponents can indeed be achieved. Details are provided in the Appendix (Section~\ref{app:typicality}).

The classical result of Lemma~\ref{lem:exponentially_classical} coupled with asymptotic pinching for the R\'enyi divergences~\cite{mosonyi_2015,hayashi_2016-1} allow us to obtain an achievability result for quantum states.
\begin{proposition}\label{prop:exponentially_quantum}
For any two quantum states $\rho$, $\sigma$ and $K, L > 0$ such that $1 - \pi_n(\rho) \sim  \exp(- n K)$ and $1 - \pi_n(\sigma) \sim \exp( - n L)$, 
the conditional errors $\palpha_n \sim \exp(-n A)$ and $\pbeta_n \sim \exp(-n B)$ are achievable if
\begin{equation}\begin{aligned}\label{eq:exponentially_quantum}
    B \leq \begin{cases} \wt{H}_K (\sigma \| \rho) & \text{ if } A \leq \rev{H}_L(\rho \| \sigma)\\
    \rev{H}_K (\sigma \| \rho) & \text{ if } A \leq \wt{H}_L(\rho \| \sigma)\\
    H_A (\sigma \| \rho) & \text{ otherwise,}\end{cases}
\end{aligned}\end{equation}
where $H_A (\sigma \| \rho)$ is the standard Hoeffding divergence and $\wt{H}_A$, $\rev{H}_A$ denote variants of it defined using the sandwiched and reverse sandwiched R\'enyi relative entropies:
\begin{equation}\begin{aligned}
    \wt{H}_A(\sigma \| \rho) &\coloneqq \sup_{s \in (0,1)} \frac{s-1}{s} \Big( A - \wt{D}_s(\sigma \| \rho) \Big),\\
\widetildearrow{H}_A(\sigma \| \rho) &\coloneqq \sup_{s \in (0,1)} \frac{s-1}{s} \Big( A - \rev{D}_s(\sigma \| \rho) \Big)
\end{aligned}\end{equation}
with 
$\rev{D}_s(\sigma\|\rho) \coloneqq \frac{s}{1-s}\wt{D}_{1-s}(\rho\|\sigma)$.
\end{proposition}

The result gives a non-trivial achievability region that exhibits advantages over the conventional Hoeffding setting even when the inconclusiveness of the protocol is exponentially low. This feature attests to a certain robustness of the advantages discussed in this work: the improvements over the Hoeffding bound are not completely broken by imposing an exponential constraint on the inconclusiveness, but instead it is possible to smoothly interpolate between the conventional setting of quantum hypothesis testing and the broader advantages with constant probability of conclusiveness. 

However, although tight for classical distributions, this result is not optimal for quantum states in general.  This is because error exponents here are obtained through asymptotic pinching, and we already observed in Section~\ref{sec:difficulties} that na\"ive pinching does not give us optimal exponents when low inconclusiveness is required. 
This is reminiscent of how pinching and the resulting sandwiched R\'enyi divergences provide tight results in the strong converse regime of conventional hypothesis testing~\cite{mosonyi_2015,hayashi_2016-1}, but in the study of error exponents it is the Petz--R\'enyi divergences that are optimal~\cite{nussbaum_2009,nagaoka_2006,audenaert_2008,tomamichel_2025}. Indeed, the only converse bound that we can obtain here is through the Petz--R\'enyi divergences (see Appendix, Section~\ref{app:typical_pinching}), and it is natural to conjecture that it is optimal. We were however not able to obtain matching achievability results. 

A difficulty in obtaining a tighter result here is that this regime is strictly more general than the previously studied setting of hypothesis testing with low probability of inconclusive outcomes (Proposition~\ref{prop:highprob_exponents}), and it should reduce to the latter in a suitable limit. 
However, the only approach we know of that is capable of achieving optimal error exponents in Proposition~\ref{prop:highprob_exponents} is through the adaptive protocol based on sequential hypothesis testing~\cite{li_2022}.
Not only is it unclear to us whether there exist sufficiently strong large-deviation--style convergence bounds that could be used to refine that result to exponentially low inconclusiveness, but more importantly, even if such convergence bounds did exist, that would still not be enough to achieve the Petz--R\'enyi converse --- the sequential protocol is explicitly based on performing adaptive measurements, and Petz--R\'enyi relative entropies cannot be achieved by measurements~\cite{mosonyi_2015,tomamichel_2025}. A complete description of this regime thus appears to require different techniques, and its resolution is an intriguing open problem.


\section{Symmetric hypothesis testing}\label{sec:symmetric}


\subsection{Conventional setting}

In symmetric hypothesis testing, instead of trying to optimise one of the two errors $\alpha_n$ and $\beta_n$ at the possible expense of the other, we aim to treat the two errors on equal footing. There are two ways to understand this problem, and it will be important to distinguish them here. We begin with a discussion of the conventional setting of hypothesis testing with no inconclusive outcomes. 

One common way to understand the asymptotic study of symmetric hypothesis testing is the Bayesian motivation, namely as the characterisation of the asymptotic behaviour of the expected, \emph{average} error
\begin{equation}\begin{aligned}
    \erravg_n(M_n, N_n) = p \,\alpha_n(M_n, N_n) + (1-p)\, \beta_n(M_n, N_n).
\end{aligned}\end{equation}
Here, $p$ and $(1-p)$ are the prior probabilities of the two hypotheses: with probability $p$ the true hypothesis is $\rho^{\otimes n}$, and with probability $(1-p)$ it is $\sigma^{\otimes n}$.

Another way to think about this setting --- which in fact is closer to Chernoff's original motivations~\cite{chernoff_1952} --- is that we intend to construct a single sequence of tests that performs well for both types of errors. Precisely, we require that the given testing procedure have an equally low chance of mistaking $\rho$ for $\sigma$ (type I error) and of mistaking $\sigma$ for $\rho$ (type II error). This corresponds to the optimisation of the \emph{largest} (worst-case) error,
\begin{equation}\begin{aligned}
    \errmin_n (M_n, N_n) = \max \left\{ \alpha_n(M_n, N_n), \, \beta_n(M_n, N_n) \right\}.
\end{aligned}\end{equation}

At the level of exponents, the two definitions of error are asymptotically fully equivalent. This follows since if $\alpha_n \sim \exp(-nA)$ and $\beta_n \sim \exp(-nB)$, then
\begin{equation}\begin{aligned}
    \erravg_n \sim \exp(-n \min \{ A , B \}) \sim \errmin_n.
\end{aligned}\end{equation}
In particular, the asymptotics of the average error are actually independent of the priors. For this reason, in the analysis of exponents of symmetric hypothesis testing $p = \frac12$ is often assumed for simplicity. 

The quantum Chernoff bound~\cite{chernoff_1952,audenaert_2007,nussbaum_2009} says that the optimal asymptotic behaviour of the errors in this scenario is that they both  decay with the same error exponent, and hence so does their average. This exponent is determined by the Chernoff divergence $\xi(\rho\|\sigma)$ as
\begin{equation}\begin{aligned}
    \erravg_n \sim \errmin_n \sim \exp(-n \, \xi(\rho \| \sigma)),
\end{aligned}\end{equation}
where
\begin{equation}\begin{aligned}\label{eq:chernoff_divergence}
    \xi(\rho\|\sigma) \coloneqq \sup_{s \in (0,1)} (1-s) \, D_s(\rho\|\sigma) = \xi(\sigma\|\rho),
\end{aligned}\end{equation}
recalling that $D_s(\rho\|\sigma) = \frac{1}{s-1} \log \Tr \rho^{s} \sigma^{1-s}$ denotes the Petz--R\'enyi divergences.

\subsection{Inconclusive outcomes in symmetric testing}

When allowing for an inconclusive measurement outcome, the two definitions of error in symmetric hypothesis testing --- average and maximal error --- are no longer asymptotically equivalent.

The average probability that a three-outcome test $(M_n, N_n, \id - M_n - N_n)$ yields a conclusive outcome, meaning that one of the first two outcomes is obtained, is
\begin{equation}\begin{aligned}
    \pavg_n(M_n, N_n) &= p \Tr \rho (M_n + N_n) + (1-p) \Tr \sigma (M_n + N_n)\\
    &= \Tr (M_n + N_n) (p \rho + (1-p) \sigma).
\end{aligned}\end{equation}
The average error in a Bayesian sense can then be obtained through a conditional expectation: it is the expected probability of error conditioned on a conclusive outcome. It corresponds to the ratio of errors (average incorrect guesses) to all conclusive outcomes, which is
\begin{equation}\begin{aligned}\label{eq:sym_err_average}
    \perravg_n (M_n, N_n) = \frac{ p \Tr \rho^{\otimes n} N_n + (1-p) \Tr \sigma^{\otimes n} M_n }{\pavg_n(M_n, N_n)}.
\end{aligned}\end{equation}
This is how conditional symmetric errors were considered e.g.\ in~\cite{fiurasek_2003,regula_2024}.

Another possible definition would be to condition each of the two errors on a conclusive outcome separately, as we did in Section~\ref{sec:asymmetric}, and minimise the worst-case error
\begin{equation}\begin{aligned}
    \perrmin_n (M_n, N_n) = \max \left\{ \palpha_n(M_n, N_n), \, \pbeta_n(M_n, N_n) \right\}.
\end{aligned}\end{equation}

These two possible definitions of error in symmetric hypothesis testing with inconclusive outcomes --- $\perravg_n$ on one hand, and $\perrmin_n$ on the other --- can be understood as describing different models of the source, with the distinction being particularly relevant to 
understanding repeated experiments, through which it affects the calculation of achievable effective error exponents (cf.~Section~\ref{subsec:setting}). 
Consider a situation where, upon performing a measurement, we obtain an inconclusive outcome, and we would want to request additional copies of the state to repeat the experiment until a conclusive outcome is obtained.  Then the following question arises: when additional copies are requested from the source, 
do the new copies have the same ground truth as the copies that have been used already, i.e.\ is the true hypothesis ($\rho$ or $\sigma$) fixed, or do the new copies have a newly generated ground truth according to the prior probabilities $p$ and $1-p$?  If we assume the first possibility to be true, then one needs either $1/\pi_n(\rho)$ or $1/\pi_n(\sigma)$ repetitions on average until a conclusive outcome is obtained, and the associated conditional error probability is either $\palpha_n$ or $\pbeta_n$.  This then motivates the study of the worse-case scenario, where the conclusive probability and conditional error probability of interest are $\pi_n^\text{(min)}\coloneqq\min\{\pi_n(\rho),\pi_n(\sigma)\}$ and $\perrmin_n$.  The effective error exponent resulting from $\pi_n^\text{(min)}$ and $\perrmin_n$ then precisely quantifies the rate at which the error probability decays with respect to the expected number of copies consumed in the worst-case scenario under this model.  On the other hand, if we assume the second possibility to be true, then one needs $1/\pavg_n$ repetitions on average to reach a conclusion, and the associated conditional error probability is $\perravg_n$.  The effective error exponent relevant to this model is thus naturally defined based on $\pavg_n$ and $\perravg_n$.  To summarize, both the average and the maximum model of symmetric hypothesis testing are operationally justified, and their difference originates from the differing assumptions made about the source in repeated experiments.  In conventional hypothesis testing, as no repeated experiment due to inconclusiveness is needed, the difference between the two models disappears. 

Viewed another way, minimising $\perrmin_n$ is asymptotically fully equivalent to minimising a variant of an average error defined as $p \palpha_n + (1-p) \pbeta_n$ rather than through the Bayesian average in~\eqref{eq:sym_err_average}.  In fact, these two possible definitions of an average error probability can be thought of as arising from two distinct interpretations of a `mixture' of quantum states~\cite{cavalcanti_2012,brun_2012}, with~\eqref{eq:sym_err_average} corresponding to the so-called `improper mixture' while the variant $p \palpha_n + (1-p) \pbeta_n$ corresponding to the `proper mixture'~\cite{espagnat_2006}.  These two interpretations of a mixture are indistinguishable in linear models of information processing, such as conventional hypothesis testing, but their distinction becomes apparent when non-linear processing is involved, such as conditioning on conclusive outcomes. 
This distinction was also observed in the study of state discrimination assisted by closed timelike curves (which likewise feature non-linearity)~\cite{brun_2009,bennett_2009,ralph_2010,cavalcanti_2012,brun_2012}, where the definition of an average error probability has been a subject of debate. Here for the sake of clarity we reserve the name `average' for the Bayesian error defined in~\eqref{eq:sym_err_average}, but we stress that our analysis of the exponents of the maximal error $\perrmin_n$ immediately carries over to the `proper mixture' average.


\subsection{Beyond Chernoff in quantum state discrimination}

We will now characterise symmetric hypothesis testing in the inconclusive setting. As before, it will be important to understand not only the errors, but also the conclusive probabilities with which they are achieved --- $\pavg_n$ in the case of the average error, and $\min\{ \pi_n(\rho), \pi_n(\sigma)\}$ in case of the maximal error. 

The understanding of symmetric discrimination under the maximal error $\perrmin_n$ follows directly from the results in asymmetric hypothesis testing that we obtained throughout Section~\ref{sec:asymmetric}. 
In particular, an immediate application of the asymmetric result in Proposition~\ref{prop:highprob_exponents} is that the exact same protocol can be used for symmetric hypothesis testing, providing an achievability result with arbitrarily high probability of conclusive outcomes. 

\begin{corollary}\label{cor:symmetric_minimal}
In the symmetric hypothesis testing of two quantum states, the error exponent of maximal error $\perrmin_n \sim \exp( - n E)$ can be achieved with an arbitrarily high conclusive probability on both hypotheses, $\pi_n(\rho) \to 1$ and $\pi_n(\sigma) \to 1$, if and only if
\begin{equation}\begin{aligned}
    E \leq \min \!\big\{ D(\rho\|\sigma),\, D(\sigma\|\rho) \big\}.
\end{aligned}\end{equation}
\end{corollary}

From the definition of the Chernoff divergence, and using the fact that $D(\rho\|\sigma) \geq D_s (\rho\|\sigma)$ for all $s \in (0,1)$, we immediately see that this yields a performance at least as good as the Chernoff bound. Indeed, the improvement is generally strict. 

Working with the average error allows us to take a different approach by introducing a certain asymmetry to symmetric hypothesis testing. Since we are only concerned with the expected conclusive probability, it does not matter if the test cannot conclusively discriminate one of the two states, as long as the other has a sufficiently high discrimination probability. To make this intuition precise, let us momentarily assume that $D(\rho\|\sigma) \geq D(\sigma\|\rho)$ without loss of generality. We can then simply use the test $(M_n, 0, \id - M_n)$, where $((M_n, \id - M_n))_n$ is a sequence of tests that achieves the conventional Stein exponent $D(\rho\|\sigma)$ of type II error for some constant type I error $\ve$. The resulting protocol satisfies 
\begin{equation}\begin{aligned}
    \perravg_n \sim \exp(-nD(\rho\|\sigma)),\quad \pavg_n \geq p (1-\ve).
\end{aligned}\end{equation}
By taking $\ve$ to 0, we can make this probability arbitrarily close to the prior probability $p$. Together with a corresponding converse bound, we can state this as follows.

\begin{proposition}\label{prop:symmetric_average}
In the symmetric hypothesis testing of two quantum states, if the exponent of the average error  $\perravg_n \sim \exp( - n E)$ can be achieved with an average probability of conclusive outcomes satisfying $\pavg_n \geq c$ for some constant $c > 0$, then
\begin{equation}\begin{aligned}
    E \leq \max \!\big\{ D(\rho\|\sigma),\, D(\sigma\|\rho) \big\}.
\end{aligned}\end{equation}
Furthermore, there exists a sequence of tests that achieves this error exponent with average probability of conclusive outcomes $\pavg_n \geq q$, where $q = p$ if $D(\rho\|\sigma) > D(\sigma\|\rho)$, $q = 1- p$ if $D(\sigma\|\rho) > D(\rho\|\sigma)$, or $q = \max\{p, 1-p\}$ otherwise.
\end{proposition}

Although seemingly stronger than Corollary~\ref{cor:symmetric_minimal} at the level of the error exponent, we see that achievability here is not realised with arbitrarily high probability: the conclusiveness explicitly depends on the prior probabilities $p$ and $(1-p)$. Because of this, accounting for the required repetitions of the protocol due to inconclusive outcomes, the effectively achieved error exponent is reduced by a factor of $q$, in the sense that the rate of decrease of the error per each copy used in the process scales as $\sim \exp\left(-n\, q \max \!\big\{ D(\rho\|\sigma),\, D(\sigma\|\rho) \big\} \right)$.

Even with this reduction, this is still often enough to ensure an improvement over the conventional Chernoff bound. In particular, from the definition of the Chernoff divergence and its symmetry in the arguments it is not difficult to obtain the relation
\begin{equation}\begin{aligned}
    \xi(\rho \| \sigma) \leq \frac12 \max \!\big\{ D(\rho\|\sigma),\, D(\sigma\|\rho) \big\}.
\end{aligned}\end{equation}
Hence, as long as $q \geq \frac12$ --- which is trivially the case in the most commonly considered case of equal priors, where $p = \frac12$ --- then the error exponent achieved through the protocol of Proposition~\ref{prop:symmetric_average} outperforms the Chernoff bound.


\subsection{Symmetric hypothesis testing with high inconclusiveness}

The case of symmetric postselected hypothesis was first addressed in~\cite{regula_2024} where it was shown that, if the inconclusiveness is not constrained, the optimal average error exponent is given by $\perravg_n \sim \exp(-n D_\Xi(\rho\|\sigma))$ where $D_\Xi(\rho\|\sigma) = \max \{ D_{\max}(\rho\|\sigma), D_{\max}(\sigma\|\rho) \}$  denotes the Thompson metric. Here we will aim to understand exactly the trade-off between achievable exponents of average error and the exponents of conclusiveness needed to achieve them. Indeed, an extension of the techniques that we used in the study of the regime of high inconclusiveness of asymmetric postselected hypothesis testing in Sec.~\ref{sec:postselected_asym} can be used to establish the following.

\begin{proposition}\label{prop:symmetric_postselected_tradeoffs}
In symmetric hypothesis testing of two quantum states with average probability of conclusive outcome satisfying $\pavg_n \sim \exp(-n Z)$, the average error  $\perravg_n \sim \exp( - n E)$ can be achieved if and only if
\begin{align}
    E&\leq\max\left\{\inf_{s>1}\left(\frac{1}{s-1}Z+\wt{D}_s\fleft(\rho\middle\|\sigma\fright)\right),\right. \notag\\
    &\hphantom{\leq\max\big\{}\left.\inf_{t>1}\left(\frac{1}{t-1}Z+\wt{D}_t\fleft(\sigma\middle\|\rho\fright)\right)\right\},
\end{align}
or equivalently
\begin{equation}\begin{aligned}\label{eq:symmetric_posts_exponent}
    Z \geq \min \left\{ \vphantom{\sup_{s>1}} \right.  & \sup_{s>1} \,(s-1) \left( E - \wt{D}_s(\rho\|\sigma)\right),\\
    & \sup_{t>1} \,(t-1) \left(E - \wt{D}_t(\sigma\|\rho)\right) \left. \vphantom{ \sup_{s>1}} \right\}.
\end{aligned}\end{equation}
\end{proposition}

This once again gives a precise characterisation of the achievable trade-offs in this setting. The quantities appearing in~\eqref{eq:symmetric_posts_exponent} are not quite the Han--Kobayashi anti-divergences that we encountered earlier, but they are closely related and have indeed already found applications: they exactly quantify the asymptotic performance of a family of quantum Neyman--Pearson tests~\cite{mosonyi_2015}, or the asymptotic error exponent of variants of smoothed max-relative entropy~\cite{li_2023,regula_2025}.


\section{Discussion}\label{sec:discussion}

We introduced a general framework for the study of the performance of quantum hypothesis testing with a prescribed probability of inconclusive outcomes, both in the asymmetric and symmetric settings. We then comprehensively characterised how the landscape of optimal quantum hypothesis testing is reshaped by an allowance for inconclusiveness.

First, this allowed us to generalise classical frameworks such as hypothesis testing with rejection and almost-fixed-length hypothesis testing, where discrimination is studied with high probability of conclusiveness. There, we showed that conventional quantum bounds of Hoeffding and Chernoff can be exceeded in a number of ways, even with an arbitrarily small or exponentially vanishing inconclusiveness. The fact that such improvements are possible with negligibly small overhead compared to conventional hypothesis testing demonstrates a potential for significant practical advantages that can be obtained by simply allowing inconclusive measurement outcomes. 

We emphasise that this does not directly contradict the fundamental character of the Hoeffding and Chernoff bounds. Their interpretation as ultimate limits of hypothesis testing is certainly valid, but only when perfect conclusiveness is demanded. Our findings show that making a concession to allow for inconclusive outcomes, even vanishingly unlikely, is enough to exceed such limitations. Whether this is an acceptable concession to make  will certainly depend on the given hypothesis testing problem --- if one forbids any flexibility in the number of copies that can be used for each test, then repetition of the protocol becomes unfeasible, and our approach may not be possible to implement in a way that would be comparable to conventional, deterministic state discrimination; however, just a little flexibility would remove this limitation. We believe that the low operational overheads associated with the inconclusive protocols make this an appealing approach for a practical treatment of many problems encountered in quantum information.

Second, investigating the other extreme of the spectrum of conclusiveness led us to a complete characterisation of the `strong converse' regime of this setting, where we precisely evaluated the least rates of decay of conclusiveness required to achieve error exponents even higher than those that are achievable with vanishing inconclusive probability. This establishes the ultimate limits that cannot be surpassed even through a significant sacrifice in conclusiveness. On a technical level, our results revealed  new applications of sandwiched R\'enyi divergences in governing achievable error exponents, conceptually distinct from --- but technically connected to --- the previous uses of these divergences in the study of strong converse exponents in conventional hypothesis testing.

Some intriguing questions are left open. Notably, there is a marked difference between how straightforwardly the optimal error exponents of the low inconclusiveness regime can be achieved in the classical case (Sections~\ref{sec:asymmetric_classical} and~\ref{sec:exponentially_good}) and how the natural extensions of this idea to quantum states --- very often sufficient for the asymptotic study of quantum information --- do not yield optimal bounds (Section~\ref{sec:difficulties}). It would be interesting to investigate whether there is a more direct way to achieve the optimal quantum exponents, in particular allowing for an evaluation of the optimal error exponents when an exponential decay of the inconclusive probability is required (Section~\ref{sec:exponentially_good}).

The techniques introduced here can be generalised in a number of ways. Perhaps the most natural extension is the study of quantum channel discrimination, which is known to be more intricate than state discrimination~\cite{harrow_2010-1,wilde_2020,salek_2022,wang_2019-4,fang_2020-2}. Sequential protocols have already been studied in that setting~\cite{li_2022-1} as has postselected hypothesis testing~\cite{regula_2024,ji_2024-1}, and we are sure that protocols with inconclusive outcomes can again yield a number of insights into the limits of channel discrimination.
Another direction is composite quantum hypothesis testing, which involves testing hypotheses against a whole set of non-i.i.d.\ quantum states~\cite{bjelakovic_2005,brandao_2010,berta_2021,mosonyi_2021}, of key relevance to quantum resource manipulation~\cite{brandao_2010-1,berta_2024,lami_2024-1}. The recent years have seen major developments in the understanding of this setting~\cite{hayashi_2024,lami_2025-1,lami_2024-1}, and constraints in the form of Hoeffding- or Chernoff-style bounds have also been studied very recently~\cite{mosonyi_2021,fang_2025,fang_2025-1}. Although such problems are often markedly harder than simple i.i.d.\ ones and generally do not yield exact single-letter solutions even classically, there are cases in which they significantly simplify, leaving the door open to an exact evaluation of the optimal exponents~\cite{hayashi_2016-1,lami_2024-1}. An application of inconclusive protocols in such contexts would be another interesting extension of our methods.



\section*{Acknowledgments}

We thank Mark M. Wilde for many useful discussions and suggestions concerning this work. B.R. is also grateful to Mario Berta, Hao-Chung Cheng, Ludovico Lami, Roberto Rubboli, Ryuji Takagi,  Marco Tomamichel, and Kaito Watanabe for helpful comments and inspiring discussions about hypothesis testing and other big problems in life.  K.J. acknowledges the Canadian rock band Rush for mental stimulation.

K.J. acknowledges support from the National Science Foundation under grant no.~2329662 and from the Cornell School of Electrical and Computer Engineering. B.R. acknowledges the support of the Japan Science and Technology Agency (JST) PRESTO grant no.\ JPMJPR25FB and the Japan Society for the Promotion of Science (JSPS) KAKENHI grant no.\ 24K16984.


\bibliographystyle{apsc}
\bibliography{main}

\let\addcontentsline\oldaddcontentsline


\onecolumngrid
\clearpage
\newgeometry{left=1.2in,right=1.2in,top=.7in,bottom=1in}

\fakepart{Appendix}
\hypertarget{app}{}
\renewcommand{\theequation}{S\arabic{equation}}
\renewcommand{\thetheorem}{S\arabic{theorem}}
\renewcommand{\thesection}{S\arabic{section}}
\renewcommand{\thefigure}{S\arabic{figure}}

\hypertarget{supp}{}
\begin{center}
\vspace*{\baselineskip}
{\textbf{\large --- Appendix ---}}\\[1pt] \quad \\
\end{center}

\setcounter{tocdepth}{0}
\tableofcontents
\vspace*{2\baselineskip}

\setlength{\epigraphwidth}{.33\textwidth}
\epigraph{If you choose not to decide,\\you still have made a choice.}{\textit{Rush, `Freewill'}}

In the Appendix, we will take a different route towards introducing our main results, working backwards from general converse bounds and investigating how they change when additional constraints on the probability of conclusiveness are imposed, then studying in which settings and in what ways they can be achieved. This presents a slightly different perspective with distinct motivations for some of our investigations.

We begin in Section~\ref{app:conclusive_bigsection} with the study of the strong-converse--style bounds that constrain the achievable error exponents in hypothesis testing even when the probability of conclusiveness is allowed to decay with fixed exponents --- that is, the regime of high inconclusiveness. 
Here we obtain a complete characterisation of the error--conclusiveness trade-offs. 
Then, in Section~\ref{app:special} we study some implications of these bounds and in particular observe that taking their extreme cases, where the conclusive probability does \emph{not} vanish, leads to curious converse bounds for hypothesis testing with lower inconclusiveness, including our results on `one-sided conclusiveness'.
Although we show all of such bounds to be tight at the level of the exponents, the immediate achievability results with non-vanishing conclusiveness are somewhat weak: at best they imply a constant probability of conclusiveness, without elucidating how high this constant is. 
This motivates us to investigate the achievability of the converse bounds also in the much more practical regime of low inconclusiveness, where the conclusive probability is required to be high. In Section~\ref{app:typicality} we then discuss the classical problem of hypothesis testing with low (including exponentially low) inconclusiveness as well as a number of quantum achievability results that follow from it. Section~\ref{app:sequential} is then concerned with the adaptive discrimination protocol based on sequential hypothesis testing that achieves the optimal exponents of quantum state discrimination with arbitrarily low probability of inconclusive outcomes.

Readers interested in the proofs of the results in the main text can find them in the following sections.
\begin{itemize}
\item Proof of Proposition~\ref{prop:highprob_exponents}: Section~\ref{app:subsec_strongconverse} (strong converse) and Section~\ref{app:sequential} (achievability)
\item Proof of Corollary~\ref{cor:single_measurement_sequence}: Section~\ref{app:typical_measurements}
\item Proof of Proposition~\ref{prop:onesided}: Section~\ref{app:subsec_onesided}
\item Proof of Proposition~\ref{prop:conclusive}: Section~\ref{app:conclusive}
\item Proof of Lemma~\ref{lem:exponentially_classical}, Proposition~\ref{prop:exponentially_quantum}, and the achievability results discussed in  Section~\ref{sec:asymmetric}: Section~\ref{app:typicality}
\item Proof of Corollary~\ref{cor:symmetric_minimal}: immediate from Proposition~\ref{prop:highprob_exponents} 
\item Proof of Proposition~\ref{prop:symmetric_average}: Section~\ref{app:subsec_symmetric}
\item Proof of Proposition~\ref{prop:symmetric_postselected_tradeoffs}: Section~\ref{app:symmetric}.
\end{itemize}

\begin{figure*}[h]
 \includegraphics[width=.75\textwidth]{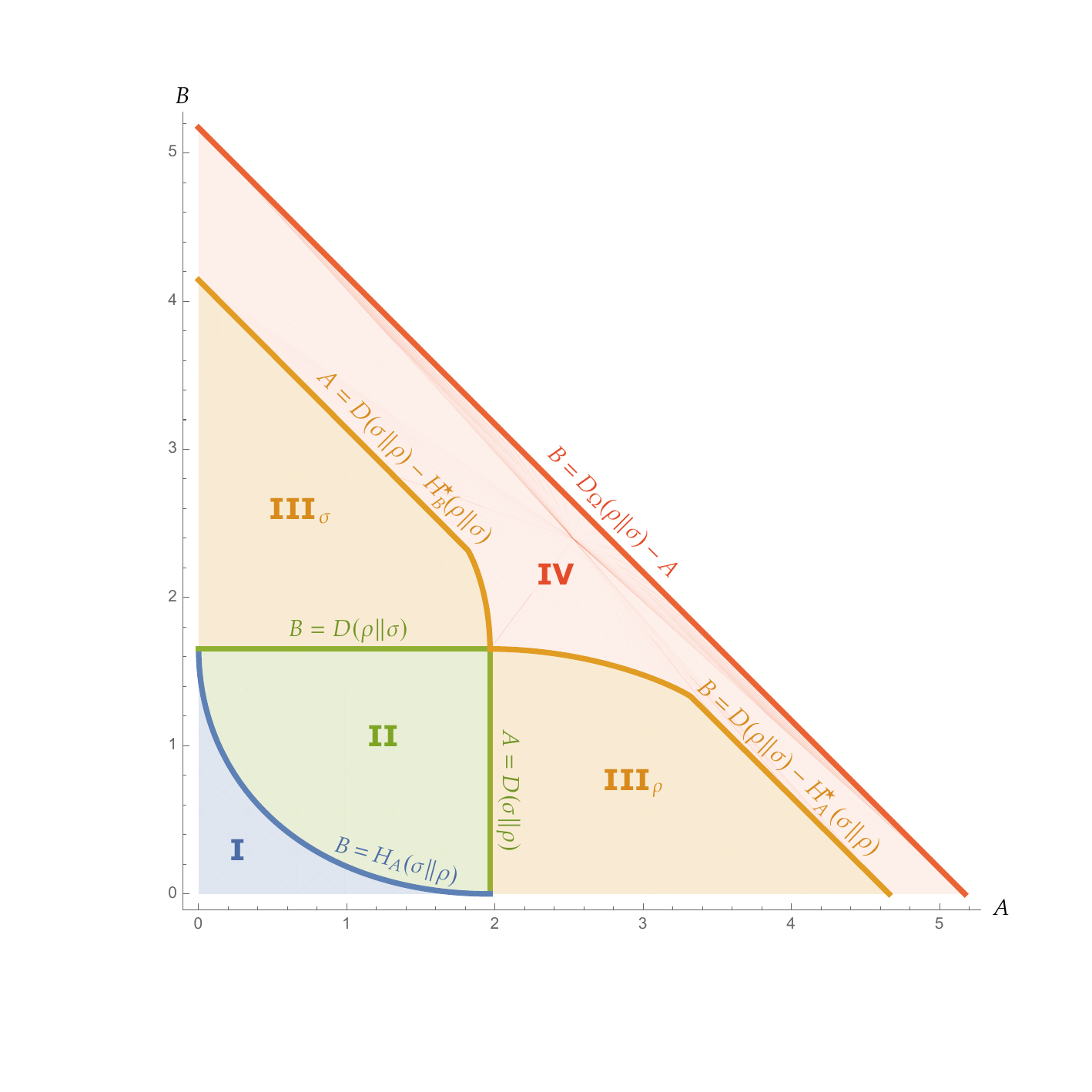}
\caption{\textbf{The regimes of asymmetric hypothesis testing.} Here we show the ranges of exponents of $\palpha_n \sim \exp(-nA)$ and $\pbeta_n \sim \exp(-nB)$ achievable in the different settings studied in this work. Region I is the conventional Hoeffding regime, achievable in hypothesis testing with no inconclusive outcomes~\cite{audenaert_2008}. Region II shows the regime of exponents achievable in inconclusive hypothesis testing with low inconclusiveness, where the asymptotic probability of inconclusive outcomes is an arbitrary constant or converges to 0 (Sections~\ref{app:typicality} and~\ref{app:sequential}). Regions III are the parameter ranges achievable when the inconclusiveness of one of the states is fixed as an arbitrary constant, while the inconclusiveness of the other is unconstrained --- region III${}_\rho$ fixes low inconclusiveness on $\rho$, while region III${}_\sigma$ on $\sigma$ (Section~\ref{app:special}). Finally, region IV represents hypothesis testing with high inconclusiveness (Section~\ref{app:conclusive_bigsection}), where the inconclusive probability converges to 1 and we study how imposing constraints on the speed of this affects the achievable exponent trade-offs.\\ 
(The states $\rho$, $\sigma$ considered in the plot are classical Bernoulli distributions with parameters $0.9$ and $0.2$, respectively, and the logarithm is to base two.)
}%
\end{figure*}

\section{High inconclusiveness: optimal exponents and trade-offs}
\label{app:conclusive_bigsection}

\subsection{Recap of notation}
\label{app:notation}

Given two states $\rho$ and $\sigma$ acting on the same, finite-dimensional Hilbert space, consider a sequence $((M_n,N_n,\id-M_n-N_n))_n$ of three-outcome tests on $n$ copies of the states, where $M_n,N_n\geq0$ are measurement operators satisfying $M_n+N_n\leq\id$ for all $n$. The \deff{conclusive probabilities} (or \emph{conclusiveness} for short) are defined as the probabilities of obtaining a conclusive outcome under hypothesis $\rho$ or $\sigma$:
\begin{align}
    \pi_n\fleft(\rho, M_n,N_n\fright)&\coloneq\Tr\fleft[\left(M_n+N_n\right)\rho^{\otimes n}\fright], \\
    \pi_n\fleft(\sigma, M_n,N_n\fright)&\coloneq\Tr\fleft[\left(M_n+N_n\right)\sigma^{\otimes n}\fright].
\end{align}
In our asymptotic analysis, we also often need to refer to \emph{inconclusiveness}, which corresponds to the probabilities of obtaining an inconclusive outcome: $ 1 - \pi_n\fleft(\rho, M_n,N_n\fright)$ and $ 1 - \pi_n\fleft(\sigma, M_n,N_n\fright)$. We do not introduce a separate notation for this concept.

The \deff{conditional error probabilities} of type I and type II are then respectively defined as
\begin{align}
    \overline{\alpha}_n\fleft(M_n,N_n\fright)&\coloneq\frac{\Tr\fleft[N_n\rho^{\otimes n}\fright]}{\pi_n\fleft(\rho,M_n,N_n\fright)}=\frac{\Tr\fleft[N_n\rho^{\otimes n}\fright]}{\Tr\fleft[\left(M_n+N_n\right)\rho^{\otimes n}\fright]}, \\
    \overline{\beta}_n\fleft(M_n,N_n\fright)&\coloneq\frac{\Tr\fleft[M_n\sigma^{\otimes n}\fright]}{\pi_n\fleft(\sigma,M_n,N_n\fright)}=\frac{\Tr\fleft[M_n\sigma^{\otimes n}\fright]}{\Tr\fleft[\left(M_n+N_n\right)\sigma^{\otimes n}\fright]}
\end{align}
for each positive integer $n$.

In the main text, we often referred to achievable exponents of the form $\palpha_n \sim \exp(-n A)$ and $\pbeta_n \sim \exp ( -n B)$; formally, $A > 0$  and $B> 0$ are called achievable (conditional) error exponents if, for all $\ve > 0$, there exists a sequence $((M_n,N_n,\id-M_n-N_n))_n$ of tests such that
\begin{equation}\begin{aligned}
    \liminf_{n\to\infty} - \frac1n \log \palpha_n(M_n,N_n) &\geq A - \ve,\\
     \liminf_{n\to\infty} - \frac1n \log \pbeta_n(M_n,N_n) &\geq B - \ve.
\end{aligned}\end{equation}
To ensure clarity of the statements of the results and also allow for stronger statements of converse bounds, we will hereafter refrain from using the notation $\sim$ and opt for more precise statements.


\subsection{Asymmetric hypothesis testing with high inconclusiveness}
\label{app:conclusive}

Recall the definition of the quantum Han--Kobayashi anti-divergence and its operational interpretation as the strong converse exponent of asymmetric hypothesis testing~\cite{mosonyi_2015}:
\begin{align}
	H_R^*\fleft(\rho\middle\|\sigma\fright)&\coloneq\sup_{s>1}\frac{s-1}{s}\left(R-\wt{D}_s\fleft(\rho\middle\|\sigma\fright)\right) \label{eq:antidivergence}\\
	&=\inf_{(M_n)_n}\left\{\begin{array}{c}
		\lim_{n\to\infty}-\frac{1}{n}\log\Tr\fleft[M_n\rho^{\otimes n}\fright]\colon \\
		\liminf_{n\to\infty}-\frac{1}{n}\log\Tr\fleft[M_n\sigma^{\otimes n}\fright]\geq R, \\
		0\leq M_n\leq\id\;\forall n
	\end{array}\right\}.
\end{align}
The quantity $H_R^*$ is also referred to as the Hoeffding anti-divergence in the literature.

We will find it particularly useful to rely on a slightly tighter formulation of the achievability of strong converse exponents that was observed in~\cite[Theorem~4.10 and Remark~4.11]{mosonyi_2015}:  given two states $\rho$ and $\sigma$ and two real numbers $R\geq0$ and $C\geq H_R^*(\rho\|\sigma)$, there exists a sequence $(M_n)_n$ of measurement operators such that
\begin{align}
    \liminf_{n\to\infty}-\frac{1}{n}\log\Tr\fleft[M_n\sigma^{\otimes n}\fright]&\geq R, \label{eq:antidivergence-1}\\
    \lim_{n\to\infty}-\frac{1}{n}\log\Tr\fleft[M_n\rho^{\otimes n}\fright]&=C. \label{eq:antidivergence-2}
\end{align}

The following proposition precisely characterises the achievable conditional error exponents of asymmetric hypothesis testing given that the conclusive probabilities decay at certain prescribed rates.   Alternatively, it can be understood as characterising the rates at which the conclusive probabilities can possibly decay in order to achieve a given pair conditional error exponents.

\begin{boxed}
\begin{proposition}[Formal statement of Proposition~\ref{prop:conclusive}]
\label{prop:conclusive-app}
Let $\rho$ and $\sigma$ be two states, and let $A,B,K,L\geq0$ be four non-negative real numbers.  The following statements are equivalent.
\begin{enumerate} 
	\item There exists a sequence $((M_n,N_n,\id-M_n-N_n))_n$ of three-outcome tests such that
	\begin{align}
		\liminf_{n\to\infty}-\frac{1}{n}\log\overline{\alpha}_n\fleft(M_n,N_n\fright)&\geq A, \label{pf:conclusive-3}\\
		\liminf_{n\to\infty}-\frac{1}{n}\log\overline{\beta}_n\fleft(M_n,N_n\fright)&\geq B, \\
		\limsup_{n\to\infty}-\frac{1}{n}\log\pi_n\fleft(\rho,M_n,N_n\fright)&=K, \\
		\limsup_{n\to\infty}-\frac{1}{n}\log\pi_n\fleft(\sigma,M_n,N_n\fright)&=L. \label{pf:conclusive-4}
	\end{align}\item There exists a sequence $((M_n,N_n,\id-M_n-N_n))_n$ of three-outcome tests such that
	\begin{align}
		\liminf_{n\to\infty}-\frac{1}{n}\log\overline{\alpha}_n\fleft(M_n,N_n\fright)&\geq A, \label{pf:conclusive-5}\\
		\liminf_{n\to\infty}-\frac{1}{n}\log\overline{\beta}_n\fleft(M_n,N_n\fright)&\geq B, \\
		\liminf_{n\to\infty}-\frac{1}{n}\log\pi_n\fleft(\rho,M_n,N_n\fright)&=K, \\
		\liminf_{n\to\infty}-\frac{1}{n}\log\pi_n\fleft(\sigma,M_n,N_n\fright)&=L. \label{pf:conclusive-6}
	\end{align} 
	\item The following inequalities are both satisfied:
	\begin{align}
		A+K&\leq\inf_{t>1}\left(\frac{t}{t-1}L+\wt{D}_t\fleft(\sigma\middle\|\rho\fright)\right), \label{pf:conclusive-1}\\
		B+L&\leq\inf_{s>1}\left(\frac{s}{s-1}K+\wt{D}_s\fleft(\rho\middle\|\sigma\fright)\right). \label{pf:conclusive-2}
	\end{align}
\end{enumerate}
\end{proposition}
\end{boxed}

\begin{proof}[Proof of 1 $\Rightarrow$ 3]
Let $((M_n,N_n,\id-M_n-N_n))_n$ be a sequence of three-outcome tests satisfying Eqs.~\eqref{pf:conclusive-3}--\eqref{pf:conclusive-4}.  For each $n$, let $\M_n$ denote the measurement channel corresponding to $(M_n,N_n,\id-M_n-N_n)$.  For all $s>1$, the data-processing inequality of $\wt{D}_s$ under $\M_n$ and the additivity of $\wt{D}_s$ imply that
\begin{align}
	&n\wt{D}_s\fleft(\rho\middle\|\sigma\fright) \notag\\
    &=\wt{D}_s\fleft(\rho^{\otimes n}\middle\|\sigma^{\otimes n}\fright) \\
	&\geq\wt{D}_s\fleft(\M_n\fleft[\rho^{\otimes n}\fright]\middle\|\M_n\fleft[\sigma^{\otimes}\fright]\fright) \\
	&=\frac{1}{s-1}\log\left(\Tr\fleft[M_n\rho^{\otimes n}\fright]^s\Tr\fleft[M_n\sigma^{\otimes n}\fright]^{1-s}+\Tr\fleft[N_n\rho^{\otimes n}\fright]^s\Tr\fleft[N_n\sigma^{\otimes n}\fright]^{1-s}\right. \notag\\
	&\quad\left.\vphantom{}+\Tr\fleft[\left(\id-M_n-N_n\right)\rho^{\otimes n}\fright]^s\Tr\fleft[\left(\id-M_n-N_n\right)\sigma^{\otimes n}\fright]^{1-s}\right) \\
	&=\frac{1}{s-1}\log\left(\left(1-\overline{\alpha}_n\fleft(M_n,N_n\fright)\right)^s\pi_n\fleft(\rho,M_n,N_n\fright)^s\overline{\beta}_n\fleft(M_n,N_n\fright)^{1-s}\pi_n\fleft(\sigma,M_n,N_n\fright)^{1-s}\right. \notag\\
	&\quad\left.\vphantom{}+\overline{\alpha}_n\fleft(M_n,N_n\fright)^s\pi_n\fleft(\rho,M_n,N_n\fright)^s\left(1-\overline{\beta}_n\fleft(M_n,N_n\fright)\right)^{1-s}\pi_n\fleft(\sigma,M_n,N_n\fright)^{1-s}\right. \notag\\
	&\quad\left.\vphantom{}+\left(1-\pi_n\fleft(\rho,M_n,N_n\fright)\right)^s\left(1-\pi_n\fleft(\sigma,M_n,N_n\fright)\right)^{1-s}\right) \\
	&\geq\frac{1}{s-1}\log\left(\left(1-\overline{\alpha}_n\fleft(M_n,N_n\fright)\right)^s\pi_n\fleft(\rho,M_n,N_n\fright)^s\overline{\beta}_n\fleft(M_n,N_n\fright)^{1-s}\pi_n\fleft(\sigma,M_n,N_n\fright)^{1-s}\right) \\
	&=-\log\overline{\beta}_n\fleft(M_n,N_n\fright)+\frac{s}{s-1}\log\pi_n\fleft(\rho,M_n,N_n\fright)-\log\pi_n\fleft(\sigma,M_n,N_n\fright)+\frac{s}{s-1}\log\left(1-\overline{\alpha}_n\fleft(M_n,N_n\fright)\right). \label{eq:DPI}
\end{align}
This implies that
\begin{align}
	&-\log\overline{\beta}_n\fleft(M_n,N_n\fright)+\frac{s}{s-1}\log\pi_n\fleft(\rho,M_n,N_n\fright) \notag\\
    &\quad\leq n\wt{D}_s\fleft(\rho\middle\|\sigma\fright)+\log\pi_n\fleft(\sigma,M_n,N_n\fright)-\frac{s}{s-1}\log\left(1-\overline{\alpha}_n\fleft(M_n,N_n\fright)\right).
\end{align}
Dividing both sides of the inequality by $n$ and taking the limit inferior as $n\to\infty$, we have that, for all $s>1$,
\begin{align}
    B-\frac{s}{s-1}K&\leq
	\wt{D}_s\fleft(\rho\middle\|\sigma\fright)-L.
\end{align}
Likewise, for all $t>1$,
\begin{align}
	A-\frac{t}{t-1}L&\leq\wt{D}_t\fleft(\sigma\middle\|\rho\fright)-K.
\end{align}
The direction 1 $\Rightarrow$ 3 of Proposition~\ref{prop:conclusive-app} then immediately follows.
\end{proof}

\begin{proof}[Proof of 2 $\Rightarrow$ 3]
Let $((M_n,N_n,\id-M_n-N_n))_n$ be a sequence of three-outcome tests satisfying Eqs.~\eqref{pf:conclusive-5}--\eqref{pf:conclusive-6}.  For each $n$, let $\M_n$ denote the measurement channel corresponding to $(M_n,N_n,\id-M_n-N_n)$.  It follows from Eq.~\eqref{eq:DPI} that, for all $s>1$,
\begin{align}
	&-\log\overline{\beta}_n\fleft(M_n,N_n\fright)-\log\pi_n\fleft(\sigma,M_n,N_n\fright) \notag\\
    &\quad\leq n\wt{D}_s\fleft(\rho\middle\|\sigma\fright)-\frac{s}{s-1}\log\pi_n\fleft(\rho,M_n,N_n\fright)-\frac{s}{s-1}\log\left(1-\overline{\alpha}_n\fleft(M_n,N_n\fright)\right).
\end{align}
Dividing both sides of the inequality by $n$ and taking the limit inferior as $n\to\infty$, we have that, for all $s>1$,
\begin{align}
    B+L&\leq\wt{D}_s\fleft(\rho\middle\|\sigma\fright)+\frac{s}{s-1}K.
\end{align}
Likewise, for all $t>1$,
\begin{align}
	A+K&\leq\wt{D}_t\fleft(\sigma\middle\|\rho\fright)+\frac{t}{t-1}L.
\end{align}
The direction 2 $\Rightarrow$ 3 of Proposition~\ref{prop:conclusive} then immediately follows.
\end{proof}

\begin{proof}[Proof of achievability (3 $\Rightarrow$ 1 and 2)]
Let $A,B,K,L\geq0$ be non-negative real numbers satisfying Eqs.~\eqref{pf:conclusive-1} and \eqref{pf:conclusive-2}.  This implies that, for all $s,t>1$,
\begin{align}
	K&\geq\frac{s-1}{s}\left(B+L-\wt{D}_s\fleft(\rho\middle\|\sigma\fright)\right), \\
	L&\geq\frac{t}{t-1}\left(A+K-\wt{D}_t\fleft(\sigma\middle\|\rho\fright)\right),
\end{align}
which, by Eq.~\eqref{eq:antidivergence}, are equivalent to
\begin{align}
	K&\geq H_{B+L}^*\fleft(\rho\middle\|\sigma\fright), \\
	L&\geq H_{A+K}^*\fleft(\sigma\middle\|\rho\fright).
\end{align}
By Eqs.~\eqref{eq:antidivergence-1} and \eqref{eq:antidivergence-2}, there exist two sequences of operators, $(M_n)_n$ and $(N_n)_n$, such that
\begin{align}
	\liminf_{n\to\infty}-\frac{1}{n}\log\Tr\fleft[M_n\sigma^{\otimes n}\fright]&\geq B+L, \label{pf:conclusive-7}\\
	\lim_{n\to\infty}-\frac{1}{n}\log\Tr\fleft[M_n\rho^{\otimes n}\fright]&=K, \label{pf:conclusive-8}\\
	\liminf_{n\to\infty}-\frac{1}{n}\log\Tr\fleft[N_n\rho^{\otimes n}\fright]&\geq A+K, \label{pf:conclusive-9}\\
	\lim_{n\to\infty}-\frac{1}{n}\log\Tr\fleft[N_n\sigma^{\otimes n}\fright]&=L, \label{pf:conclusive-10}\\
	0\leq M_n,N_n&\leq\id\quad\forall n.
\end{align}
Consider the following sequence of three-outcome tests: $((M_n/2,N_n/2,\id-M_n/2-N_n/2))_n$.  It follows from Eqs.~\eqref{pf:conclusive-9} and \eqref{pf:conclusive-8} that
\begin{align}
	\liminf_{n\to\infty}-\frac{1}{n}\log\overline{\alpha}_n\fleft(M_n/2,N_n/2\fright)&=\liminf_{n\to\infty}-\frac{1}{n}\log\frac{\Tr\fleft[N_n\rho^{\otimes n}\fright]}{\Tr\fleft[\left(M_n+N_n\right)\rho^{\otimes n}\fright]} \\
	&\geq\liminf_{n\to\infty}-\frac{1}{n}\log\Tr\fleft[N_n\rho^{\otimes n}\fright]-\limsup_{n\to\infty}-\frac{1}{n}\log\Tr\fleft[M_n\rho^{\otimes n}\fright] \\
	&\geq A+K-K \\
	&\geq A.
\end{align}
Likewise, it follows from Eqs.~\eqref{pf:conclusive-7} and \eqref{pf:conclusive-10} that
\begin{align}
	\liminf_{n\to\infty}-\frac{1}{n}\log\overline{\beta}_n\fleft(M_n/2,N_n/2\fright)&=\liminf_{n\to\infty}-\frac{1}{n}\log\frac{\Tr\fleft[M_n\sigma^{\otimes n}\fright]}{\Tr\fleft[\left(M_n+N_n\right)\sigma^{\otimes n}\fright]} \\
	&\geq\liminf_{n\to\infty}-\frac{1}{n}\log\Tr\fleft[M_n\sigma^{\otimes n}\fright]-\limsup_{n\to\infty}-\frac{1}{n}\log\Tr\fleft[N_n\sigma^{\otimes n}\fright] \\
	&\geq B+L-L \\
	&\geq B.
\end{align}
Moreover, it follows from Eqs.~\eqref{pf:conclusive-8} and \eqref{pf:conclusive-9} that there exists an integer $n_0$ such that $-\frac{1}{n}\log\Tr[M_n\rho^{\otimes n}]\leq-\frac{1}{n}\log\Tr[N_n\rho^{\otimes n}]$ for all $n\geq n_0$.  This implies that
\begin{align}
	\lim_{n\to\infty}-\frac{1}{n}\log\pi_n\fleft(\rho,M_n/2,N_n/2\fright)&=\lim_{n\to\infty}-\frac{1}{n}\log\Tr\fleft[\left(M_n+N_n\right)\rho^{\otimes n}\fright] \\
	&=\lim_{n\to\infty}\min\left\{-\frac{1}{n}\log\Tr\fleft[M_n\rho^{\otimes n}\fright],-\frac{1}{n}\log\Tr\fleft[N_n\rho^{\otimes n}\fright]\right\} \\
    &=\lim_{n\to\infty}-\frac{1}{n}\log\Tr\fleft[M_n\rho^{\otimes n}\fright] \\
    &=K.
\end{align}
Likewise, it follows from Eqs.~\eqref{pf:conclusive-7} and \eqref{pf:conclusive-10} that
\begin{align}
	\lim_{n\to\infty}-\frac{1}{n}\log\pi_n\fleft(\sigma,M_n/2,N_n/2\fright)&=\lim_{n\to\infty}-\frac{1}{n}\log\Tr\fleft[\left(M_n+N_n\right)\sigma^{\otimes n}\fright] \\
	&=\lim_{n\to\infty}\min\left\{-\frac{1}{n}\log\Tr\fleft[M_n\sigma^{\otimes n}\fright],-\frac{1}{n}\log\Tr\fleft[N_n\sigma^{\otimes n}\fright]\right\} \\
    &=\lim_{n\to\infty}-\frac{1}{n}\log\Tr\fleft[N_n\sigma^{\otimes n}\fright] \\
	&=L.
\end{align}
This proves the direction 3 $\Rightarrow$ 1 and 2 of Proposition~\ref{prop:conclusive}.
\end{proof}


\subsection{Symmetric hypothesis testing with high inconclusiveness}
\label{app:symmetric}

For two states $\rho$ and $\sigma$ with prior probabilities $p$ and $1-p$, respectively, and for a sequence $((M_n,N_n,\id-M_n-N_n))_n$ of three-outcome tests, where $M_n,N_n\geq0$ and $M_n+N_n\leq\id$ for all $n$, the \deff{average conclusive probability} for each positive integer $n$ is defined as
\begin{align}
	\pavg_n\fleft(M_n,N_n\fright)&\coloneq p\pi_n\fleft(\rho,M_n,N_n\fright)+\left(1-p\right)\pi_n\fleft(\sigma,M_n,N_n\fright)=\Tr\fleft[\left(M_n+N_n\right)\left(p\rho^{\otimes n}+\left(1-p\right)\sigma^{\otimes n}\right)\fright],
\end{align}
and the \deff{conditional average error probability} is defined as
\begin{align}
	\perravg_n\fleft(M_n,N_n\fright)&\coloneq\frac{p\,\overline{\alpha}_n\fleft(M_n,N_n\fright)\,\pi_n\fleft(\rho,M_n,N_n\fright)+\left(1-p\right)\overline{\beta}_n\fleft(M_n,N_n\fright)\pi_n\fleft(\sigma,M_n,N_n\fright)}{\pavg_n\fleft(M_n,N_n\fright)} \\
	&=\frac{\Tr\fleft[p\,N_n\rho^{\otimes n}+\left(1-p\right)M_n\sigma^{\otimes n}\fright]}{\Tr\fleft[\left(M_n+N_n\right)\left(p\rho^{\otimes n}+\left(1-p\right)\sigma^{\otimes n}\right)\fright]}.
\end{align}

The following proposition precisely characterises the achievable conditional error exponent of symmetric hypothesis testing given the rate at which the conclusive probability decays.   Alternatively, it can be understood as characterising the minimum rate at which the conclusive probability must decay in order to achieve a given conditional error exponent.

\begin{boxed}
\begin{proposition}[Formal statement of Proposition~\ref{prop:symmetric_postselected_tradeoffs}]
\label{prop:symmetric}
Let $\rho$ and $\sigma$ be two states, and let $E,Z\geq0$ be two non-negative real numbers.  There exists a sequence $((M_n,N_n,\id-M_n-N_n))_n$ of three-outcome tests such that
\begin{align}
	\liminf_{n\to\infty}-\frac{1}{n}\log\perravg_n\fleft(M_n,N_n\fright)&\geq E, \label{pf:symmetric-2}\\
	\limsup_{n\to\infty}-\frac{1}{n}\log\pavg_n\fleft(M_n,N_n\fright)&\leq Z \label{pf:symmetric-3}
\end{align} 
if and only if
\begin{align}
 	E&\leq\max\left\{\inf_{s>1}\left(\frac{1}{s-1}Z+\wt{D}_s\fleft(\rho\middle\|\sigma\fright)\right),\;\inf_{t>1}\left(\frac{1}{t-1}Z+\wt{D}_t\fleft(\sigma\middle\|\rho\fright)\right)\right\}. \label{pf:symmetric-1}
\end{align}
\end{proposition}
\end{boxed}

\begin{proof}[Proof of converse]
Let $((M_n,N_n,\id-M_n-N_n))_n$ be a sequence of three-outcome tests satisfying Eqs.~\eqref{pf:symmetric-2} and \eqref{pf:symmetric-3}.  For each $n$, let $\M_n$ denote the measurement channel corresponding to $(M_n,N_n,\id-M_n-N_n)$.  For a given $n$, suppose that $p\pi_n(\rho,M_n,N_n)\geq(1-p)\pi_n(\sigma,M_n,N_n)$.  In this case, for all $s>1$, it follows from Eq.~\eqref{eq:DPI} that
\begin{align}
	&n\wt{D}_s\fleft(\rho\middle\|\sigma\fright) \notag\\
    &\geq-\log\overline{\beta}_n\fleft(M_n,N_n\fright)+\frac{s}{s-1}\log\perravg_n\fleft(M_n,N_n\fright)-\log\lambda_n\fleft(M_n,N_n\fright)+\frac{s}{s-1}\log\left(1-\overline{\alpha}_n\fleft(M_n,N_n\fright)\right) \\
	&=-\log\left(\left(1-p\right)\overline{\beta}_n\fleft(M_n,N_n\fright)\,\lambda_n\fleft(M_n,N_n\fright)\right)+\frac{s}{s-1}\log\left(p\,\perravg_n\fleft(M_n,N_n\fright)\right) \notag\\
	&\quad+\frac{s}{s-1}\log\left(1-\overline{\alpha}_n\fleft(M_n,N_n\fright)\right)+\log\left(1-p\right)-\frac{s}{s-1}\log p \\
	&\geq-\log\left(p\,\overline{\alpha}_n\fleft(M_n,N_n\fright)\,\perravg_n\fleft(M_n,N_n\fright)+\left(1-p\right)\overline{\beta}_n\fleft(M_n,N_n\fright)\,\lambda_n\fleft(M_n,N_n\fright)\right) \notag\\
	&\quad+\frac{s}{s-1}\log\left(p\,\perravg_n\fleft(M_n,N_n\fright)+\left(1-p\right)\lambda_n\fleft(M_n,N_n\fright)\right)+\frac{s}{s-1}\log\left(1-\overline{\alpha}_n\fleft(M_n,N_n\fright)\right)+\log\left(1-p\right) \notag\\
	&\quad-\frac{s}{s-1}\log p-\frac{s}{s-1} \\
	&=-\log\left(\perravg_n\fleft(M_n,N_n\fright)\,\pavg_n\fleft(M_n,N_n\fright)\right)+\frac{s}{s-1}\log\pavg_n\fleft(M_n,N_n\fright)+\frac{s}{s-1}\log\left(1-\overline{\alpha}_n\fleft(M_n,N_n\fright)\right) \notag\\
	&\quad+\log\left(1-p\right)-\frac{s}{s-1}\log p-\frac{s}{s-1} \\
	&=-\log\perravg_n\fleft(M_n,N_n\fright)+\frac{1}{s-1}\log\pavg_n\fleft(M_n,N_n\fright)+\frac{s}{s-1}\log\left(1-\overline{\alpha}_n\fleft(M_n,N_n\fright)\right)+\log\left(1-p\right) \notag\\
	&\quad-\frac{s}{s-1}\log p-\frac{s}{s-1}. \label{pf:symmetric-4}
\end{align}
Otherwise, we have that $p\perravg_n(M_n,N_n)\leq(1-p)\lambda_n(M_n,N_n)$.  In this case, likewise, for all $t>1$, we have that
\begin{align}
	n\wt{D}_t\fleft(\sigma\middle\|\rho\fright)&\geq-\log\perravg_n\fleft(M_n,N_n\fright)+\frac{1}{t-1}\log\pavg_n\fleft(M_n,N_n\fright)+\frac{t}{t-1}\log\left(1-\overline{\beta}_n\fleft(M_n,N_n\fright)\right) \notag\\
	&\quad+\log p-\frac{t}{t-1}\log\left(1-p\right)-\frac{t}{t-1}. \label{pf:symmetric-5}
\end{align}
Equations~\eqref{pf:symmetric-4} and \eqref{pf:symmetric-5} imply that, for all $s,t>1$,
\begin{align}
	&-\log\perravg_n\fleft(M_n,N_n\fright)\\
	&\leq\max\left\{n\wt{D}_s\fleft(\rho\middle\|\sigma\fright)-\frac{1}{s-1}\log\pavg_n\fleft(M_n,N_n\fright)-\frac{s}{s-1}\log\left(1-\overline{\alpha}_n\fleft(M_n,N_n\fright)\right)-\log\left(1-p\right)+\frac{s}{s-1}\log p+\frac{s}{s-1},\right. \notag\\
	&\quad\left.n\wt{D}_t\fleft(\sigma\middle\|\rho\fright)-\frac{1}{t-1}\log\pavg_n\fleft(M_n,N_n\fright)-\frac{t}{t-1}\log\left(1-\overline{\beta}_n\fleft(M_n,N_n\fright)\right)-\log p+\frac{t}{t-1}\log\left(1-p\right)+\frac{t}{t-1}\right\}\notag.
\end{align}
Dividing both sides of the inequality by $n$, taking the limit inferior as $n\to\infty$ on the left-hand side and taking the limit superior on the right-hand side, we have that, for all $s,t>1$,
\begin{align}
	E&\leq\max\left\{\frac{1}{s-1}Z+\wt{D}_s\fleft(\rho\middle\|\sigma\fright),\;\frac{1}{t-1}Z+\wt{D}_t\fleft(\sigma\middle\|\rho\fright)\right\}.
\end{align}
The converse (i.e., the ``only if'' direction) then immediately follows.
\end{proof}

\begin{proof}[Proof of achievability]
Let $E,Z\geq0$ be non-negative real numbers satisfying Eq.~\eqref{pf:symmetric-1}.  Without loss of generality, we assume that
\begin{align}
	\inf_{s>1}\left(\frac{1}{s-1}Z+\wt{D}_s\fleft(\rho\middle\|\sigma\fright)\right)&\geq\inf_{t>1}\left(\frac{1}{t-1}Z+\wt{D}_t\fleft(\sigma\middle\|\rho\fright)\right).
\end{align}
This implies that, for all $s>1$,
\begin{align}
	E&\leq\frac{1}{s-1}Z+\wt{D}_s\fleft(\rho\middle\|\sigma\fright),
\end{align}
which is equivalent to
\begin{align}
	Z&\geq H_{E+Z}^*\fleft(\rho\middle\|\sigma\fright).
\end{align}
By Eqs.~\eqref{eq:antidivergence-1} and \eqref{eq:antidivergence-2}, there exists a sequence $(M_n)_n$ of operators such that
\begin{align}
	\liminf_{n\to\infty}-\frac{1}{n}\log\Tr\fleft[M_n\sigma^{\otimes n}\fright]&\geq E+Z, \\
	\limsup_{n\to\infty}-\frac{1}{n}\log\Tr\fleft[M_n\rho^{\otimes n}\fright]&=H_{E+Z}^*\fleft(\rho\middle\|\sigma\fright), \\
	0\leq M_n&\leq\id\quad\forall n.
\end{align}
Consider the following sequence of three-outcome tests: $((M_n,0,\id-M_n))_n$.  We have that
\begin{align}
	\liminf_{n\to\infty}-\frac{1}{n}\log\perravg_n\fleft(M_n,0\fright)&=\liminf_{n\to\infty}-\frac{1}{n}\log\frac{\Tr\fleft[M_n\sigma^{\otimes n}\fright]}{\Tr\fleft[M_n\left(\rho^{\otimes n}+\sigma^{\otimes n}\right)\fright]} \\
	&\geq\liminf_{n\to\infty}-\frac{1}{n}\log\Tr\fleft[M_n\sigma^{\otimes n}\fright]-\limsup_{n\to\infty}-\frac{1}{n}\log\Tr\fleft[M_n\rho^{\otimes n}\fright] \\
	&= E+Z-H_{E+Z}^*\fleft(\rho\middle\|\sigma\fright) \\
	&\geq E,
\end{align}
and
\begin{align}
	\limsup_{n\to\infty}-\frac{1}{n}\log\pavg_n\fleft(M_n,0\fright)&=\limsup_{n\to\infty}-\frac{1}{n}\log\Tr\fleft[M_n\left(\rho^{\otimes n}+\sigma^{\otimes n}\right)\fright] \\
	&\leq\limsup_{n\to\infty}-\frac{1}{n}\log\Tr\fleft[M_n\rho^{\otimes n}\fright] \\
	&\leq H_{E+Z}^*\fleft(\rho\middle\|\sigma\fright) \\
	&\leq Z.
\end{align}
This proves the achievability (i.e., the ``if'' direction).
\end{proof}

\begin{remark}
In the main text we also considered the \deff{maximal probability of error} in symmetric hypothesis testing, namely
\begin{equation}\begin{aligned}
    \perrmin_n (M_n, N_n) = \max \left\{ \palpha_n(M_n, N_n), \, \pbeta_n(M_n, N_n) \right\}.
\end{aligned}\end{equation}
The trade-offs concerning this setting can be derived directly from the asymmetric trade-offs of Proposition~\ref{prop:conclusive-app}. Namely, in this context we are concerned with the exponent of $\perrmin_n$ and the exponent of $\min\{ \pi_n(\rho), \pi_n(\sigma)\}$, which corresponds to the trade-off between $E'=\min\{A,B\}$ and $Z'=\max\{K,L\}$.  Their characterisation follows directly from Proposition~\ref{prop:conclusive-app} by setting $E'=A=B$ and $Z'=K=L$, leading to the following achievable region:
\begin{align}
	E'&\leq\min\left\{\inf_{s>1}\left(\frac{1}{s-1}Z'+\wt{D}_s\fleft(\rho\middle\|\sigma\fright)\right), \;\inf_{t>1}\left(\frac{1}{t-1}Z'+\wt{D}_t\fleft(\sigma\middle\|\rho\fright)\right)\right\},
\end{align}
which is only different from Proposition~\ref{prop:symmetric} by replacing the latter's maximum with a minimum.
\end{remark}

\section{One-sided constraints, strong converses, and other implications}
\label{app:special}

In this section, we present and discuss several consequences and corollaries of Propositions~\ref{prop:conclusive-app} and \ref{prop:symmetric} in various regimes of quantum hypothesis testing.

\subsection{High inconclusiveness regime}\label{app:subsec_Kdecay}

As can be inferred from Proposition~\ref{prop:conclusive-app}, the following corollary formalises the optimal trade-off between the conditional error exponents and the rate at which the conclusive probability under hypothesis $\rho$ decays, given that the conclusive probability under the other hypothesis $\sigma$ is unrestricted.

\begin{boxed}
\begin{corollary}
\label{cor:onesided-conclusive}
Let $\rho$ and $\sigma$ be two states, and let $A,B,K\geq0$ be three non-negative real numbers.  there exists a sequence $((M_n,N_n,\id-M_n-N_n))_n$ of three-outcome tests such that
\begin{align}
	\liminf_{n\to\infty}-\frac{1}{n}\log\overline{\alpha}_n\fleft(M_n,N_n\fright)&\geq A, \label{pf:special-1}\\
	\liminf_{n\to\infty}-\frac{1}{n}\log\overline{\beta}_n\fleft(M_n,N_n\fright)&\geq B, \\
	\limsup_{n\to\infty}-\frac{1}{n}\log\pi_n\fleft(\rho,M_n,N_n\fright)&\leq K \label{pf:special-2}
\end{align}
if and only if
\begin{align}\label{eq:conclusive_condensed}
	B&\leq\inf_{s,t>1}\left(\left(\frac{s}{s-1}-\frac{t-1}{t}\right)K+\wt{D}_s\fleft(\rho\middle\|\sigma\fright)-\frac{t-1}{t}\left(A-\wt{D}_t\fleft(\sigma\middle\|\rho\fright)\right)\right).
\end{align}
\end{corollary}
\end{boxed}

\begin{proof}
According to Proposition~\ref{prop:conclusive-app}, the existence of a sequence $((M_n,N_n,\id-M_n-N_n))_n$ of three-outcome tests satisfying Eqs.~\eqref{pf:special-1}--\eqref{pf:special-2} is equivalent to the existence of a non-negative real number $L\geq0$ satisfying Eqs.~\eqref{pf:conclusive-1} and \eqref{pf:conclusive-2}, restated as follows:
\begin{align}
	A+K&\leq\inf_{t>1}\left(\frac{t}{t-1}L+\wt{D}_t\fleft(\sigma\middle\|\rho\fright)\right), \label{pf:special-3}\\
	B+L&\leq\inf_{s>1}\left(\frac{s}{s-1}K+\wt{D}_s\fleft(\rho\middle\|\sigma\fright)\right). \label{pf:special-4}
\end{align}
The existence of such an $L$ is equivalent to
\begin{align}
	A&\leq\inf_{t>1}\left(\frac{t}{t-1}\left(\inf_{s>1}\left(\frac{s}{s-1}K+\wt{D}_s\fleft(\rho\middle\|\sigma\fright)\right)-B\right)+\wt{D}_t\fleft(\sigma\middle\|\rho\fright)\right)-K, \label{pf:special-5}\\
	B&\leq\inf_{s>1}\left(\frac{s}{s-1}K+\wt{D}_s\fleft(\rho\middle\|\sigma\fright)\right), \label{pf:special-6}
\end{align}
where Eq.~\eqref{pf:special-5} follows from combining Eqs.~\eqref{pf:special-3} and \eqref{pf:special-4}; Eq.~\eqref{pf:special-6} follows from $L\geq0$.  Equation~\eqref{pf:special-5} is further equivalent to
\begin{align}
	B&\leq\inf_{s,t>1}\left(\left(\frac{s}{s-1}-\frac{t-1}{t}\right)K+\wt{D}_s\fleft(\rho\middle\|\sigma\fright)-\frac{t-1}{t}\left(A-\wt{D}_t\fleft(\sigma\middle\|\rho\fright)\right)\right). \label{pf:special-7}
\end{align}
Since Eq.~\eqref{pf:special-7} implies Eq.~\eqref{pf:special-6}, it is equivalent to both Eqs.~\eqref{pf:special-5} and \eqref{pf:special-6} being satisfied.  Consequently, Eq.~\eqref{pf:special-7} is equivalent to the existence of a sequence $((M_n,N_n,\id-M_n-N_n))_n$ of three-outcome tests satisfying Eqs.~\eqref{pf:special-1}--\eqref{pf:special-2}.
\end{proof}

\subsection{One-sided conclusiveness regime}\label{app:subsec_onesided}

By investigating the special case of Corollary~\ref{cor:onesided-conclusive} for when $K=0$, the following corollary establishes an optimal trade-off between the type I and type II conditional error exponents when the conclusive probability under hypothesis $\rho$ is arbitrarily large and that under hypothesis $\sigma$ is unrestricted.

\begin{boxed}
\begin{corollary}[Formal statement of Proposition~\ref{prop:onesided}]\label{cor:onesided_app}
Let $\rho$ and $\sigma$ be two states, and let $A,B\geq0$ be two non-negative real numbers.  The following statements are equivalent.
\begin{enumerate}
    \item There exists a sequence $((M_n,N_n,\id-M_n-N_n))_n$ of three-outcome tests such that
    \begin{align}
    	\liminf_{n\to\infty}-\frac{1}{n}\log\overline{\alpha}_n\fleft(M_n,N_n\fright)&\geq A, \label{pf:special-9}\\
    	\liminf_{n\to\infty}-\frac{1}{n}\log\overline{\beta}_n\fleft(M_n,N_n\fright)&\geq B, \\
    	\liminf_{n\to\infty}\pi_n\fleft(\rho,M_n,N_n\fright)&>0. \label{pf:special-10}
    \end{align} 
    \item There exists a sequence $((M_n,N_n,\id-M_n-N_n))_n$ of three-outcome tests such that
    \begin{align}
    	\liminf_{n\to\infty}-\frac{1}{n}\log\overline{\alpha}_n\fleft(M_n,N_n\fright)&\geq A, \\
    	\liminf_{n\to\infty}-\frac{1}{n}\log\overline{\beta}_n\fleft(M_n,N_n\fright)&\geq B, \\
    	\liminf_{n\to\infty}\pi_n\fleft(\rho,M_n,N_n\fright)&=1.
    \end{align}
    \item The following inequality holds:
    \begin{align}
    	B&\leq D\fleft(\rho\middle\|\sigma\fright)-\Hsc_A\fleft(\sigma\middle\|\rho\fright). \label{pf:special-8}
    \end{align}
\end{enumerate}
\end{corollary}
\end{boxed}

\begin{proof}[Proof of 1 $\Rightarrow$ 3]
Let $((M_n,N_n,\id-M_n-N_n))_n$ be a sequence of three-outcome tests satisfying Eqs.~\eqref{pf:special-9}--\eqref{pf:special-10}.  Note that
\begin{align}
	\limsup_{n\to\infty}-\frac{1}{n}\log\pi_n\fleft(\rho,M_n,N_n\fright)&=0.
\end{align}
It thus follows from Corollary~\ref{cor:onesided-conclusive} by plugging in $K=0$ that
\begin{align}
    B&\leq\inf_{s,t>1}\left(\wt{D}_s\fleft(\rho\middle\|\sigma\fright)-\frac{t-1}{t}\left(A-\wt{D}_t\fleft(\sigma\middle\|\rho\fright)\right)\right) \\
	&=D\fleft(\rho\middle\|\sigma\fright)-\sup_{t>1}\frac{t-1}{t}\left(A-\wt{D}_t\fleft(\sigma\middle\|\rho\fright)\right) \\
	&=D\fleft(\rho\middle\|\sigma\fright)-\Hsc_A\fleft(\sigma\middle\|\rho\fright).
\end{align}
This proves the direction 1 $\Rightarrow$ 3 of Corollary~\ref{cor:onesided_app}.
\end{proof}

\begin{proof}[Proof of 3 $\Rightarrow$ 2]
Let $A,B\geq0$ be non-negative real numbers satisfying Eq.~\eqref{pf:special-8}.  It follows from the achievability part of the quantum Stein's lemma~\cite{hiai_1991} that, there exists a sequence $(M_n)_n$ of operators such that
\begin{align}
	\liminf_{n\to\infty}-\frac{1}{n}\log\Tr\fleft[M_n\sigma^{\otimes n}\fright]&\geq D\fleft(\rho\middle\|\sigma\fright), \\
	\liminf_{n\to\infty}\Tr\fleft[M_n\rho^{\otimes n}\fright]&=1, \\
	0\leq M_n&\leq\id\quad\forall n.
\end{align}
On the other hand, from the achievability of the strong converse exponents of quantum hypothesis testing~\cite{mosonyi_2015}, and in particular by Eqs.~\eqref{eq:antidivergence-1} and \eqref{eq:antidivergence-2}, there exists another sequence $(N_n)_n$ of operators such that
\begin{align}
	\liminf_{n\to\infty}-\frac{1}{n}\log\Tr\fleft[N_n\rho^{\otimes n}\fright]&\geq A, \\
	\limsup_{n\to\infty}-\frac{1}{n}\log\Tr\fleft[N_n\sigma^{\otimes n}\fright]&=\Hsc_A\fleft(\sigma\middle\|\rho\fright), \\
	0\leq N_n&\leq\id\quad\forall n.
\end{align}
Consider the following sequence of three-outcome tests: $(((1-\delta_n)M_n,\delta_n N_n,\id-(1-\delta_n)M_n-\delta_n N_n))_n$, where\ $(\delta_n)_n$ with $\delta_n\in(0,1)$ for all $n$ is an arbitrary sequence of real numbers decaying at a subexponential rate:
\begin{align}
    \lim_{n\to\infty}\delta_n=0, \qquad \lim_{n\to\infty}-\frac{1}{n}\log\delta_n=0.
\end{align}
Then we have that
\begin{align}
	&\liminf_{n\to\infty}-\frac{1}{n}\log\overline{\alpha}_n\fleft(\left(1-\delta_n\right)M_n,\delta_n N_n\fright) \notag\\
    &=\liminf_{n\to\infty}-\frac{1}{n}\log\frac{\delta_n\Tr\fleft[N_n\rho^{\otimes n}\fright]}{\Tr\fleft[\left(\left(1-\delta_n\right)M_n+\delta_nN_n\right)\rho^{\otimes n}\fright]} \\
    &\geq\lim_{n\to\infty}-\frac{1}{n}\log\delta_n+\liminf_{n\to\infty}-\frac{1}{n}\log\Tr\fleft[N_n\rho^{\otimes n}\fright]-\lim_{n\to\infty}-\frac{1}{n}\log\left(1-\delta_n\right)-\limsup_{n\to\infty}-\frac{1}{n}\log\Tr\fleft[M_n\rho^{\otimes n}\fright] \nonumber\\
	&=\liminf_{n\to\infty}-\frac{1}{n}\log\Tr\fleft[N_n\rho^{\otimes n}\fright]-\limsup_{n\to\infty}-\frac{1}{n}\log\Tr\fleft[M_n\rho^{\otimes n}\fright] \\
	&\geq A,
\end{align}
and
\begin{align}
	&\liminf_{n\to\infty}-\frac{1}{n}\log\overline{\beta}_n\fleft(\left(1-\delta_n\right)M_n,\delta_nN_n\fright) \notag\\
    &=\liminf_{n\to\infty}-\frac{1}{n}\log\frac{\left(1-\delta_n\right)\Tr\fleft[M_n\sigma^{\otimes n}\fright]}{\Tr\fleft[\left(\left(1-\delta_n\right)M_n+\delta_nN_n\right)\sigma^{\otimes n}\fright]} \\
    &\geq\lim_{n\to\infty}-\frac{1}{n}\log\left(1-\delta_n\right)+\liminf_{n\to\infty}-\frac{1}{n}\log\Tr\fleft[M_n\sigma^{\otimes n}\fright]-\lim_{n\to\infty}-\frac{1}{n}\log\delta_n-\limsup_{n\to\infty}-\frac{1}{n}\log\Tr\fleft[N_n\sigma^{\otimes n}\fright] \nonumber\\
	&=\liminf_{n\to\infty}-\frac{1}{n}\log\Tr\fleft[M_n\sigma^{\otimes n}\fright]-\limsup_{n\to\infty}-\frac{1}{n}\log\Tr\fleft[N_n\sigma^{\otimes n}\fright] \\
	&\geq D\fleft(\rho\middle\|\sigma\fright)-\Hsc_A\fleft(\sigma\middle\|\rho\fright) \\
	&\geq B,
\end{align}
and
\begin{align}
	\liminf_{n\to\infty}\pi_n\fleft(\rho,\left(1-\delta_n\right)M_n,\delta_n N_n\fright)&=\left(\lim_{n\to\infty}\left(1-\delta_n\right)\right)\liminf_{n\to\infty}\Tr\fleft[M_n\rho^{\otimes n}\fright]+\left(\lim_{n\to\infty}\delta_n\right)\liminf_{n\to\infty}\Tr\fleft[N_n\rho^{\otimes n}\fright] \\
	&\geq\liminf_{n\to\infty}\Tr\fleft[M_n\rho^{\otimes n}\fright] \\
	&=1.
\end{align}
This shows the direction 3 $\Rightarrow$ 2 of Corollary~\ref{cor:onesided_app}.
\end{proof}

\begin{proof}[Proof of 2 $\Rightarrow$ 1]
Trivial.
\end{proof}

It is also natural here to ask about the largest exponent of type I error achievable when the type II error is only required to vanish asymptotically but not at a particular rate,
while the constraint of high conclusiveness is only imposed on $\rho$. This corresponds to the point of intersection of the plot in Figure~\ref{fig:hoeffding} with the $A$-axis. Fixing $B=0$ and solving~\eqref{pf:special-8} for $A$, we can express this as follows.
\begin{boxed}
\begin{corollary}
For any nonnegative real number $A$ such that
\begin{equation}\begin{aligned}
A < D_+(\sigma\|\rho),
\end{aligned}\end{equation}
where
\begin{equation}\begin{aligned}
D_+(\sigma\|\rho) \coloneqq \inf_{s > 1}\left(\frac{s}{s-1}D\fleft(\rho\middle\|\sigma\fright)+\wt{D}_s\fleft(\sigma\middle\|\rho\fright)\right),
\end{aligned}\end{equation}
there exists a sequence $((M_n,N_n,\id-M_n-N_n))_n$ of three-outcome tests such that
\begin{align}
    \liminf_{n\to\infty}-\frac{1}{n}\log\overline{\alpha}_n\fleft(M_n,N_n\fright)&\geq A, \label{eq:dplus1} \\
    \lim_{n\to\infty}\overline{\beta}_n\fleft(M_n,N_n\fright)& = 0, \label{pf:dplus}\\
    \lim_{n\to\infty}\pi_n\fleft(\rho,M_n,N_n\fright)&=1. \label{eq:dplus3}
\end{align} 
Conversely, if there exists a sequence of tests satisfying~\eqref{eq:dplus1}--\eqref{eq:dplus3}, then
\begin{equation}\begin{aligned}
A \leq D_+(\sigma\|\rho).
\end{aligned}\end{equation}
\end{corollary}
\end{boxed}

As $\frac{s}{s-1} \geq 1$, we see in particular that $D_+(\sigma\|\rho)\geq D(\rho\|\sigma)+D(\sigma\|\rho)$, so there is always a strict advantage over conventional hypothesis testing, where only the exponents $A \leq D(\sigma\|\rho)$ can be achieved.

We now give a simple sufficient condition that allows for a major simplification of the expression for $D_+(\sigma\|\rho)$ for many pairs of quantum states.
\begin{boxed}
\begin{lemma}\label{lem:intercept}
Let $\Pi$ denote the projector onto the eigenspace of the largest eigenvalue of $\rho^{-1/2}\sigma\rho^{-1/2}$. If
\begin{equation}\begin{aligned}\label{eq:condition}
    D(\rho\|\sigma) + D_{\max}(\sigma\|\rho) \geq - \log \Tr \Pi \exp\!\big(\Pi \log(\rho) \Pi\big),
\end{aligned}\end{equation}
then the quantity $D_+$ takes the simplified form
\begin{equation}\begin{aligned}\label{eq:Dplus_simplified}
    D_+(\sigma\|\rho) = D(\rho\|\sigma) + D_{\max}(\sigma\|\rho).
\end{aligned}\end{equation}
Here, $D_{\max}$ denotes the max-relative entropy
\begin{align}
 D_{\max}(\rho\|\sigma) \coloneqq \inf \lset \lambda \bar \rho \leq \exp(\lambda)\, \sigma \rset.
\end{align} 
\end{lemma}
\end{boxed}
\begin{remark}
In the classical (commuting) case, the condition~\eqref{eq:condition} reduces to $D(P\|Q) + D_{\max}(Q\|P) \geq - \log \Tr \Pi P$ or equivalently $D(P\|Q) \geq - \log \Tr \Pi Q$. In this case, this condition is also necessary for $D_+$ to simplify as in~\eqref{eq:Dplus_simplified}.
\end{remark}
\begin{proof}
Recall that $D_+(\sigma\|\rho)$ corresponds to the value of $A$ for which $D\fleft(\rho\middle\|\sigma\fright)-\Hsc_A\fleft(\sigma\middle\|\rho\fright) = 0$. 
According to~\cite[Lemma~IV.9]{mosonyi_2015}, 
\begin{equation}\begin{aligned}
    \Hsc_A(\sigma\|\rho) = A - D_{\max}(\sigma \| \rho)
\end{aligned}\end{equation}
if and only if
\begin{equation}\begin{aligned}\label{eq:milan_inequality}
 A \geq \sup_{s \geq 1} \,(s-1) \left( D_{\max}(\sigma \| \rho) - \wt{D}_s(\sigma\|\rho)\right) + D_{\max}(\sigma\|\rho).
\end{aligned}\end{equation}
Assuming that this is true and solving $D\fleft(\rho\middle\|\sigma\fright)-\Hsc_A\fleft(\sigma\middle\|\rho\fright) = 0$ for $A$ gives
\begin{equation}\begin{aligned}
    A = D(\rho\|\sigma) + D_{\max}(\sigma\|\rho).
\end{aligned}\end{equation}
which is precisely the claimed expression. 
We thus need to investigate when~\eqref{eq:milan_inequality} is satisfied.

Defining $f(s) \coloneqq (s-1) \left( D_{\max}(\sigma \| \rho) - \wt{D}_s(\sigma\|\rho)\right)$, we first show that the supremum of $f(s)$ over $s \geq 1$ is always achieved in the limit $s \to \infty$. To see this, defining $\lambda = \exp(D_{\max}(\sigma\|\rho))$, $\Delta = \rho^{-1/2}\sigma\rho^{-1/2}$ and $\wt{\Delta} = \Delta / \lambda$, rewrite
\begin{equation}\begin{aligned}
    f(s) &= (s-1) \log \lambda - \log \Tr \left( \rho^{\frac{1-s}{2s}} \sigma \rho^{\frac{1-s}{2s}}\right)^s\\
    &= (s-1) \log \lambda - \log \Tr \left(  \rho^{\frac{1}{2s}} \Delta \rho^{\frac{1}{2s}} \right)^s\\
    &= (s-1) \log \lambda - \log \Tr \left(  \rho^{\frac{1}{2s}} \wt\Delta \rho^{\frac{1}{2s}} \right)^s - s \log \lambda\\
    &= - \log \lambda - \log \Tr \left(  \rho^{\frac{1}{2s}} \wt\Delta \rho^{\frac{1}{2s}} \right)^s.
\end{aligned}\end{equation}
We would like to show that $f(s)$ is non-decreasing in $s$, which now reduces to showing that
\begin{equation}\begin{aligned}
    g(s) \coloneqq \log \Tr \left(  \rho^{\frac{1}{2s}} \wt\Delta \rho^{\frac{1}{2s}} \right)^s
\end{aligned}\end{equation}
is non-increasing in $s$. Since this can be written as
\begin{equation}\begin{aligned}
    g(s) = s \log \norm{\rho^{\frac{1}{2s}} \wt\Delta \rho^{\frac{1}{2s}}}{s}
\end{aligned}\end{equation}
where $\norm{\cdot}{s}$ denotes the Schatten $p$-norm, an application of~\cite[Corollary~3]{beigi_2013} with the choice of parameters $p_0 = s$, $p_1 = \infty$, and $p_\theta = \s > s$ tells us that
\begin{equation}\begin{aligned}
    \norm{\rho^{\frac{1}{2\s}} \wt\Delta \rho^{\frac{1}{2\s}}}{\s} \leq \norm{\rho^{\frac{1}{2s}} \wt\Delta \rho^{\frac{1}{2s}}}{s}^{\frac{s}{\s}} \, \norm{\rho^{\frac{1}{2s}} \wt\Delta \rho^{\frac{1}{2s}}}{\infty}^{\frac{\s-s}{s}}
\end{aligned}\end{equation}
Noting that $\lambda = \norm{\Delta}{\infty}$, it holds that $\|\wt\Delta\|_\infty = 1$, and hence $\norm{\rho^{\frac{1}{2s}} \wt\Delta \rho^{\frac{1}{2s}}}{\infty} \leq \norm{\rho^{\frac{1}{s}}}{\infty} \leq 1$ since $\norm{\rho}{\infty}\leq 1$. Then
\begin{equation}\begin{aligned}
    g({\s}) = \log \norm{\rho^{\frac{1}{2\s}} \wt\Delta \rho^{\frac{1}{2\s}}}{\s}^{\s} \leq \log \norm{\rho^{\frac{1}{2s}} \wt\Delta \rho^{\frac{1}{2s}}}{s}^{s} = g(s)
\end{aligned}\end{equation}
for any $\s > s$, which shows that $f(s)$ is non-decreasing in $s$ as desired. 

It thus remains to bound the limit $s \to \infty$ of $f(s)$. Write $\Delta$ in its spectral decomposition as $\Delta = \lambda \Pi + \sum_{j>1} \lambda_j \Pi_j$ where $\lambda_j < \lambda$, recalling that $\Pi$ denotes the projector onto the eigenspace of the largest eigenvalue of $\Delta$, which is $\lambda$. 
Using the fact that $\Delta \geq \lambda \Pi$, the unitary invariance of Schatten norms implies that
\begin{equation}\begin{aligned}
\Tr \left(  \rho^{\frac{1}{2s}} \Delta \rho^{\frac{1}{2s}} \right)^s &\geq \lambda^s \Tr \left(  \rho^{\frac{1}{2s}} \Pi \rho^{\frac{1}{2s}} \right)^s\\
&= \lambda^s \Tr \left(  \Pi \rho^{\frac{1}{s}} \Pi \right)^s\\
&= \lambda^s \Tr \left( \Pi \exp\!\left( \frac{\log(\rho)}{s}\right) \Pi\right)^s.
\end{aligned}\end{equation}
The evaluation of the limit of this expression as $s\to\infty$ follows from a Lie--Trotter argument. While the presence of the projection $\Pi$ prevents a direct application of the standard formulation of this result (e.g.~\cite[Theorem~IX.1.3]{bhatia_1996}), a minor modification of the standard proof suffices. 
Let us for simplicity momentarily assume that $\exp$ and $\log$ are to the natural base $e$. 
Using the Taylor series expansion of $\exp(x)$ gives
\begin{equation}\begin{aligned}
     \exp\!\left(\frac{\log(\rho)}{s}\right) = \id + \frac{\log(\rho)}{s} + O\!\left( \frac{1}{s^2} \right).
\end{aligned}\end{equation}
Applying the projection $\Pi$, the resulting expression --- understood as an operator acting only on $\supp\Pi$ --- is
\begin{equation}\begin{aligned}
    \left.\Pi \exp\!\left(\frac{\log(\rho)}{s}\right)\Pi \;\right|_{\supp\Pi}  = \id + \frac{\Pi\log(\rho)\Pi}{s} + O\!\left( \frac{1}{s^2} \right).
\end{aligned}\end{equation}
To understand what happens when exponentiating this expression, use the Taylor series of $\log(1+x)$ to write
\begin{equation}\begin{aligned}
    \log\left[ \left(\Pi \exp\!\left(\frac{\log(\rho)}{s}\right)\Pi \right)^s \right] &= s \log \left[ \id + \frac{\Pi\log(\rho)\Pi}{s} + O\!\left( \frac{1}{s^2} \right) \right]\\
    \\&= s \left[ \frac{\Pi\log(\rho)\Pi}{s} + O\!\left( \frac{1}{s^2} \right) + O\!\left( \frac{1}{s^2} \right) \right]\\
    &= \Pi\log(\rho)\Pi + O\!\left( \frac{1}{s} \right),
\end{aligned}\end{equation}
which by the continuity of the exponential function ensures that
\begin{equation}\begin{aligned}
    \lim_{s\to\infty} \left(\Pi \exp\!\left(\frac{\log(\rho)}{s}\right)\Pi  \right)^s = \exp\left( \Pi\log(\rho)\Pi \,\big|_{\supp\Pi} \right).
\end{aligned}\end{equation}
Putting everything together, we get
\begin{align}
    \sup_{s \geq 1} \,(s-1) \left( D_{\max}(\sigma \| \rho) - \wt{D}_s(\sigma\|\rho)\right) &= \lim_{s\to\infty} f(s)\nonumber\\
    &= \lim_{s\to\infty} \left[ (s-1) \log \lambda - \log \Tr \left(  \rho^{\frac{1}{2s}} \Delta \rho^{\frac{1}{2s}} \right)^s \right]\\
    &\leq \lim_{s\to\infty} \left[ (s-1) \log \lambda - s \log \lambda - \log \Tr \left(\Pi \exp\!\left(\frac{\log(\rho)}{s}\right)\Pi  \right)^s \right]\nonumber\\
    &= - D_{\max}(\sigma\|\rho) - \log \Tr \Pi \exp\!\big( \Pi\log(\rho)\Pi\big).\nonumber
\end{align}
This then shows that if $A \geq - \log \Tr \Pi \exp\!\big( \Pi\log(\rho)\Pi\big)$, then~\eqref{eq:milan_inequality} is satisfied, ensuring that $D_+(\sigma\|\rho) = D(\rho\|\sigma) + D_{\max}(\sigma\|\rho)$.
\end{proof}

We expect that the statement of Lemma~\ref{lem:intercept} can be tightened to an `if and only if', that is, that
\begin{equation}\begin{aligned}\label{eq:conjectured_legendre}
    \sup_{s \geq 1} \,(s-1) \left( D_{\max}(\sigma \| \rho) - \wt{D}_s(\sigma\|\rho)\right) \texteq{?} - \log \Tr \Pi \exp\!\big( \Pi\log(\rho)\Pi\big) - D_{\max}(\sigma\|\rho).
\end{aligned}\end{equation}
We were not able to verify this in general, although a lower bound can be obtained as follows. An application of the Araki--Lieb--Thirring inequality tells us that
\begin{equation}\begin{aligned}
   \Tr \left(  \rho^{\frac{1}{2s}} \Delta \rho^{\frac{1}{2s}} \right)^s &\leq \Tr \Delta^s \rho\\
   &= \lambda^s \Tr \Pi \rho + \sum_{j>1} \lambda_j^s \Tr \Pi_j \rho\\
   &= \lambda^s \Big( \Tr \Pi \rho + o(1)\Big).
\end{aligned}\end{equation}
Hence,
\begin{equation}\begin{aligned}
    \lim_{s\to\infty} f(s) &\geq \lim_{s\to\infty} \left[ (s-1) \log \lambda - s \log \lambda - \log \Big( \Tr \Pi \rho + o(1)\Big) \right]\\
    &= - \log (\lambda \Tr \Pi \rho )\\
    &= - \log \Tr \Pi \sigma.
\end{aligned}\end{equation}
This in particular shows that equality in~\eqref{eq:conjectured_legendre} is true for commuting states.

\subsection{`Strong converse' for the low inconclusiveness regime}\label{app:subsec_strongconverse}

The following corollary establishes an exponential strong-converse--like statement for Proposition~\ref{prop:highprob_exponents} based on Corollary~\ref{cor:onesided-conclusive}, showing that exceeding the error exponents given by the relative entropy necessarily leads to an asymptotically vanishing probability of conclusiveness.

\begin{boxed}
\begin{corollary}[`Strong converse' for Proposition~\ref{prop:highprob_exponents}]\label{cor:strong_converse_for_highconc}
Let $\rho$ and $\sigma$ be two states.  If there exists a sequence $((M_n,N_n,\id-M_n-N_n))_n$ of three-outcome tests such that
\begin{align}
	\liminf_{n\to\infty}-\frac{1}{n}\log\overline{\beta}_n\fleft(M_n,N_n\fright)& > D(\rho\|\sigma), \label{pf:strong_converse-2}
\end{align}
then
\begin{align}
	\liminf_{n\to\infty}-\frac{1}{n}\log\pi_n\fleft(\rho,M_n,N_n\fright)&>0.
\end{align}
Analogously, if there exists a sequence $((M_n,N_n,\id-M_n-N_n))_n$ of three-outcome tests such that
\begin{align}
	\liminf_{n\to\infty}-\frac{1}{n}\log\overline{\alpha}_n\fleft(M_n,N_n\fright)&>D(\sigma\|\rho), \label{pf:strong_converse-1}
\end{align}
then
\begin{align}
	\liminf_{n\to\infty}-\frac{1}{n}\log\pi_n\fleft(\sigma,M_n,N_n\fright)&>0.
\end{align}
\end{corollary}
\end{boxed}

\begin{proof}
Suppose that there exists a sequence $((M_n,N_n,\id-M_n-N_n))_n$ of three-outcome tests satisfying Eq.~\eqref{pf:strong_converse-2}.  Denoting
\begin{align}
	B&=\liminf_{n\to\infty}-\frac{1}{n}\log\overline{\beta}_n\fleft(M_n,N_n\fright), \\
	K&=	\liminf_{n\to\infty}-\frac{1}{n}\log\pi_n\fleft(\rho,M_n,N_n\fright), \\
\end{align}
it follows from Corollary~\ref{cor:onesided-conclusive} that, for all $s>1$,
\begin{align}
    B&\leq\inf_{t>1}\left(\left(\frac{s}{s-1}-\frac{t-1}{t}\right)K+\wt{D}_s\fleft(\rho\middle\|\sigma\fright)-\frac{t-1}{t}\left(A-\wt{D}_t\fleft(\sigma\middle\|\rho\fright)\right)\right) \\
    &\leq\frac{s}{s-1}K+\wt{D}_s\fleft(\rho\middle\|\sigma\fright),
\end{align}
which implies that
\begin{align}
    K&\geq\sup_{s>1}\frac{s-1}{s}\left(B-\wt{D}_s\fleft(\rho\middle\|\sigma\fright)\right) \\
    &=H_B^*\fleft(\rho\middle\|\sigma\fright).
\end{align}
Since $B>D(\rho\|\sigma)$, we necessarily have that $K>0$.  Likewise, if there exists a sequence $((M_n,N_n,\id-M_n-N_n))_n$ of three-outcome tests satisfying Eq.~\eqref{pf:strong_converse-1}, then $L>0$,
where
\begin{align}
	L&=\liminf_{n\to\infty}-\frac{1}{n}\log\pi_n\fleft(\sigma,M_n,N_n\fright).
\end{align}
This completes the proof of the desired statement.
\end{proof}

\subsection{Postselected hypothesis testing}\label{app:subsec_postsel}

This setting, introduced in~\cite{regula_2024}, is concerned with the study of achievable conditional error exponents when the inconclusiveness is not constrained whatsoever. However, the work~\cite{regula_2024} only studied the case where one of the two conditional errors is constant, i.e.\ either $A=0$ or $B=0$, and in the symmetric setting only the average error $\perravg$ was considered.

 An immediate application of Proposition~\ref{prop:conclusive-app} recovers the asymptotic result of~\cite{regula_2024} in asymmetric hypothesis testing and extends it to show that, when exponential decay of both conditional errors is imposed, there is a simple linear trade-off between them.

\begin{boxed}
\begin{corollary}\label{cor:postsel_asym_app}
Let $\rho$ and $\sigma$ be two states, and let $A,B\geq0$ be non-negative real numbers. There exists a sequence $((M_n,N_n,\id-M_n-N_n))_n$ of three-outcome tests such that
\begin{align}
	\liminf_{n\to\infty}-\frac{1}{n}\log\overline{\alpha}_n\fleft(M_n,N_n\fright)&\geq A, \\
	\liminf_{n\to\infty}-\frac{1}{n}\log\overline{\beta}_n\fleft(M_n,N_n\fright)&\geq B
\end{align} 
if and only if
\begin{align}
	A + B \leq D_{\Omega}(\rho\|\sigma) = D_{\max}(\rho\|\sigma) + D_{\max}(\sigma||\rho).
\end{align} 
\end{corollary}
\end{boxed}
\begin{proof}
Observe that taking $K \to \infty$ in Corollary~\ref{cor:onesided-conclusive} forces the parameters $s,t$ to both diverge to infinity so that the term $\frac{s}{s-1}-\frac{t-1}{t}$ can vanish in order to minimise Eq.~\eqref{eq:conclusive_condensed}. The statement then follows since
$\lim_{s\to\infty} \wt{D}_s (\rho \| \sigma) = D_{\max}(\rho\|\sigma)$~\cite{muller-lennert_2013}.
\end{proof}

From the above, we also obtain a characterisation of symmetric postselected hypothesis testing under the maximal error $ \perrmin_n (M_n, N_n) = \max \left\{ \palpha_n(M_n, N_n), \, \pbeta_n(M_n, N_n) \right\}$.

\begin{boxed}
\begin{corollary}
Let $\rho$ and $\sigma$ be two states, and let $E\geq0$. There exists a sequence $((M_n,N_n,\id-M_n-N_n))_n$ of three-outcome tests such that
\begin{align}
	\liminf_{n\to\infty}-\frac{1}{n}\log\perrmin_n\fleft(M_n,N_n\fright)&\geq E
\end{align} 
if and only if
\begin{align}
	E \leq \frac12 D_{\Omega}(\rho\|\sigma).
\end{align} 
\end{corollary}
\end{boxed}
\begin{proof}
Follows since maximising $\min \{ A, B \}$ under the constraint $A + B \leq D_{\Omega}(\rho\|\sigma)$ imposed by Corollary~\ref{cor:postsel_asym_app} gives $A = B = \frac12 D_{\Omega}(\rho\|\sigma)$.
\end{proof}
If the average error $\perravg_n$ is used instead, the necessary and sufficient condition for achievability in this regime is instead $E \leq D_\Xi(\rho\|\sigma) = \max \{ D_{\max}(\rho\|\sigma) ,\, D_{\max}(\sigma||\rho) \}$~\cite{regula_2024}. 

\subsection{Symmetric hypothesis testing with low inconclusiveness}\label{app:subsec_symmetric}

Finally, by investigating the special case of Proposition~\ref{prop:symmetric} for when $Z=0$, the following corollary precisely characterises the achievable conditional error exponent of symmetric hypothesis testing with a constant conclusive probability.

\begin{boxed}
\begin{corollary}[Formal statement of Proposition~\ref{prop:symmetric_average}]\label{cor:symmetric-average}
Let $\rho$ and $\sigma$ be two states, and let $E\geq0$ be a non-negative real number.  The following statements are equivalent.
\begin{enumerate}
	\item There exists a sequence $((M_n,N_n,\id-M_n-N_n))_n$ of three-outcome tests such that
	\begin{align}
		\liminf_{n\to\infty}-\frac{1}{n}\log\perravg_n\fleft(M_n,N_n\fright)&\geq E, \label{pf:special-12}\\
		\liminf_{n\to\infty}\pavg_n\fleft(M_n,N_n\fright)&>0. \label{pf:special-13}
	\end{align}
	\item There exists a sequence $((M_n,N_n,\id-M_n-N_n))_n$ of three-outcome tests such that
	\begin{align}
		\liminf_{n\to\infty}-\frac{1}{n}\log\perravg_n\fleft(M_n,N_n\fright)&\geq E, \label{pf:special-14}\\
		\liminf_{n\to\infty}\pavg_n\fleft(M_n,N_n\fright)&\geq q, \label{pf:special-15}
	\end{align}
	where $q = p$ if $D(\rho\|\sigma) > D(\sigma\|\rho)$, $q = 1- p$ if $D(\sigma\|\rho) > D(\rho\|\sigma)$, or $q = \max\{p, 1-p\}$ otherwise.
	\item The following inequality holds:
	\begin{align}
		E&\leq\max\left\{D\fleft(\rho\middle\|\sigma\fright),D\fleft(\sigma\middle\|\rho\fright)\right\}. \label{pf:special-11}
	\end{align}
\end{enumerate}
\end{corollary}
\end{boxed}

\begin{proof}[Proof of 1 $\Rightarrow$ 3]
Let $((M_n,N_n,\id-M_n-N_n))_n$ be a sequence of three-outcome tests satisfying Eqs.~\eqref{pf:special-12}--\eqref{pf:special-13}.  Note that
\begin{align}
    \limsup_{n\to\infty}-\frac{1}{n}\log\pavg_n\fleft(M_n,N_n\fright)&=0.
\end{align}
It thus follows from Proposition~\ref{prop:symmetric} by plugging in $Z=0$ that
\begin{align}
	 E&\leq\max\left\{\inf_{s>1}\wt{D}_s\fleft(\rho\middle\|\sigma\fright),\inf_{t>1}\wt{D}_t\fleft(\sigma\middle\|\rho\fright)\right\} \\
	 &=\max\left\{D\fleft(\rho\middle\|\sigma\fright),D\fleft(\sigma\middle\|\rho\fright)\right\}. \label{pf:special-16}
\end{align}
This proves the direction 1 $\Rightarrow$ 3 of Corollary~\ref{cor:symmetric-average}.
\end{proof}

\begin{proof}[Proof of 3 $\Rightarrow$ 2]
Let $E\geq0$ be a non-negative real number satisfying Eq.~\eqref{pf:special-16}.  Without loss of generality, we assume that
\begin{align}
	D\fleft(\rho\middle\|\sigma\fright)&\geq D\fleft(\sigma\middle\|\rho\fright).
\end{align}
It follows from the achievability part of the quantum Stein's lemma~\cite{hiai_1991} that, there exists a sequence $(M_n)_n$ of operators such that
\begin{align}
	\liminf_{n\to\infty}-\frac{1}{n}\log\Tr\fleft[M_n\sigma^{\otimes n}\fright]&\geq D\fleft(\rho\middle\|\sigma\fright), \\
	\liminf_{n\to\infty}\Tr\fleft[M_n\rho^{\otimes n}\fright]&=1, \\
	0\leq M_n&\leq\id\quad\forall n.
\end{align}
Consider the following sequence of three-outcome tests: $((M_n,0,\id-M_n))_n$.  We have that
\begin{align}
	\liminf_{n\to\infty}-\frac{1}{n}\log\perravg_n\fleft(M_n,0\fright)&=\liminf_{n\to\infty}-\frac{1}{n}\log\frac{\Tr\fleft[M_n\sigma^{\otimes n}\fright]}{\Tr\fleft[M_n\left(\rho^{\otimes n}+\sigma^{\otimes n}\right)\fright]} \\
	&\geq\liminf_{n\to\infty}-\frac{1}{n}\log\Tr\fleft[M_n\sigma^{\otimes n}\fright]-\limsup_{n\to\infty}-\frac{1}{n}\log\Tr\fleft[M_n\rho^{\otimes n}\fright] \\
	&\geq D\fleft(\rho\middle\|\sigma\fright),
\end{align}
and
\begin{align}
	\liminf_{n\to\infty}\pavg_n\fleft(M_n,0\fright)&=\liminf_{n\to\infty}\Tr\fleft[M_n\left(p\rho^{\otimes n}+\left(1-p\right)\sigma^{\otimes n}\right)\fright] \\
	&\geq\liminf_{n\to\infty}p\Tr\fleft[M_n\rho^{\otimes n}\fright] \\
	&\geq p.
\end{align}
Likewise, if we assume that $D(\rho\|\sigma)\leq D(\sigma\|\rho)$, then one can construct a sequence $((0,N_n,\id-N_n))_n$ of three-outcome tests such that $\liminf_{n\to\infty}-\frac{1}{n}\log\perravg_n\fleft(0,N_n\fright)\geq D(\sigma\|\rho)$ and $\liminf_{n\to\infty}\pavg_n\fleft(0,N_n\fright)\geq 1-p$.  The direction 3 $\Rightarrow$ 2 then follows immediately.
\end{proof}

\begin{proof}[Proof of 2 $\Rightarrow$ 1]
Trivial.
\end{proof}

\section{Low inconclusiveness: achievability from typicality}\label{app:typicality}

The converse bounds of Sections~\ref{app:conclusive_bigsection}--\ref{app:special} impose restrictions on the achievable conditional error exponents of quantum hypothesis testing. Of particular interest to us is the strong converse bound of Corollary~\ref{cor:strong_converse_for_highconc}, which tells us that the error exponents $A \leq D(\sigma\|\rho)$ and $B \leq D(\rho\|\sigma)$ cannot be exceeded when the probability of conclusiveness is non-vanishing. However, due to their inherent strong converse character, the bounds do not tell us whether those exponents can be achieved with \emph{high} conclusiveness. This section investigates their achievability by using the method of types~\cite{csiszar_2011}, starting with the special case of classical probability distributions (or commuting quantum states).


\subsection{Properties of types}

Let $\X$ be a finite alphabet and $\P(\X)$ denote the set of probability distributions thereon. The \deff{type} $\type{x^n} \in \P(X)$ of a sequence $x^n = (x_1, \ldots, x_n) \in \mathcal{X}^n$ is defined as the probability distribution
\begin{equation}\begin{aligned}
    \type{x^n} \coloneqq \frac1n \sum_{i=1}^n \boldsymbol{1}[x_i]
\end{aligned}\end{equation}
where $\boldsymbol{1}[x_k]$ denotes the indicator function, i.e.\ $\boldsymbol{1}[x_k](x)$ equals $1$ if $x_k = x$ and $0$ otherwise. In other words, $n \type{x^n}(x)$ is the number of occurrences of the symbol $x$ in $x^n$.

Let $\mathcal{T}_n(\X)$ denote the set of all types of denominator $n$. 
For any type $t \in \mathcal{T}_n(\X)$, the \deff{type class} $T^n_{t}$ is the set of all sequences which have this type, that is,
\begin{equation}\begin{aligned}
T^n_{t} \coloneqq \lsetr x^n \in \X^n \barr t_{x^n} = t \rsetr.    
\end{aligned}\end{equation}
For any distribution $P \in \P(X)$, the set of (strongly) \deff{$\boldsymbol{\delta}$-typical sequences} is defined as
\begin{equation}\begin{aligned}
    T^n_{P, \delta} &\coloneqq \lset x^n \in \X^n \bar \norm{\type{x^n} - P}{\infty} \leq \delta \rset\\
    &\hphantom{:}= \bigcup_{\wt{P} : \norm{\wt{P} - P}{\infty} \!\leq \delta} T^n_{\wt{P}}.
\end{aligned}\end{equation}

Some of the standard results regarding types are as follows (see e.g.~\cite[Chapter~2]{csiszar_2011}). 
The type-counting lemma says that there are only polynomially many types: 
\begin{equation}\begin{aligned}
    |\T_n(\X)| &\leq (n+1)^{|\X|}.
\end{aligned}\end{equation}
However, the best bound we can give on the size of any type class is exponential in $n$: for any distribution $Q$ on $\X$ and any $ x^n \in T^n_P$, it holds that
\begin{equation}\begin{aligned}
    \Pr_{Q^n}(x^n) &= \exp \left( - n [ D(P \| Q) + H(P) ]\right),
\end{aligned}\end{equation}
and as a consequence $|T^n_P| \leq \exp(n H(P))$. Extending this insight to $\delta$-typical sequences, the continuity of the KL divergence implies that
\begin{equation}\begin{aligned}\label{eq:types_Qprob}
    \Pr_{Q^n} \!\left(T^n_{P,\delta}\right) &\leq \!\sum_{\substack{\wt{P} \in \mathcal{T}_n(\X) : \\ \norm{\wt{P} - P}{\infty} \!\leq \delta}} \sum_{x^n \in T_{\wt{P}}^n} \exp\left(- n D(\wt{P} \| Q) - n H(\wt{P}) \right)\\
    &\leq \!\sum_{\substack{\wt{P} \in \mathcal{T}_n(\X) : \\ \norm{\wt{P} - P}{\infty} \!\leq \delta}} \exp\left(- n D(\wt{P} \| Q) \right)\\
    &\leq (n+1)^{|\X|} \exp\left(- n \left[\min_{\wt{P} : \norm{\wt{P} - P}{\infty}\!\leq \delta} D(\wt{P} \| Q) \right]\right)\\
    &\leq (n+1)^{|\X|} \exp\left(- n \left[D(P \| Q) - g(\delta) \right]\right),
\end{aligned}\end{equation}
for some function $g$ such that $g(\delta) \to 0$ as $\delta \to 0$. This tells us that if $P \neq Q$, then any i.i.d.\ sequence drawn from $Q$ is exponentially unlikely to be $\delta$-typical for $P$, with the exponent approximately governed by $D(P \| Q)$. 
Conversely, an estimate using Hoeffding's (concentration) inequality gives
\begin{equation}\begin{aligned}\label{eq:types_Pprob}
    1 -  \Pr_{P^n}\!\left(T^n_{P,\delta}\right) &\leq 2 |X| \exp\left( - 2n \delta^2 \log e \right),
\end{aligned}\end{equation}
which says that a sequence drawn from $P$ itself is very highly likely to be $\delta$-typical for $P$.

Sanov's theorem~\cite{sanov_1957} extends the above analysis to more general sets of types. Namely, for any closed set $\S \subseteq \P(\X)$ of non-empty interior, it holds that
\begin{equation}\begin{aligned}\label{eq:dense}
    \lim_{n\to\infty} - \frac1n \log \Pr_{Q^n}\!\left( \lset x^n \in \X^n \bar \type{x^n} \in \S \rset\right) =  \min_{P \in \S} D(P \| Q).
\end{aligned}\end{equation}
An important application of this is to sets defined as
\begin{equation}\begin{aligned}
    \wt{T}_{P,A} \coloneqq \bigcup_{\wt{P} :\,  D(\wt{P}\|P) \leq A} T^n_{\wt{P}}
\end{aligned}\end{equation}
for some parameter $A > 0$, 
which gives
\begin{equation}\begin{aligned}\label{eq:sanov_for_hoeffding}
    \lim_{n\to\infty} - \frac1n \log \Pr_{Q^n} \!\left( \wt{T}_{P,A} \right) &= \min_{\wt{P} :\, D(\wt{P}\|P) \leq A} D(\wt{P} \| Q)\\
    &= \sup_{s \in (0,1)} \frac{s-1}{s} \Big( A - D_s(Q\|P)\Big)\\
    &= H_A(Q\|P).
\end{aligned}\end{equation}
The second line here is a standard argument based on the tilted distribution $P_s (x) = \frac{P(x)^{1-s} Q(x)^s}{\sum_y P(y)^{1-s} Q(y)^s}$ (see again~\cite[Chapter~2]{csiszar_2011}).


\subsection{Asymptotically optimal tests for classical distributions}

Combining Eq.~\eqref{eq:types_Pprob} and~\eqref{eq:types_Qprob} gives that, for any $\ve > 0$, one can choose $\delta$ small enough so that
\begin{equation}\begin{aligned}
     \liminf_{n\to\infty} \Pr_{P^n} \!\left(T^n_{P,\delta}\right) &\geq 1 -\ve,\\
     \liminf_{n\to\infty} -\frac1n \log \Pr_{Q^n} \!\left(T^n_{P,\delta}\right) &\geq D(P \| Q) - \ve.
\end{aligned}\end{equation}
Indeed, by choosing $(\delta_n)_n$ to be a sequence such that $\delta_n \to 0$ but $n \delta_n^2 \to \infty$ and using $T^n_{P,\delta_n}$ instead of $T^n_{P,\delta}$, we can even take $\ve = 0$.

The above immediately gives an idea for how to construct a feasible test for hypothesis testing between $P$ and $Q$ with high conclusiveness. To keep our notation consistent between the quantum and classical cases, we can understand POVM elements in the measurement $(M_n, N_n, \id - M_n - N_n)$ as randomised classical test functions $M_n, N_n : \X^n \to [0,1]$ such that $M_n(x^n) + N_n(x^n) \leq 1$ for all $x^n \in \X^n$. We can then take
\begin{equation}\begin{aligned}\label{eq:classical_tests}
    M_n \coloneqq \boldsymbol{1}\!\left[T^n_{P,\delta_n}\right], \qquad
    N_n \coloneqq \boldsymbol{1}\!\left[T^n_{Q,\delta_n}\right],
\end{aligned}\end{equation}
where $\boldsymbol{1}$ stands for the indicator function of the set (in quantum terminology, this can be identified with a projection) and $\delta_n$ is as before. The reason why this defines a valid measurement for all sufficiently large $n$ is because the two sets are eventually disjoint provided that $P \neq Q$: were it the case that $x^n \in T^n_{P,\delta} \cap T^n_{Q,\delta}$, it would follow that
\begin{equation}\begin{aligned}
    \norm{\type{x^n} - P}{\infty} \leq \delta,\quad \norm{\type{x^n} - Q}{\infty} \leq \delta,
\end{aligned}\end{equation}
so for any $\delta < \frac12 \norm{P-Q}{\infty}$ an application of the triangle inequality would lead to a contradiction. Combining everything, we have shown the following lemma.

\begin{boxed}
\begin{lemma}\label{lem:classical_achievability_stein}
The sequence of tests defined in Eq.~\eqref{eq:classical_tests} satisfies
\begin{equation}\begin{aligned}
     \liminf_{n\to\infty} -\frac1n \log \palpha_n (M_n, N_n) &\geq D(Q \| P),\\
     \liminf_{n\to\infty} -\frac1n \log \pbeta_n (M_n, N_n) &\geq D(P \| Q),\\
     \lim_{n\to\infty} \pi_n (P, M_n, N_n) & = 1,\\
     \lim_{n\to\infty} \pi_n (Q, M_n, N_n) & = 1.
 \end{aligned}\end{equation}
\end{lemma}
\end{boxed}
Here we recall the notation
\begin{equation}\begin{aligned}
    \palpha_n(M_n, N_n) = \frac{\EE_{P^n}(N_n)}{\pconc(P, M_n, N_n)}, \qquad  \pbeta_n(M_n, N_n) &= \frac{\EE_{Q^n}(M_n)}{\pconc(Q, M_n, N_n)},
\end{aligned}\end{equation}
where 
\begin{equation}\begin{aligned}
    \pconc(P, M_n, N_n) = \EE_{P^n} (M_n + N_n), \qquad  \pconc(Q, M_n, N_n) = \EE_{Q^n} (M_n + N_n).
\end{aligned}\end{equation}

An extension of the above idea can be used to show the result stated in the main text as Lemma~\ref{lem:exponentially_classical}, closely related to the prior studies of classical hypothesis testing with rejection~\cite{gutman_1989,grigoryan_2011,sason_2012,lalitha_2016}, thus providing a self-contained proof and slight generalisation thereof.

\begin{boxed}
\begin{lemma}[Formal statement of Lemma~\ref{lem:exponentially_classical}]\label{lem:classical_achievability_exponentiallygood}
For any $K \in (0, D(Q\|P))$, $L \in (0, D(P \| Q))$, and any $A, B >0$ such that
\begin{equation}\begin{aligned}\label{eq:exponentially_good_max_constraints}
    A &\leq \max \{ H_{B} (P\|Q),\, H_{L} (P\|Q) \},\\
B &\leq \max \{ H_A(Q\|P),\, H_K (Q\|P) \},
\end{aligned}\end{equation}
or equivalently that
\begin{equation}\begin{aligned}\label{eq:exponentially_good_B_constraints}
    B \leq \begin{cases} H_A (Q\| P) & \text{ if } A < K \text{ or } A > H_L(P \| Q)\\
                      H_K (Q \| P) & \text{ if } K \leq A \leq H_L(P \| Q),\end{cases}
\end{aligned}\end{equation}
there exists a sequence of three-outcome hypothesis tests $(M_n, N_n, \id - M_n - N_n)$ such that
\begin{equation}\begin{aligned}\label{eq:exponentiall_good_exponents_achiev}
         \liminf_{n\to\infty} -\frac1n \log \palpha_n (M_n, N_n) &\geq A,\\
     \liminf_{n\to\infty} -\frac1n \log \pbeta_n (M_n, N_n) &\geq B,\\
     \liminf_{n\to\infty} - \frac1n \log \left(1 - \pi_n (P, M_n, N_n)\right) &\geq K,\\
     \liminf_{n\to\infty} - \frac1n \log \left(1 - \pi_n (Q, M_n, N_n)\right) &\geq L.
\end{aligned}\end{equation}

Conversely, any sequence of tests $(M_n, N_n, \id - M_n - N_n)$ with exponents satisfying Eq.~\eqref{eq:exponentiall_good_exponents_achiev} must also satisfy Eq.~\eqref{eq:exponentially_good_max_constraints}/\eqref{eq:exponentially_good_B_constraints}.
\end{lemma}
\end{boxed}
\begin{proof}
For the achievability, notice first that any exponent where $A \leq H_B(P\|Q)$ or, equivalently, $B \leq H_A(Q\|P)$ is achievable deterministically, making the statement trivially true. We therefore do not need to consider the range of exponents where $H_B(P\|Q) > H_L(P\|Q)$ or where $H_A(Q\|P) > H_K (Q\|P)$.

Let us then assume that $H_B(P\|Q) \leq H_L(P\|Q)$ and $H_A(Q\|P) \leq H_K (Q\|P)$. 
By hypothesis, we then have that
\begin{equation}\begin{aligned}
 A \leq H_L(P\|Q), \qquad B \leq H_K(Q\|P).
 \end{aligned}\end{equation}
Using the fact that $H_{H_{R}(Q\|P)}(P\|Q) = R$ together with the easily verifiable property that $H_{R}(Q\|P)$ is monotonically non-increasing in $R$, this gives $H_A(Q\|P) \geq L$, establishing that 
\begin{equation}\begin{aligned}
H_K(Q\|P) \geq H_A(Q\|P) \geq L
 \end{aligned}\end{equation}
in particular.

Recall now the notation for the sets
\begin{equation}\begin{aligned}
    \wt{T}_{P,A} = \bigcup_{\wt{P} :\,  D(\wt{P}\|P) \leq A} T^n_{\wt{P}}
\end{aligned}\end{equation}
and let us define their complements as
\begin{equation}\begin{aligned}
    \wt{T}_{P,A}^c \coloneqq \X^n \setminus \wt{T}_{P,A} = \!\bigcup_{\wt{P} : D(\wt{P}\|P) > A} T^n_{\wt{P}}.
\end{aligned}\end{equation}
Consider then the functions
\begin{equation}\begin{aligned}
M_n \coloneqq \boldsymbol{1}\!\left[\wt{T}^c_{Q,H_K(Q\|P)}\right],\qquad N_n \coloneqq \boldsymbol{1}\!\left[\wt{T}_{Q,L}\right].
\end{aligned}\end{equation}
The property that $H_K(Q\|P) \geq H_A(Q\|P) \geq L$ immediately implies that $\wt{T}_{Q,L} \subseteq \wt{T}_{Q,H_K(Q\|P)}$, meaning that the relevant sets are disjoint and hence that $(M_n, N_n, \id - M_n- N_n)$ defines a valid test.

We then compute
\begin{equation}\begin{aligned}
     \liminf_{n\to\infty} -\frac1n \log \palpha_n (M_n, N_n) &= \liminf_{n\to\infty} -\frac1n \log \Pr_{P^n} \!\left( \wt{T}_{Q, L} \right) \texteq{(i)} H_L(P\|Q) \geq A,\\
     \liminf_{n\to\infty} -\frac1n \log \pbeta_n (M_n, N_n) &= \liminf_{n\to\infty} -\frac1n \log \Pr_{Q^n} \!\left( \wt{T}^c_{Q,H_K(Q\|P)} \right) \textgeq{(ii)} H_K(Q\|P) \geq B,\\
     \liminf_{n\to\infty} - \frac1n \log \left(1 - \pconc (P, M_n, N_n)\right) &\geq \liminf_{n\to\infty} - \frac1n \log \Pr_{P^n}\!\left(\wt{T}_{Q,H_K(Q\|P)}\right) \texteq{(iii)} H_{H_K(Q\|P)}(P\|Q) = K,\\
     \liminf_{n\to\infty} - \frac1n \log \left(1 - \pconc (Q, M_n, N_n)\right) &\geq \liminf_{n\to\infty} - \frac1n \log \Pr_{Q^n}\!\left(\wt{T}^c_{Q,L}\right) \textgeq{(iv)} L,
     \end{aligned}\end{equation}
where (i) and (iii) are an application of Sanov's theorem (see Eq.~\eqref{eq:sanov_for_hoeffding}), while (ii) and (iv) follow from the simple bound 
\begin{equation}\begin{aligned}\label{eq:simplebound}
   \Pr_{P^n} ( \wt{T}^c_{P,A}) \leq (n+1)^{|\X|} \exp\left(-n \!\inf_{\wt{P} : D(\wt{P}\|P) > A}\! D(\wt{P} \| P)\right) \leq (n+1)^{|\X|} \exp(-nA).
\end{aligned}\end{equation}

For the converse, consider that any feasible protocol $(M_n, N_n, \id - M_n - N_n)$ can be turned into a deterministic (two-outcome) protocol in two different ways: either as $\big(M_n, N_n + (\id - M_n - N_n)\big)$ or as $\big(M_n + (\id - M_n - N_n), N_n\big)$. From \eqref{eq:exponentiall_good_exponents_achiev} we can see that the first one of the deterministic protocols has a type II error exponent of at least $B$ and a type I error exponent of at least $\min \{ A, K\}$; the second one has a type I error exponent lower bounded by $A$ and a type II error exponent of at least $\min \{ B, L \}$. Applying the standard Hoeffding's converse bound to these two deterministic protocols gives $\min \{ A, K \} \leq H_B ( P \| Q)$ and $\min \{ B, L \} \leq H_A (Q \| P)$.  Using again the fact that $H_{H_{R}(P\|Q)}(Q\|P)=R$, the former condition then implies that $B \leq \max \{ H_A (Q \| P), H_K(Q\| P) \}$, while the latter that $A \leq \max\{ H_B(P\|Q), H_L(P\|Q) \}$, i.e.\ that $B \leq H_A(Q \| P)$ when $A > H_L(P\|Q)$. 
\end{proof}


\subsection{Quantum achievability from measurements}\label{app:typical_measurements}

Let us begin with the case of arbitrarily high  probability of conclusiveness. This can be understood as the probability being arbitrarily close to $1$, or indeed converging to $1$ in the limit $n\to\infty$.

From the classical result of Lemma~\ref{lem:classical_achievability_stein}, it is easy to see a straightforward achievability result for quantum: simply measure both states with some measurement, then apply the classical result. Perhaps less obviously, this also gives a converse.

\begin{boxed}
\begin{proposition}\label{lem:blocking_lemma}
For all states $\rho$ and $\sigma$, for all sequences of measurement channels $(\M_k)_k$, and for all $\ve >0$, there exists a sequence of three-outcome test $(M_n, N_n, \id - M_n- N_n)$ such that 
\begin{equation}\begin{aligned}
     \liminf_{n\to\infty} -\frac1n \log \palpha_n (M_n, N_n) &\geq \liminf_{k\to\infty} \frac1k D\!\left(\left.\M_k(\sigma^{\otimes k}) \right\| \M_k(\rho^{\otimes k})\right) - \ve,\\
     \liminf_{n\to\infty} -\frac1n \log \pbeta_n (M_n, N_n) &\geq \liminf_{k\to\infty} \frac1k D\!\left(\left.\M_k(\rho^{\otimes k}) \right\| \M_k(\sigma^{\otimes k})\right) - \ve,\\
     \lim_{n\to\infty} \pi_n (\rho, M_n, N_n) & = 1,\\
     \lim_{n\to\infty} \pi_n (\sigma, M_n, N_n) & = 1.
 \end{aligned}\end{equation}

 Conversely, if $(M_n, N_n, \id - M_n- N_n)$ satisfies that 
\begin{equation}\begin{aligned}
     \lim_{n\to\infty} \pi_n (\rho, M_n, N_n)  = 1 \qquad 
     \lim_{n\to\infty} \pi_n (\sigma, M_n, N_n)  = 1,
 \end{aligned}\end{equation}
 then there exists a sequence of measurement channels $(\M_k)_k$ such that
\begin{equation}\begin{aligned}
    \liminf_{n\to\infty} -\frac1n \log \palpha_n (M_n, N_n) &\leq \liminf_{k\to\infty}\frac1k  D\!\left(\left. \M_k(\sigma^{\otimes k}) \right\| \M_k(\rho^{\otimes k})\right),\\
     \liminf_{n\to\infty} -\frac1n \log \pbeta_n (M_n, N_n) &\leq \liminf_{k\to\infty} \frac1k D\!\left(\left. \M_k(\rho^{\otimes k}) \right\| \M_k(\sigma^{\otimes k})\right).\\
\end{aligned}\end{equation}
\end{proposition}
\end{boxed}

This result tells us that the pair of exponents $(A,B)$ is achievable with $\lim_{n\to\infty} \pi_n(\rho, M_n, N_n) = \lim_{n\to\infty} \pi_n(\sigma, M_n, N_n) = 1$ if and only if there exists a sequence $(\M_k)_k$ of measurements such that 
\begin{equation}\begin{aligned}
    A \leq \liminf_{k\to\infty} \frac1k D\!\left(\left. \M_k(\sigma^{\otimes k}) \right\| \M_k(\rho^{\otimes k})\right), \qquad B \leq \liminf_{k\to\infty} \frac1k D\!\left(\left. \M_k(\rho^{\otimes k}) \right\| \M_k(\sigma^{\otimes k})\right).
\end{aligned}\end{equation}
Combined with Proposition~\ref{prop:highprob_exponents}, whose proof we will consider in the next section as Proposition~\ref{prop:fullstatement_sequential}, this implies that the existence of such a sequence is equivalent to the condition $A \leq D(\sigma\|\rho)$, $B \leq D(\rho\|\sigma)$.

\begin{proof}
The construction of the test follows a standard double-blocking argument. For any $k \in \NN$, from Lemma~\ref{lem:classical_achievability_stein}, we know that there exists a sequence of tests that distinguishes the classical distributions $\M_k(\rho^{\otimes k})$ and $\M_k(\sigma^{\otimes k})$ with the exponents
\begin{equation}\begin{aligned}\label{eq:postmeasurement_classical}
     \liminf_{m\to\infty} -\frac1m \log \palpha_m \! \left( \M_k(\rho^{\otimes k}), M'_m, N'_m\right) &\geq  D\!\left( \left.\M_k(\sigma^{\otimes k}) \right\| \M_k(\rho^{\otimes k})\right),\\
     \liminf_{m\to\infty} -\frac1m \log \pbeta_m \!\left( \M_k(\sigma^{\otimes k}), M'_m, N'_m\right) &\geq  D\!\left(\left. \M_k(\rho^{\otimes k}) \right\| \M_k(\sigma^{\otimes k})\right),\\
     \lim_{m\to\infty} \pi_m\!\left(\M_k(\rho^{\otimes k}), M'_m, N'_m\right) & = 1,\\
     \lim_{m\to\infty} \pi_m\!\left(\M_k(\sigma^{\otimes k}), M'_m, N'_m\right) & = 1.
 \end{aligned}\end{equation}
 Here, for clarity we have incorporated the hypotheses (states) to be distinguished into the notation for $\palpha_m$ and $\pbeta_m$.
 
Given $n$ copies of the unknown quantum state $\rho$ or $\sigma$, our measurement strategy $(M_n, N_n, \id - M_n - N_n)$ is then defined as follows: (1) divide the $n$ copies into $\floor{\frac{n}{k}}$ blocks of $k$ copies and discard the rest; (2) apply the measurement $\M_k$ on each of the $\floor{\frac{n}{k}}$ blocks; (3) use the discrimination strategy from~\eqref{eq:postmeasurement_classical}. This gives
\begin{equation}\begin{aligned}
    \liminf_{n\to\infty} -\frac1n \log \palpha_n (\rho, M_n, N_n) &= \liminf_{n\to\infty} -\frac1n \log \Tr N'_{\floor{n/k}} \M_k(\rho^{\otimes k})^{\otimes \floor{n/k}}\\
    &= \liminf_{n\to\infty} -\frac{\floor{n/k}}{n} \frac{1}{\floor{n/k}} \log \Tr N'_{\floor{n/k}} \M_k(\rho^{\otimes k})^{\otimes \floor{n/k}}\\
    &\geq \left(\lim_{n\to\infty} \frac{\floor{n/k}}{n}\right) \left( \liminf_{m\to\infty} - \frac{1}{m} \log \Tr N'_{m} \M_k(\rho^{\otimes k})^{\otimes m} \right)\\
    &\geq \frac{1}{k} D\!\left( \M_k(\sigma^{\otimes k}) \| \M_k(\rho^{\otimes k})\right)
\end{aligned}\end{equation}
and analogously for $\pbeta_n(\sigma, M_n, N_n)$. The probability of conclusiveness here is clearly
\begin{equation}\begin{aligned}
    \liminf_{n\to\infty} \pi_n(\rho, M_n, N_n) &= \liminf_{n\to\infty} \,\Tr \Big[ \left(M'_{\floor{n/k}}+N'_{\floor{n/k}}\right) \M_k(\rho^{\otimes k})^{\otimes \floor{n/k}} \Big]\\
    &\geq \liminf_{m\to\infty}\, \pi_m \!\left( \M_k(\rho^{\otimes k}), M'_m, N'_m\right)\\
    &= 1.
\end{aligned}\end{equation}
Taking $k$ large enough concludes the achievability proof.

For the converse, we adapt the weak converse of the quantum Stein's lemma of Hiai and Petz~\cite{hiai_1991}, which is sufficient for our purposes here. Let $(M_n, N_n, Z_n)$ be any feasible test (i.e.\ $Z_n = \id - M_n- N_n$) with error exponents
\begin{equation}\begin{aligned}
    \liminf_{n\to\infty} -\frac1n \log \palpha_n (M_n, N_n) = A, \qquad  \liminf_{n\to\infty} -\frac1n \log \pbeta_n (M_n, N_n) = B.
\end{aligned}\end{equation}
Let $\wt\M_n$ denote the corresponding measurement channel. Then
\begin{equation}\begin{aligned}
   D\left(\wt\M_n(\sigma^{\otimes n}) \middle\| \wt\M_n(\rho^{\otimes n})\right) &= - H\!\left( \wt\M_n(\sigma^{\otimes n}) \right) + \Tr M_n \sigma^{\otimes n} \log \frac{1}{\Tr M_n \rho^{\otimes n}} \\
   & \hphantom{=} + \Tr N_n \sigma^{\otimes n} \log \frac{1}{\Tr N_n \rho^{\otimes n}} + \Tr Z_n \sigma^{\otimes n} \log \frac{1}{\Tr Z_n \rho^{\otimes n}}\\
   &\geq - \log 3 - \Tr N_n \sigma^{\otimes n} \log \Tr N_n \rho^{\otimes n},
\end{aligned}\end{equation}
where the last line follows since the size of the alphabet (number of measurement outcomes) is 3. Since $\Tr M_n \sigma^{\otimes n}$ decays to 0 while we know by assumption that $\lim_{n\to\infty} \Tr (M_n+N_n) \sigma^{\otimes n} = 1$, it must hold that
\begin{equation}\begin{aligned}
    \lim_{n\to\infty} \Tr N_n \sigma^{\otimes n} = 1.
\end{aligned}\end{equation}
Hence
\begin{equation}\begin{aligned}
    \liminf_{n\to\infty} \frac1n D\left(\wt\M_n(\sigma^{\otimes n}) \| \wt\M_n(\rho^{\otimes n})\right) &\geq \liminf_{n\to\infty} -\frac1n \log \Tr N_n \rho^{\otimes n} = A.
\end{aligned}\end{equation}
Reversing the roles of $\rho$ and $\sigma$ and repeating the argument with the same sequence of measurements $(\wt\M_n)_n$ gives
\begin{equation}\begin{aligned}
    \liminf_{n\to\infty} \frac1n D\left(\wt\M_n(\rho^{\otimes n}) \| \wt\M_n(\sigma^{\otimes n})\right) &\geq \liminf_{n\to\infty} -\frac1n \log \Tr M_n \sigma^{\otimes n} = B.
\end{aligned}\end{equation}
This is exactly the claimed statement.
\end{proof}


\subsection{Quantum achievability from pinching}\label{app:typical_pinching}

Instead of working with general measurements, it is known that in quantum information a remarkable technique known as asymptotic spectral pinching~\cite{hiai_1991,hayashi_2002} suffices to obtain optimal asymptotic results in a broad range of settings. Given a state $\sigma$, write it in its spectral decomposition as $\sigma = \sum_{i=1}^{\left|\operatorname{spec}(\sigma)\right|} \lambda_i P_i$ where $P_i$ are projectors corresponding to distinct eigenvalues $\lambda_i$. The \deff{pinching map} is defined as
\begin{equation}\begin{aligned}
    \E_\sigma (X) \coloneqq \sum_{i=1}^{\left|\operatorname{spec}(\sigma)\right|} P_i \, X \, P_i.
\end{aligned}\end{equation}
The key properties of it are that $\E_\sigma(\sigma) = \sigma$, $\E_\sigma (\rho)$ commutes with $\sigma$, and that for any state $\rho$ it holds that
\begin{equation}\begin{aligned}
    \E_{\sigma^{\otimes k}} (\rho^{\otimes k}) \leq \left|\operatorname{spec}(\sigma^{\otimes k})\right| \, \E_{\sigma^{\otimes k}} (\rho^{\otimes k}) \leq (k+1)^{d} \, \E_{\sigma^{\otimes k}} (\rho^{\otimes k}),
\end{aligned}\end{equation}
where the last inequality is essentially a consequence of the type-counting lemma~\cite{hiai_1991,hayashi_2002}, with $d$ denoting the underlying dimension that $\sigma$ acts on. This inequality implies that approximating a state with its pinched version incurs only a polynomially large scaling factor, and this is sufficient to show that, asymptotically, many entropic quantities can be evaluated exactly by pinching a state and then using classical results between the distributions resulting from pinching~\cite{hiai_1991,mosonyi_2015,hayashi_2016-1}. Most importantly for us, for all $s \in (0,2]$ we have~\cite{hayashi_2016-1}
\begin{equation}\begin{aligned}
   D_s\!\left( \left. \E_{\sigma^{\otimes k}} (\rho^{\otimes k}) \right\| \sigma^{\otimes k}\right) \leq  \wt{D}_s(\rho^{\otimes k} \| \sigma^{\otimes k}) \leq D_s\!\left( \left. \E_{\sigma^{\otimes k}} (\rho^{\otimes k}) \right\| \sigma^{\otimes k}\right) + d \log(k+1)
\end{aligned}\end{equation}
which directly implies that
\begin{equation}\begin{aligned}\label{eq:pinching_gives_sandwiched}
    \lim_{k\to\infty} \frac1k D_s \left( \left. \E_{\sigma^{\otimes k}} (\rho^{\otimes k}) \right\| \sigma^{\otimes k} \right) = \wt{D}_s (\rho \| \sigma),
\end{aligned}\end{equation}
including also the case of $s= 1$ with $\wt{D}_1$ understood as the Umegaki relative entropy. However, here it is important to notice that the pinching is applied to the first state. If we instead pinch the second state, we have for all $s \in (0,1)$ that~\cite{audenaert_2015,lipka-bartosik_2024}
\begin{equation}\begin{aligned}\label{eq:pinching_gives_reverse}
    \lim_{k\to\infty} \frac1k D_s \left( \rho^{\otimes k} \left\|  \E_{\rho^{\otimes k}} (\sigma^{\otimes k}) \right. \right) = \rev{D}_s (\rho \| \sigma),
\end{aligned}\end{equation}
where we recall that $\rev{D}_s(\sigma\|\rho) \coloneqq \frac{s}{1-s}\wt{D}_{1-s}(\rho\|\sigma)$. This is because the sandwiched R\'enyi divergences do not satisfy the natural symmetry $(1-s) D_s(\rho\|\sigma) = s D_{1-s} (\sigma \| \rho)$ which is obeyed by the Petz--R\'enyi divergences. Crucially, the quantity
\begin{equation}\begin{aligned}
   D^\star(\rho \| \sigma) \coloneqq \lim_{s\to1^-} \rev{D}_s (\rho \| \sigma)
\end{aligned}\end{equation}
may, in general, be strictly smaller than the Umegaki relative entropy $D(\rho \| \sigma)$~\cite{audenaert_2015}.

A simple achievability bound follows immediately by double blocking.
\begin{boxed}
\begin{lemma}
For all states $\rho$, $\sigma$ and for all $\ve >0$,  there exists a sequence of three-outcome test $(M_n, N_n, \id - M_n- N_n)$ such that
\begin{equation}\begin{aligned}
     \liminf_{n\to\infty} -\frac1n \log \palpha_n (M_n, N_n) &\geq D^\star(\sigma\|\rho) - \ve\\
     \liminf_{n\to\infty} -\frac1n \log \pbeta_n (M_n, N_n) &\geq D(\rho\|\sigma) - \ve,\\
     \lim_{n\to\infty} \pi_n (\rho, M_n, N_n) & = 1,\\
     \lim_{n\to\infty} \pi_n (\sigma, M_n, N_n) & = 1.
 \end{aligned}\end{equation} 
 Alternatively, the roles of $D^\star$ and $D$ can be interchanged.
\end{lemma}
\end{boxed}
\begin{proof}
For any fixed $s \in (0,1)$ and any $k \in \NN$, by the commuting result of Lemma~\ref{lem:classical_achievability_stein} there exists a testing protocol $(M'_m, N'_m, \id - M'_m- N'_m)$ for the states $\E_{\sigma^{\otimes k}}(\rho^{\otimes k})$ and $\sigma^{\otimes k}$ such that
\begin{equation}\begin{aligned}
     \liminf_{m\to\infty} -\frac1m \log \palpha_m \!\left( \E_{\sigma^{\otimes k}}(\rho^{\otimes k}),  M'_m, N'_m\right) &\geq D(\sigma^{\otimes k} \| \E_{\sigma^{\otimes k}}(\rho^{\otimes k})) \geq D_s(\sigma^{\otimes k} \| \E_{\sigma^{\otimes k}}(\rho^{\otimes k})) \\
     \liminf_{m\to\infty} -\frac1m \log \pbeta_m (\sigma^{\otimes k}, M'_m, N'_m) &\geq D(\E_{\sigma^{\otimes k}}(\rho^{\otimes k}) \| \sigma^{\otimes k} ) \geq D_s(\E_{\sigma^{\otimes k}}(\rho^{\otimes k}) \| \sigma^{\otimes k} ),\\
     \lim_{m\to\infty} \pi_m\!\left(\E_{\sigma^{\otimes k}}(\rho^{\otimes k}), M'_m, N'_m\right) & = 1,\\
     \lim_{m\to\infty} \pi_m (\sigma^{\otimes k}, M'_m, N'_m) & = 1.
 \end{aligned}\end{equation} 
 As in the proof of Proposition~\ref{lem:blocking_lemma}, we then employ a blocking approach with $\floor{n/k}$ blocks of size $k$ to get an achievable protocol $(M_n, N_n, \id - M_n - N_n)$ such that
\begin{equation}\begin{aligned}
    \liminf_{n\to\infty} -\frac1n \log \palpha_n (\rho, M_n, N_n) &\geq \frac1k D_s(\sigma^{\otimes k} \| \E_{\sigma^{\otimes k}}(\rho^{\otimes k}))\\
    \liminf_{n\to\infty} -\frac1n \log \pbeta_n (\sigma, M_n, N_n) &\geq \frac1k  D_s(\E_{\sigma^{\otimes k}}(\rho^{\otimes k}) \| \sigma^{\otimes k}),\\
         \lim_{n\to\infty} \pi_n (\rho, M_n, N_n) & = 1,\\
     \lim_{n\to\infty} \pi_n (\sigma, M_n, N_n) & = 1.
\end{aligned}\end{equation}
Taking $k$ sufficiently large and using Eqs.~\eqref{eq:pinching_gives_sandwiched}--\eqref{eq:pinching_gives_reverse} then gives a protocol that achieves the exponents
\begin{equation}\begin{aligned}
     \liminf_{n\to\infty} -\frac1n \log \palpha_n (\rho, M_n, N_n) &\geq \rev{D}_s(\sigma \| \rho) - \frac{\ve}{2}\\
     \liminf_{n\to\infty} -\frac1n \log \pbeta_n (\sigma, M_n, N_n) &\geq \wt{D}_s(\rho \| \sigma) - \frac{\ve}{2}.
 \end{aligned}\end{equation}
 Taking $s$ sufficiently close to $1$ leads to the stated result.

Repeating the proof with the choice of states $\rho^{\otimes k}$ and $\E_{\rho^{\otimes k}}(\sigma^{\otimes k})$ leads to an analogous achievability result but with 
\begin{equation}\begin{aligned}
     \liminf_{n\to\infty} -\frac1n \log \palpha_n (M_n, N_n) &\geq D(\sigma \| \rho) - \ve\\
     \liminf_{n\to\infty} -\frac1n \log \pbeta_n (M_n, N_n) &\geq D^\star(\rho \| \sigma) - \ve.
 \end{aligned}\end{equation}
\end{proof}

A modification of this idea can also be used to turn the `exponentially conclusive' classical achievability result of Lemma~\ref{lem:classical_achievability_exponentiallygood} into a quantum achievability result.

To establish this, we start with a lemma that clarifies the behaviour of the regularisation of the sandwiched Hoeffding quantities
\begin{equation}\begin{aligned}
    \wt{H}_A(\sigma \| \rho) = \sup_{s \in (0,1)} \frac{s-1}{s} \Big( A - \wt{D}_s(\sigma \| \rho) \Big), \qquad
\rev{H}_A(\sigma \| \rho) = \sup_{s \in (0,1)} \frac{s-1}{s} \Big( A - \rev{D}_s(\sigma \| \rho) \Big).
\end{aligned}\end{equation}
The result can be deduced from~\cite[Theorem~4.8]{hiai_2008}, or more directly from~\cite[Proposition~3.5]{hiai_2009}; we give a proof for completeness.

\begin{boxed}
\begin{lemma}[\cite{hiai_2009}]\label{lem:effective_uniform_convergence}
For all states $\rho$, $\sigma$ and all $K \geq 0$, it holds that
\begin{equation}\begin{aligned}\label{eq:regularised_hoeffding_sand}
    \lim_{k\to\infty} \frac1k H_{Kk} (\E_{\sigma^{\otimes k}}(\rho^{\otimes k}) \| \sigma^{\otimes k}) = \wt{H}_{K} (\rho \| \sigma).
\end{aligned}\end{equation}
Similarly, for all $L \geq 0$, it holds that
\begin{equation}\begin{aligned}\label{eq:regularised_hoeffding_rev}
    \lim_{k\to\infty} \frac1k H_{Lk} (\rho^{\otimes k} \| \E_{\rho^{\otimes k}}(\sigma^{\otimes k})) = \rev{H}_{L} (\rho \| \sigma).
\end{aligned}\end{equation}
\end{lemma}
\end{boxed}
\begin{proof}
We consider here the first equality, with the second being completely analogous. Observe first that, for any $\s \in (0,1)$, we have
\begin{equation}\begin{aligned}
    \liminf_{k\to\infty} \frac1k H_{Kk} (\E_{\sigma^{\otimes k}}(\rho^{\otimes k}) \| \sigma^{\otimes k}) &= \liminf_{k\to\infty} \frac1k \sup_{s \in (0,1)} \frac{1-s}{s} \left( D_s (\E_{\sigma^{\otimes k}}(\rho^{\otimes k}) \| \sigma^{\otimes k}) - K k \right)\\
    &\geq \liminf_{k\to\infty} \frac{1-\s}{\s} \left(\frac1k D_{\s} (\E_{\sigma^{\otimes k}}(\rho^{\otimes k}) \| \sigma^{\otimes k}) - K \right)\\
    &= \frac{1-\s}{\s} \Big( \wt{D}_{\s}(\sigma \| \rho) - K\Big),
\end{aligned}\end{equation}
where we used~\eqref{eq:pinching_gives_sandwiched}. 
Taking the supremum over $\s$ gives
\begin{equation}\begin{aligned}
    \liminf_{k\to\infty} \frac1k H_{Kk} (\E_{\sigma^{\otimes k}}(\rho^{\otimes k}) \| \sigma^{\otimes k}) \geq \wt{H}_{K} (\rho \| \sigma).
    \end{aligned}\end{equation}
For the other direction of the inequality, consider that
\begin{equation}\begin{aligned}
    \wt{H}_K(\rho \| \sigma) &= \sup_{s \in (0,1)} \frac{1-s}{s} \left( \wt{D}_s(\rho\|\sigma) - K \right)\\
    &\texteq{(i)} \sup_{s \in (0,1)} \frac{1-s}{s} \left( \lim_{k\to\infty} \frac1k D_s (\E_{\sigma^{\otimes k}}(\rho^{\otimes k}) \| \sigma^{\otimes k}) - K \right)\\
    &\texteq{(ii)} \sup_{s \in (0,1)} \frac{1-s}{s} \left( \sup_{k \in \NN} \frac1k D_s (\E_{\sigma^{\otimes k}}(\rho^{\otimes k}) \| \sigma^{\otimes k}) - K \right)\\
    &= \sup_{k \in \NN} \frac1k \sup_{s \in (0,1)} \frac{1-s}{s} \left( D_s (\E_{\sigma^{\otimes k}}(\rho^{\otimes k}) \| \sigma^{\otimes k}) - Kk \right)\\
    &\geq \limsup_{k\to\infty} \frac1k  H_{Kk} (\E_{\sigma^{\otimes k}}(\rho^{\otimes k}) \| \sigma^{\otimes k}).
\end{aligned}\end{equation}
Here, (i) is again by~\eqref{eq:pinching_gives_sandwiched}, and the only other non-trivial step here is (ii), which equates the limit over $k$ with the supremum. This follows by Fekete's lemma, as $\left( D_s (\E_{\sigma^{\otimes k}}(\rho^{\otimes k}) \| \sigma^{\otimes k})\right)_k$ can be seen to be a superadditive sequence in $k$: for any $m, n \in \NN$, the data processing inequality and additivity of the Petz--R\'enyi divergences~\cite{petz_1986} give
\begin{equation}\begin{aligned}
    D_s (\E_{\sigma^{\otimes m+n}}(\rho^{\otimes m+n}) \| \sigma^{\otimes m+n}) &\geq D_s \Big( \left[\E_{\sigma^{\otimes m}} \otimes \E_{\sigma^{\otimes n}}\right] (\rho^{\otimes m+n}) \Big\| \sigma^{\otimes m+n}\Big) \\
    &= D_s ( \E_{\sigma^{\otimes m}} (\rho^{\otimes m}) \| \sigma^{\otimes m}) + D_s ( \E_{\sigma^{\otimes n}} (\rho^{\otimes n}) \| \sigma^{\otimes n}),
\end{aligned}\end{equation}
which is a consequence of the fact that the commutant of $\sigma^{\otimes m+n}$ contains the tensor product of the commutants of $\sigma^{\otimes m}$ and $\sigma^{\otimes n}$ --- in other words, the pinching $\E_{\sigma^{\otimes m+n}}$ is coarser than $\E_{\sigma^{\otimes m}} \otimes \E_{\sigma^{\otimes n}}$, and hence the latter can be obtained by following the former with a suitable dephasing channel.
\end{proof}

\begin{boxed}
\begin{proposition}[Formal statement of Proposition~\ref{prop:exponentially_quantum}]
For all quantum states $\rho$ and $\sigma$, all $\ve > 0$, and all $A, B, K, L >0$ such that either the inequalities
\begin{equation}\begin{aligned}\label{eq:exponentially_quantum_sm}
   A &\leq \max \{ H_{B} (\rho\|\sigma),\, \rev{H}_{L} (\rho\|\sigma) \},\\
B &\leq \max \{ H_A(\sigma\|\rho),\, \wt{H}_K (\sigma\|\rho) \}
\end{aligned}\end{equation}
both hold or the inequalities
\begin{equation}\begin{aligned}\label{eq:exponentially_quantum_sm2}
   A &\leq \max \{ H_{B} (\rho\|\sigma),\, \wt{H}_{L} (\rho\|\sigma) \},\\
B &\leq \max \{ H_A(\sigma\|\rho),\, \rev{H}_K (\sigma\|\rho) \}
\end{aligned}\end{equation}
both hold,
there exists a sequence of three-outcome hypothesis tests $(M_n, N_n, \id - M_n - N_n)$ such that
\begin{equation}\begin{aligned}
     \liminf_{n\to\infty} -\frac1n \log \palpha_n (M_n, N_n) &\geq A - \ve,\\
     \liminf_{n\to\infty} -\frac1n \log \pbeta_n (M_n, N_n) &\geq B - \ve,\\
     \liminf_{n\to\infty} - \frac1n \log \left(1 - \pi_n (\rho, M_n, N_n)\right) &\geq K,\\
     \liminf_{n\to\infty} - \frac1n \log \left(1 - \pi_n (\sigma, M_n, N_n)\right) &\geq L.
\end{aligned}\end{equation}
\end{proposition}
\end{boxed}
\begin{proof}
The achievability of error exponents in the range $A \leq H_B(\rho\|\sigma)$, which is equivalent to $B \leq H_A(\sigma\|\rho)$, follows from the known deterministic results~\cite{hayashi_2007,audenaert_2008}. 

Let us now focus on the range of $A \leq \rev{H}_L(\rho \| \sigma)$ and $B \leq \wt{H}_K (\sigma \| \rho)$. These assumptions imply that
\begin{equation}\begin{aligned}
  A &\leq \rev{H}_L(\rho \| \sigma)\\
  &= \lim_{k\to\infty} \frac1k H_{Lk} (\rho^{\otimes k} \| \E_{\rho^{\otimes k}}(\sigma^{\otimes k}))
\end{aligned}\end{equation}
by Lemma~\ref{lem:effective_uniform_convergence}. Hence, for any $\ve > 0$ we can find $k$ large enough so that
\begin{equation}\begin{aligned}
     k (A-\ve) \leq H_{Lk}( \rho^{\otimes k} \| \E_{\rho^{\otimes k}}(\sigma^{\otimes k}) ).
\end{aligned}\end{equation}
By an analogous argument and using Lemma~\ref{lem:effective_uniform_convergence} again, we can take $k$ even larger if need be to ensure that
\begin{equation}\begin{aligned}
    k (B - \ve) \leq H_{Kk} ( \E_{\rho^{\otimes k}}(\sigma^{\otimes k}) \| \rho^{\otimes k}).
\end{aligned}\end{equation}
Using then the classical achievability in Lemma~\ref{lem:classical_achievability_exponentiallygood}, 
there exists a sequence of tests $(M'_m, N'_m, \id - M'_m - N'_m)$ such that
\begin{equation}\begin{aligned}
     \liminf_{m\to\infty} -\frac1m \log \palpha_m (\rho^{\otimes k}, M'_m, N'_m) &\geq k (A - \ve),\\
     \liminf_{m\to\infty} -\frac1m \log \pbeta_m \!\left( \E_{\rho^{\otimes k}}(\sigma^{\otimes k}), M'_m, N'_m\right)) &\geq k (B- \ve),\\
     \liminf_{m\to\infty}  -\frac1m \log \left(1 - \pi_m(\rho^{\otimes k}, M'_m, N'_m)\right) & \geq k K,\\
     \liminf_{m\to\infty}  -\frac1m \log \left( 1 - \pi_m (\E_{\rho^{\otimes k}}(\sigma^{\otimes k}), M'_m, N'_m)\right) & \geq k L .
 \end{aligned}\end{equation} 
 The double-blocking argument as in Proposition~\ref{lem:blocking_lemma} defines a sequence $(M_n, N_n, \id - M_n - N_n)$ for the discrimination of $\rho$ and $\sigma$ that satisfies
 \begin{equation}\begin{aligned}
     \liminf_{n\to\infty} -\frac1n \log \palpha_n (\rho, M_n, N_n) &\geq A - \ve ,\\
     \liminf_{n\to\infty} -\frac1n \log \pbeta_n (\sigma, M_n, N_n) &\geq B - \ve,\\
     \liminf_{n\to\infty} - \frac1n \log \left(1 - \pi_n (\rho, M_n, N_n)\right) &\geq K,\\
     \liminf_{n\to\infty} - \frac1n \log \left(1 - \pi_n (\sigma, M_n, N_n)\right) &\geq L.
\end{aligned}\end{equation}
The exact same reasoning works also if we replace $\rev{H}_L(\rho\|\sigma)$ and $\wt{H}_K(\sigma\|\rho)$ with $\wt{H}_L(\rho\|\sigma)$ and $\rev{H}_K(\sigma\|\rho)$, respectively.
\end{proof}

A converse bound can be obtained in the same way as the classical result of Lemma~\ref{lem:classical_achievability_exponentiallygood}, invoking the deterministic converse of the quantum Hoeffding bound~\cite{nagaoka_2006,audenaert_2008}.
\begin{boxed}
\begin{proposition}\label{cor:converse_quantum_exponentiallygood}
For any sequence of tests $(M_n, N_n, \id - M_n- N_n)$ satisfying
\begin{equation}\begin{aligned}\label{eq:conj2}
         \liminf_{n\to\infty} -\frac1n \log \palpha_n (M_n, N_n) &\geq A, \qquad
     \liminf_{n\to\infty} -\frac1n \log \pbeta_n (M_n, N_n) \geq B,\\
     \liminf_{n\to\infty} - \frac1n \log \left(1 - \pi_n (\rho, M_n, N_n)\right) &\geq K, \qquad
     \liminf_{n\to\infty} - \frac1n \log \left(1 - \pi_n (\sigma, M_n, N_n)\right) \geq L,
\end{aligned}\end{equation}
it must hold that
\begin{equation}\begin{aligned}\label{eq:conj1}
   A &\leq \max \{ H_{B} (\rho\|\sigma),\, H_{L} (\rho\|\sigma) \},\\
B &\leq \max \{ H_A(\sigma\|\rho),\, H_K(\sigma\|\rho) \}.
\end{aligned}\end{equation}
\end{proposition}
\end{boxed}
\noindent One observes a discrepancy between the type of R\'enyi relative entropies that appear in the achievability and converse bounds --- a phenomenon that has been a bane of existence of quantum information theorists since the early days of the field~\cite{ogawa_2004,hayashi_2007,tomamichel_2025}.

We conjecture that this converse bound is tight, although it is at this point not clear to us whether the standard tools used to derive the optimal achievability bounds for error exponents in conventional quantum hypothesis testing~\cite{audenaert_2008,cheng_2023-1} can be adapted to show this.

\begin{conjecture}
For all quantum states $\rho$ and $\sigma$ and all $A, B, K, L >0$ such that Eq.~\eqref{eq:conj1} is satisfied, there exists a sequence of tests satisfying Eq.~\eqref{eq:conj2}.
\end{conjecture}

Indeed, it may very well be the case that a stronger converse bound than Proposition~\ref{cor:converse_quantum_exponentiallygood} holds, entailing that the exponents of Eq.~\eqref{eq:conj1} cannot always be achieved, and the optimal ones are instead governed by a combination of Petz--R\'enyi and sandwiched R\'enyi divergences. Although this would be a fascinating development, the standard converse proofs in quantum hypothesis testing~\cite{nussbaum_2009,nagaoka_2006,audenaert_2008} are rather delicate and it is not clear to us how they could be adapted and extended in this direction. The resolution of this question is an exciting open problem.


\section{Low inconclusiveness: achievability from sequential hypothesis testing}\label{app:sequential}


Although fully sufficient in the classical case, the results discussed so far still did not allow us to achieve the error exponents $A = D(\sigma\|\rho)$ and $B = D(\rho\|\sigma)$ for general quantum states. Here we investigate the achievability of these exponents by adapting the result of Li, Tan, and Tomamichel~\cite{li_2022}.

\subsection{Claims}

The main result here can be summarised as follows.

\begin{boxed}
\begin{proposition}[Achievability of Proposition~\ref{prop:highprob_exponents};~\cite{li_2022}]\label{prop:fullstatement_sequential}
For any full-rank quantum states $\rho$ and $\sigma$ and for any $\ve > 0$, there exists a sequence of three-outcome tests $(M_n, N_n, \id - M_n- N_n)$ such that
\begin{equation}\begin{aligned}
     \liminf_{n\to\infty} -\frac1n \log \palpha_n (M_n, N_n) &\geq  D(\sigma \| \rho) - \ve,\\
     \liminf_{n\to\infty} -\frac1n \log \pbeta_n (M_n, N_n) &\geq D(\rho\|\sigma) - \ve,\\
     \lim_{n\to\infty} \pi_n (\rho, M_n, N_n) &= 1,\\
     \lim_{n\to\infty} \pi_n (\sigma, M_n, N_n) &= 1.
 \end{aligned}\end{equation}
\end{proposition}
\end{boxed}

The full statement of Proposition~\ref{prop:highprob_exponents} in the main text will then follow by combining the achievability result above with the converse shown in Corollary~\ref{cor:strong_converse_for_highconc},
which tells us that for any sequence of tests satisfying
\begin{equation}\begin{aligned}
     \lim_{n\to\infty} \pi_n (\rho, M_n, N_n) > 0, \qquad \lim_{n\to\infty} \pi_n (\sigma, M_n, N_n) > 0,
\end{aligned}\end{equation}
it must hold that
\begin{equation}\begin{aligned}
    \liminf_{n\to\infty} -\frac1n \log \palpha_n (M_n, N_n) &\leq D(\sigma\|\rho)\\
    \liminf_{n\to\infty} -\frac1n \log \pbeta_n (M_n, N_n) &\leq D(\rho\|\sigma).
\end{aligned}\end{equation}
This can also be thought of as a consequence of the strong converse property of the quantum Stein's lemma~\cite{ogawa_2000}. The achievability by measurements discussed in Corollary~\ref{cor:single_measurement_sequence} is shown in Proposition~\ref{lem:blocking_lemma}.

The construction of the tests in Proposition~\ref{prop:fullstatement_sequential} is based on the adaptive sequential hypothesis testing protocol of Ref.~\cite{li_2022}. We discuss the definition of the protocol and investigate its structure below. For completeness, we provide a self-contained proof of its asymptotic properties that is based on~\cite{li_2022}.

We will in particular follow~\cite{li_2022} in proving the following statement.
\begin{boxed}
\begin{lemma}\label{lem:achievability_sequential}
For any full-rank quantum states $\rho$ and $\sigma$ and for any $\ve > 0$, there exists a sequence of three-outcome tests $(M_n, N_n, \id - M_n- N_n)$ such that
\begin{equation}\begin{aligned}
     \liminf_{n\to\infty} -\frac1n \log \palpha_n (M_n, N_n) &\geq  D_\MM(\sigma \| \rho) - \ve,\\
     \liminf_{n\to\infty} -\frac1n \log \pbeta_n (M_n, N_n) &\geq D_\MM(\rho\|\sigma) - \ve,\\
     \lim_{n\to\infty} \pi_n (\rho, M_n, N_n) &= 1,\\
     \lim_{n\to\infty} \pi_n (\sigma, M_n, N_n) &= 1,
 \end{aligned}\end{equation}
where
\begin{equation}\begin{aligned}
    D_\MM(\rho\|\sigma) = \max_{\M \in \MM} D(\M(\rho) \| \M(\sigma))
\end{aligned}\end{equation}
is the measured relative entropy.
\end{lemma}
\end{boxed}
Proposition~\ref{prop:fullstatement_sequential} will then follow by using the fact that $\lim_{k\to\infty} \frac1k D_{\MM}(\rho^{\otimes k} \| \sigma^{\otimes k}) = D(\rho\|\sigma)$~\cite{hiai_1991} and a standard double-blocking argument as in Proposition~\ref{lem:blocking_lemma}.


\subsection{Protocol construction}\label{sec:protocol_definition}

Let $\M_\rho$ denote an optimal measurement such that $D_\MM(\rho\|\sigma) = D(\M_\rho(\rho) \| \M_\rho(\sigma))$, and let $\M_\sigma$ denote an optimal measurement such that $D_\MM(\sigma\|\rho) = D(\M_\sigma(\sigma) \| \M_\sigma(\rho))$. In a mild abuse of notion, we will use $\M_\rho$ to denote both the measurement channel and the collection of POVMs $\M_\rho = \{ M_{\rho}(x) \}_{x=1}^{|\X|}$ that defines the given measurement. Note that the measurements can be assumed w.l.o.g.\ to have a finite number of outcomes, in fact $|\X| \leq d$~\cite{berta_2017-1}.

In the considered setting, we have $n$ i.i.d.\ copies of an unknown state that could be $\rho$ or $\sigma$. We will process the copies individually, one by one, through a sequence $(\M_k)_{k=1}^n$ of measurements defined adaptively as
\begin{equation}
\M_k = \begin{cases}
\M_\rho & \text{if } S_{k-1} \geq 0 \\
\M_\sigma & \text{if } S_{k-1} < 0
\end{cases}
\end{equation}
where
\begin{equation}\begin{aligned}
   S_k \coloneqq \sum_{j=1}^k Z_j, \qquad  Z_j \coloneqq \log\frac{\Pr_{\rho,\M_j}(X_j)}{\Pr_{\sigma,\M_j}(X_j)}, \qquad \Pr_{\rho,\M_j}(x) \coloneqq \Tr \rho M_j(x),
\end{aligned}\end{equation}
and $S_0 = 0$. 
For some sequence of outcomes $x^n = (x_1, \ldots, x_n)$,  under hypothesis $\rho$ we denote $\Pr_\rho(x^n) = \prod_{i=1}^n \Pr_{\rho,\M_i}(x_i)$ and analogously under $\sigma$.

Finally, define the thresholds
\begin{align}
A_n &= n(D_\MM(\sigma\|\rho) - \ve) \\
B_n &= n(D_\MM(\rho\|\sigma) - \ve).
\end{align}

Now, the actual definition of the test is as follows. Given $n$ copies of the unknown state, we apply the above $n$-copy adaptive protocol and make a guess as follows:
\begin{equation}
 \begin{cases}
\text{the state is } \rho & \text{if } S_n \geq B_n \\
\text{the state is } \sigma & \text{if } S_n \leq -A_n \\
\text{inconclusive } & \text{if } -A_n < S_n < B_n.
\end{cases}
\end{equation}


\subsection{Analysis}

\subsubsection{Error exponents}\label{app:seq_analysis1}

The error exponents of the protocol can be shown using a simple change-of-measure argument. 
By definition, the type I error of this procedure satisfies
\begin{equation}\begin{aligned}
    \alpha_n &= \Pr_\rho(S_n \leq -A_n)\\
    &= \sum_{x^n: S_n(x^n) \leq -A_n} \Pr_\rho(x^n)\\
    &= \sum_{x^n: S_n(x^n) \leq -A_n} \Pr_\sigma(x^n) \,\exp(S_n(x^n))\\
    &\leq \exp(-A_n) \sum_{x^n: S_n(x^n) \leq -A_n} \Pr_\sigma(x^n)\\
    &\leq \exp(- A_n),
\end{aligned}\end{equation}
where the third line follows from the definition of $S_n$. Analogously, the type II error probability is
\begin{equation}\begin{aligned}
    \beta_n &= \Pr_\sigma(S_n \geq B_n)\\
    &= \sum_{x^n: S_n(x^n) \leq -A_n} \Pr_\rho(x^n) \,\exp(- S_n(x^n))\\
    &\leq \exp(- B_n).
\end{aligned}\end{equation}
It thus remains to show that the probability of obtaining an inconclusive outcome is asymptotically vanishing, which is the main difficulty of the proof.

\subsubsection{Probability of conclusiveness}

We write
$\EE_\rho[Z_{j} \,|\, X_1, \ldots, X_{j-1}]$
for the conditional expectation of $Z_j$ conditioned on the $j-1$ prior observations under hypothesis $\rho$. 
A sequence of random variables $(M_k)_{k=0}^n$ is a \deff{martingale} w.r.t.\ $(X_1, \ldots, X_n)$ if
\begin{equation}\begin{aligned}
    \EE_\rho[M_{k} \,|\, X_1, \ldots, X_{k-1}] = M_{k-1} \qquad \forall k \in \{ 1 , \ldots, n \}.
\end{aligned}\end{equation}
The key property of martingales that we will use is the Azuma--Hoeffding inequality (see e.g.~\cite[Chapter~7]{alon_1992}), which says that if the given martingale has differences bounded by a constant $C$, i.e.\ $|M_k - M_{k-1}| \leq C$ for all $k$, then $\Pr(|M_k - M_0| \geq t ) \leq  2 \exp\left( - \frac{t^2 \log e}{2 k C^2}\right)$. 

We would like to apply this to understand what happens to the sequence $\frac1n S_n$ for large $n$. It is not difficult to realise, however, that this sequence does not define a martingale. Consider then the so-called Doob decomposition as
\begin{equation}\begin{aligned}\label{eq:doob}
    S_k = M_k + D_k , \qquad M_k \coloneqq S_k - \sum_{i=1}^k \EE_\rho [ Z_i \,|\, X_1, \ldots, X_{i-1}], \qquad D_k \coloneqq \sum_{i=1}^k \EE_\rho [ Z_i \,|\, X_1, \ldots, X_{i-1}].
\end{aligned}\end{equation}
The sequence $(M_k)_k$ here is indeed a martingale since
\begin{equation}\begin{aligned}
    \EE_\rho[M_k\,|\, X_1, \ldots, X_{k-1}] &= S_{k-1} + \EE_\rho[Z_k\,|\, X_1, \ldots, X_{k-1}] - \sum_{i=1}^{k}\EE_\rho[Z_i\,|\, X_1, \ldots, X_{i-1}] \\
    &= S_{k-1} - \sum_{i=1}^{k-1}\EE_\rho[Z_i\,|\, X_1, \ldots, X_{i-1}]\\
    &= M_{k-1}.
\end{aligned}\end{equation}
Since we are dealing with full-rank states on a finite-dimensional space, we have $|Z_k| \leq L$ for some constant $L$, and hence the martingale has bounded differences as $|M_k - M_{k-1}| = \left|Z_k - \EE_\rho[Z_k\,|\,X_1, \ldots, X_{k-1}]\right| \leq 2L$. The Azuma--Hoeffding inequality then gives
\begin{equation}\begin{aligned}
    \Pr_\rho(|M_n| \geq n \ve ) \leq 2 \exp\left( - n \frac{\ve^2 \log e}{8 L^2} \right),
\end{aligned}\end{equation}
immediately implying that $\frac1n M_n$ tends to zero in probability under $\rho$, which we denote $\frac1n M_n \tendsp{\rho} 0$. The argument for probability under $\sigma$ is completely analogous.

The difficulty in characterising the asymptotics of $\frac1n S_n$ thus lies purely in understanding the behaviour of the variable $D_n$ in the Doob decomposition~\eqref{eq:doob}, which governs the drift of the random walk associated with the adaptive testing strategy.
We investigate it in the lemma below.

\begin{boxed}
\begin{lemma}\label{lem:lemma}
It holds that
\begin{equation}\begin{aligned}
   \frac1n  D_n  \tendsp{\rho} D_\MM(\rho \| \sigma), \qquad \frac1n  D_n  \tendsp{\sigma}  - D_\MM(\sigma \| \rho),
   \end{aligned}\end{equation}
   and hence
\begin{equation}\begin{aligned}
     \frac1n S_n \tendsp{\rho} D_\MM(\rho \| \sigma), \qquad \frac1n S_n \tendsp{\sigma} - D_\MM(\sigma \| \rho).
\end{aligned}\end{equation}
\end{lemma}
\end{boxed}
\begin{proof}
We will consider the case of $\rho$, with the derivation for $\sigma$ following analogously.
Start by noticing that if $S_{k-1} \geq 0$, then $\M_k = \M_\rho$ by definition. So
\begin{equation}\begin{aligned}
    \EE_\rho[ Z_k \,|\, X_1, \ldots, X_{k-1},\; S_{k-1} \geq 0] = \sum_x \Pr_{\rho,\M_\rho}(x) \log \frac{\Pr_{\rho,\M_\rho}(x)}{\Pr_{\sigma,\M_\rho}(x)} = D_\MM(\rho \| \sigma).
\end{aligned}\end{equation}
Together with the analogous property when $S_{k-1} < 0$, we have
\begin{equation}\begin{aligned}\label{eq:expected}
     \EE_\rho[ Z_k \,|\, X_1, \ldots, X_{k-1}] = \begin{cases}
D_\MM(\rho\|\sigma) & \text{if } S_{k-1} \geq 0 \\
D(\M_\sigma(\rho)\|\M_\sigma(\sigma)) & \text{if } S_{k-1} < 0.
\end{cases}
\end{aligned}\end{equation}
The second term here does not spark joy, but luckily we will now show that asymptotically only the first term survives.

Take some $\lambda > 0$ and consider the moment generating function of $Z_k$ conditioned on the $k-1$ prior outcomes, namely 
\begin{equation}\begin{aligned}
   f(\lambda) \coloneqq \EE_\rho [ \exp(- \lambda Z_k) \,|\,  X_1, \ldots, X_{k-1} ] = \sum_x \Pr_{\rho,\M_k}(x)^{1-\lambda} \Pr_{\sigma,\M_k}(x)^\lambda.
\end{aligned}\end{equation}
We then see that $f(0) = 1$ while $f'(0) = -\EE_\rho [ Z_k \,|\,  X_1, \ldots, X_{k-1}] = - D(\M_k(\rho)\|\M_k(\sigma)) < 0$,  regardless of whether $\M_k = \M_\rho$ or $\M_k = \M_\sigma$. By continuity, $f(\lambda) \leq c$ for some $c <1$ and sufficiently small $\lambda > 0$.  
Now, the law of iterated expectation (tower property) says that $\EE[\EE[X | Y]] = \EE[X]$, which here tells us that
\begin{equation}\begin{aligned}\label{eq:tower}
   \EE_\rho [ \exp(-\lambda S_k)] &=  \EE_\rho[ \EE_\rho [ \exp(-\lambda S_{k-1})\, \exp(-\lambda Z_k) \,|\, X_1, \ldots, X_{k-1}] ]\\
   &= \EE_\rho[ \exp(- \lambda S_{k-1}) \, \EE_\rho [ \exp(-\lambda Z_k) \,|\, X_1, \ldots, X_{k-1}]]\\
   &\leq \EE_\rho [ \exp(- \lambda S_{k-1})]\, c,
\end{aligned}\end{equation}
where we used that $S_{k-1}$ is completely determined once we condition on $X_1, \ldots, X_{k-1}$. Iterating, we have
\begin{equation}\begin{aligned}
     \EE_\rho [ \exp(-\lambda S_k)] \leq c^k.
\end{aligned}\end{equation}
By Markov's inequality,
\begin{equation}\begin{aligned}
    \Pr_\rho(S_k < 0) = \Pr_\rho (  \exp(-\lambda S_k) > 1 ) \leq \EE_\rho [\exp(-\lambda S_k)] \leq c^k.
\end{aligned}\end{equation}
Because of this,
\begin{equation}\begin{aligned}
    \sum_{i=1}^n \Pr_\rho(S_{i-1} < 0) \leq \sum_{i=1}^n c^{i-1} < \frac{1}{1-c}.
\end{aligned}\end{equation}

To conclude, let us introduce the shorthand notation $\boldsymbol{1}_n^{-} = \sum_{i=1}^{n} \boldsymbol{1}[\{S_{i-1} < 0\}]$.
Using~\eqref{eq:expected}, we can write
\begin{equation}\begin{aligned}
    D_n &=   D_\MM(\rho\|\sigma) \, \left(n - \boldsymbol{1}_n^{-}\right)\, + D(\M_\sigma(\rho)\|\M_\sigma(\sigma)) \,\boldsymbol{1}_n^{-}.
\end{aligned}\end{equation}
But since
\begin{equation}\begin{aligned}
    \EE_\rho \left[ \boldsymbol{1}_n^{-} \right] = \sum_{i=1}^{n} \Pr_\rho(S_{i-1} < 0) \leq \frac{1}{1-c},
\end{aligned}\end{equation}
Markov's inequality gives
\begin{equation}\begin{aligned}
    \Pr_\rho\left( \boldsymbol{1}_n^{-} \geq n \ve \right) \leq \frac{1}{(1-c) n \ve}
\end{aligned}\end{equation}
and hence that $\frac1n \boldsymbol{1}_n^{-} \to 0$, implying that $\frac1n D_n \to D_\MM(\rho\|\sigma)$ in probability under $\rho$.
\end{proof}

We are now ready to prove the main statement.

\begin{proof}[{Proof of Lemma~\ref{lem:achievability_sequential}}]
As we already showed in Section~\ref{app:seq_analysis1}, the adaptive protocol defined in Section~\ref{sec:protocol_definition} satisfies
\begin{equation}\begin{aligned}
     \liminf_{n\to\infty} -\frac1n \log \palpha_n  &\geq\liminf_{n\to\infty} \frac1n A_n = D_\MM(\sigma \| \rho) - \ve,\\
     \liminf_{n\to\infty} -\frac1n \log \pbeta_n  &\geq \liminf_{n\to\infty}\frac1n B_n = D_\MM(\rho\|\sigma) - \ve.
 \end{aligned}\end{equation}
Now, it clearly holds that
\begin{equation}\begin{aligned}
    D_\MM(\rho \| \sigma) > D_\MM(\rho\|\sigma) - \ve = \frac1n B_n,
\end{aligned}\end{equation}
and since from Lemma~\ref{lem:lemma} we now know that $\frac1n S_n \tendsn D_\MM(\rho \| \sigma)$ in probability under $\rho$, this implies that $\Pr_\rho (S_n < B_n) \tendsn 0$. The probability of an inconclusive outcome thus satisfies
\begin{equation}\begin{aligned}
  \lim_{n\to\infty} \pi_n (\rho) =  \lim_{n\to\infty} \left( 1 - \Pr_\rho (- A_ n < S_n < B_n ) \right) \geq  \lim_{n\to\infty} \left( 1 - \Pr_\rho ( S_n < B_n ) \right) = 1,
\end{aligned}\end{equation}
which is what was to be shown. An analogous argument shows that $1 - \pi_n(\sigma) = \Pr_\sigma (- A_ n < S_n < B_n ) \leq \Pr_\sigma (- A_ n < S_n ) \tendsn 0$.
\end{proof}


\end{document}